\documentclass[acmsmall,nonacm]{acmart}
\usepackage{graphicx}
\usepackage{amsthm}
\usepackage{longtable}

\AtBeginDocument{%
  }

\setcopyright{acmlicensed}
\copyrightyear{2025}
\acmYear{2025}
\acmJournal{JACM}

\begin{document}

\title{Hardware Acceleration for Neural Networks: A Comprehensive Survey}

% (Optional) Add authors back later if needed:
% \author{...}
\author{Bin Xu}
\orcid{0009-0001-2639-7283}
\affiliation{%
  \institution{School of Electrical, Computer and Energy Engineering, Arizona State University}
  \city{Tempe}
  \state{Arizona}
  \country{USA}
}
\email{binxu4@asu.edu}

\author{Ayan Banerjee}
\affiliation{%
  \institution{School of Computing and Augmented Intelligence, Arizona State University}
  \city{Tempe}
  \country{USA}}
\email{abanerj3@asu.edu}

\author{Sandeep Gupta}
\affiliation{%
  \institution{School of Computing and Augmented Intelligence, Arizona State University}
  \city{Tempe}
  \country{USA}
}
\email{Sandeep.Gupta@asu.edu}

\begin{abstract}
Neural networks have become a dominant computational workload across cloud and edge platforms, but their rapid growth in model size and deployment diversity has exposed hardware bottlenecks that are increasingly dominated by memory movement, communication, and irregular operators rather than peak arithmetic throughput. This survey reviews the current technology landscape for hardware acceleration of deep learning, spanning Graphics Processing Units (GPUs) and tensor-core architectures, domain-specific accelerators (e.g., Tensor Processing Units (TPUs)/Neural Processing Units (NPUs)), Field-Programmable Gate Array (FPGA)-based designs, Application-Specific Integrated Circuit (ASIC) inference engines, and emerging Large Language Model (LLM)-serving accelerators such as Language Processing Units (LPUs), alongside in-/near-memory computing and neuromorphic/analog approaches. We organize the survey using a unified taxonomy across (i) workloads (Convolutional Neural Networks (CNNs), Recurrent Neural Networks (RNNs), Graph Neural Networks (GNNs), Transformers/Large Language Models (LLMs)), (ii) execution settings (training vs.\ inference; datacenter vs.\ edge), and (iii) optimization levers (reduced precision, sparsity and pruning, operator fusion, compilation and scheduling, and memory-system/interconnect design). We synthesize key architectural ideas such as systolic arrays, vector and Single Instruction, Multiple Data (SIMD) engines, specialized attention and softmax kernels, quantization-aware datapaths, and high-bandwidth memory, and we discuss how software stacks and compilers bridge model semantics to hardware. Finally, we highlight open challenges—including efficient long-context LLM inference (Key-Value (KV)-cache management), robust support for dynamic and sparse workloads, energy- and security-aware deployment, and fair benchmarking—pointing to promising directions for the next generation of neural acceleration.
\end{abstract}


\maketitle

\section{Introduction}

\begin{figure}[t]
\centering
\includegraphics[width=0.9\linewidth]{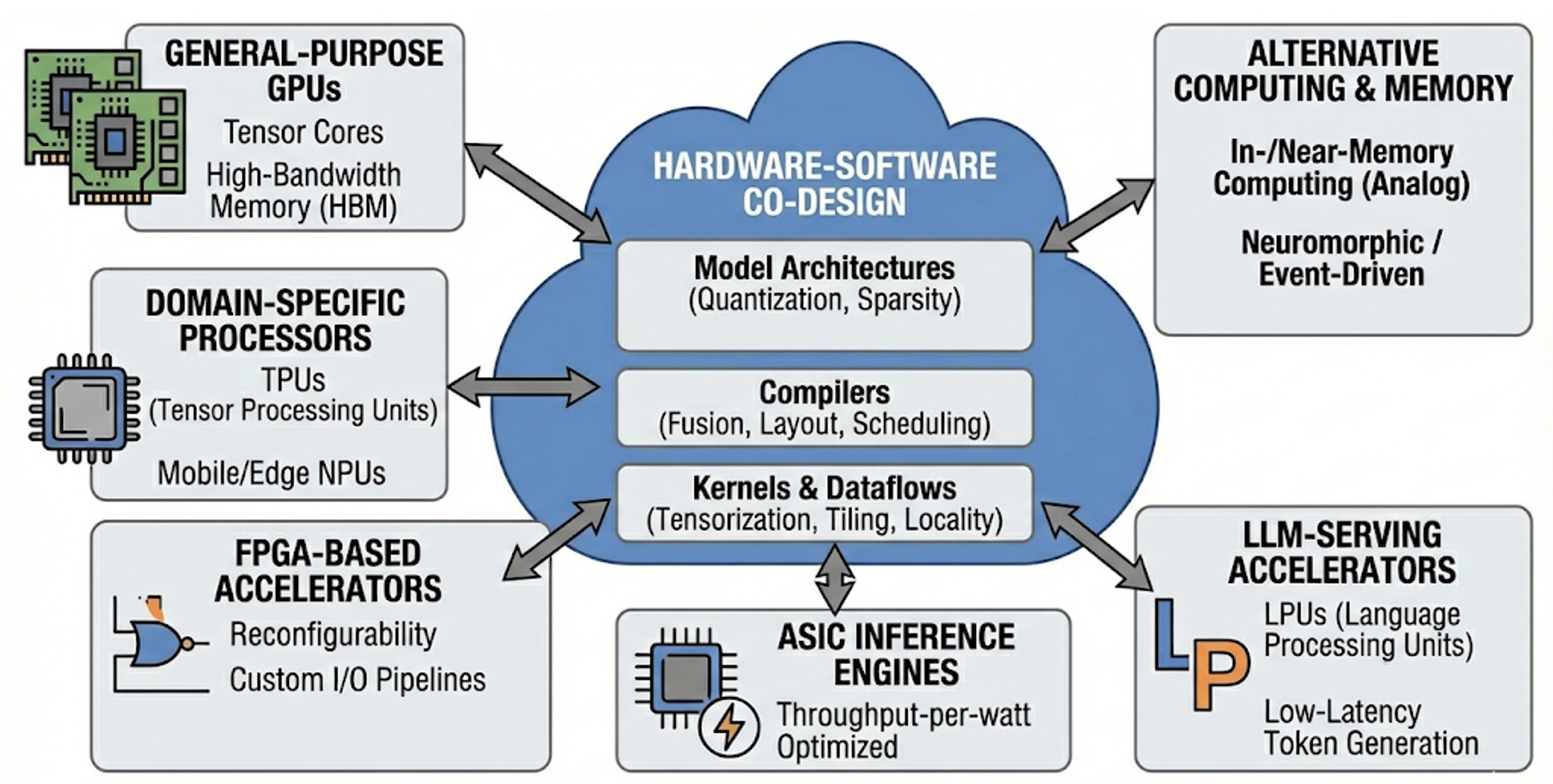}
\caption{High-level overview of the hardware acceleration landscape, illustrating the spectrum from general-purpose GPUs to domain-specific TPUs, FPGAs, and ASICs, along with the interaction between compute datapaths, memory hierarchies, and the software stack.}
\label{fig:section1}
\end{figure}

Deep neural networks now underpin a wide range of applications—from perception and control in autonomous and cyber-physical systems \cite{xu2025model,xu2025enabling} to search, recommendation, and natural language assistants \cite{krizhevsky2012imagenet,vaswani2017attention}. At the same time, modern models have expanded rapidly in scale (parameters, activation sizes, and context lengths), diversity (Convolutional Neural Networks (CNNs), Graph Neural Networks (GNNs), Transformers, diffusion models), and deployment environments (datacenter training clusters, latency-critical edge inference, and battery-powered on-device intelligence) \cite{abadi2016tensorflow,gunter2024apple}. These trends have elevated hardware acceleration from a performance optimization to an enabling technology: without specialized compute, memory systems, and software co-design, many state-of-the-art models would be impractical due to cost, latency, or energy constraints \cite{hennessy2019hwds,jouppi2017tpu}. 
% Figure~\ref{fig:section1} provides a high-level overview of the hardware-acceleration landscape covered in this survey.

Why acceleration is harder than “more Floating Point Operations (FLOPs)” is that, although matrix multiplication remains central, end-to-end efficiency is increasingly governed by data movement rather than arithmetic \cite{williams2009roofline}. Training and inference pipelines stress memory capacity and bandwidth (e.g., activation storage and optimizer state during training; Key-Value (KV)-cache and batching dynamics during Large Language Model (LLM) inference) \cite{micikevicius2018mixedprecision,kwon2023vllm}, interconnect bandwidth and collectives for distributed execution, and irregular operators arising from sparsity, dynamic control flow, and Mixture-of-Experts (MoE) routing \cite{gale2019state,shazeer2017outrageously}. As a result, accelerator design must address the full system stack: compute datapaths, memory hierarchies, on-chip networks, off-chip Dynamic Random Access Memory (DRAM)/High-Bandwidth Memory (HBM), host-device interfaces, and multi-device interconnects, together with compilers and runtime systems that map high-level models onto hardware efficiently \cite{abadi2016tensorflow}.

The technology landscape shown in Figure~\ref{fig:section1} spans general-purpose Graphics Processing Units (GPUs) enhanced with tensor cores and high-bandwidth memory; domain-specific processors such as Tensor Processing Units (TPUs) and mobile/edge Neural Processing Units (NPUs) \cite{jouppi2017tpu,jouppi2021tpuv4}; Field-Programmable Gate Array (FPGA)-based accelerators offering reconfigurability and tight integration with custom I/O pipelines \cite{nurvitadhi2017can,umuroglu2017finn,venieris2017fpgaconvnet,boutros2024fpga,yan2024fpga}; Application-Specific Integrated Circuit (ASIC) inference engines optimized for throughput-per-watt \cite{chen2014dadiannao}; and LLM-serving accelerators such as Language Processing Units (LPUs) that target predictable low-latency token generation \cite{groq2024lpu}. Beyond digital Complementary Metal-Oxide-Semiconductor (CMOS) datapaths, in-/near-memory computing and analog approaches attempt to reduce the fundamental cost of memory movement \cite{shafiee2016isaac}, while neuromorphic and event-driven designs explore alternative execution models for sparse, low-power workloads \cite{davies2018loihi}. In practice, the most capable systems emerge from hardware--software co-design: model architectures and training recipes that are amenable to quantization and sparsity \cite{jacob2018quantization,han2016deepcompression}; compilers that perform fusion, layout transformations, and scheduling; and kernels and dataflows that exploit tensorization, tiling, and locality \cite{kung1982systolic,williams2009roofline}.

In this survey, we focus on current technology and architectural principles that generalize across devices and workloads. We structure the discussion along three axes: (i) \emph{workloads} (CNNs, Recurrent Neural Networks (RNNs), GNNs, and Transformers; and key operators such as attention, convolution, normalization, and sampling)
\cite{vaswani2017attention}, (ii) \emph{execution settings} (training vs.\ inference; offline vs.\ online; cloud vs.\ edge), and (iii) \emph{optimization levers} (reduced precision, structured and unstructured sparsity, compression and pruning, operator fusion, compilation and scheduling, and memory/interconnect design) \cite{micikevicius2018mixedprecision,jacob2018quantization,gale2019state,han2016deepcompression,williams2009roofline,wang2024art,li2024comprehensive,fang2024maskllm,muralidharan2024compact}. Rather than treating hardware in isolation, we highlight how these levers interact—for example, how low precision shifts bottlenecks toward memory, how sparsity challenges utilization and load balance, and how long-context inference amplifies memory-capacity constraints \cite{kwon2023vllm}.

The main contributions of this survey are:
\begin{itemize}
  \item a unified taxonomy connecting neural workloads to accelerator architectures and system constraints;
  \item a synthesis of design patterns used in modern accelerators (tensor-core/tensorization, systolic arrays, tiling and dataflow, heterogeneous memory hierarchies, and scalable interconnects);
  \item a discussion of software stacks (frameworks, compilers, and runtimes) that determine real-world performance and portability; and
  \item an overview of open challenges and research directions, including efficient LLM serving (KV-cache and memory management), robust support for dynamic and sparse models, energy-aware edge deployment, and reproducible benchmarking.
\end{itemize}

The remainder of the paper reviews the background and workload primitives, surveys accelerator architectures and memory/interconnect designs, summarizes software ecosystems and compilation techniques, and concludes with benchmarking considerations and open problems.

\section{Accelerator Challenges}

\begin{figure}[thbp]
\centering
\includegraphics[width=0.9\linewidth]{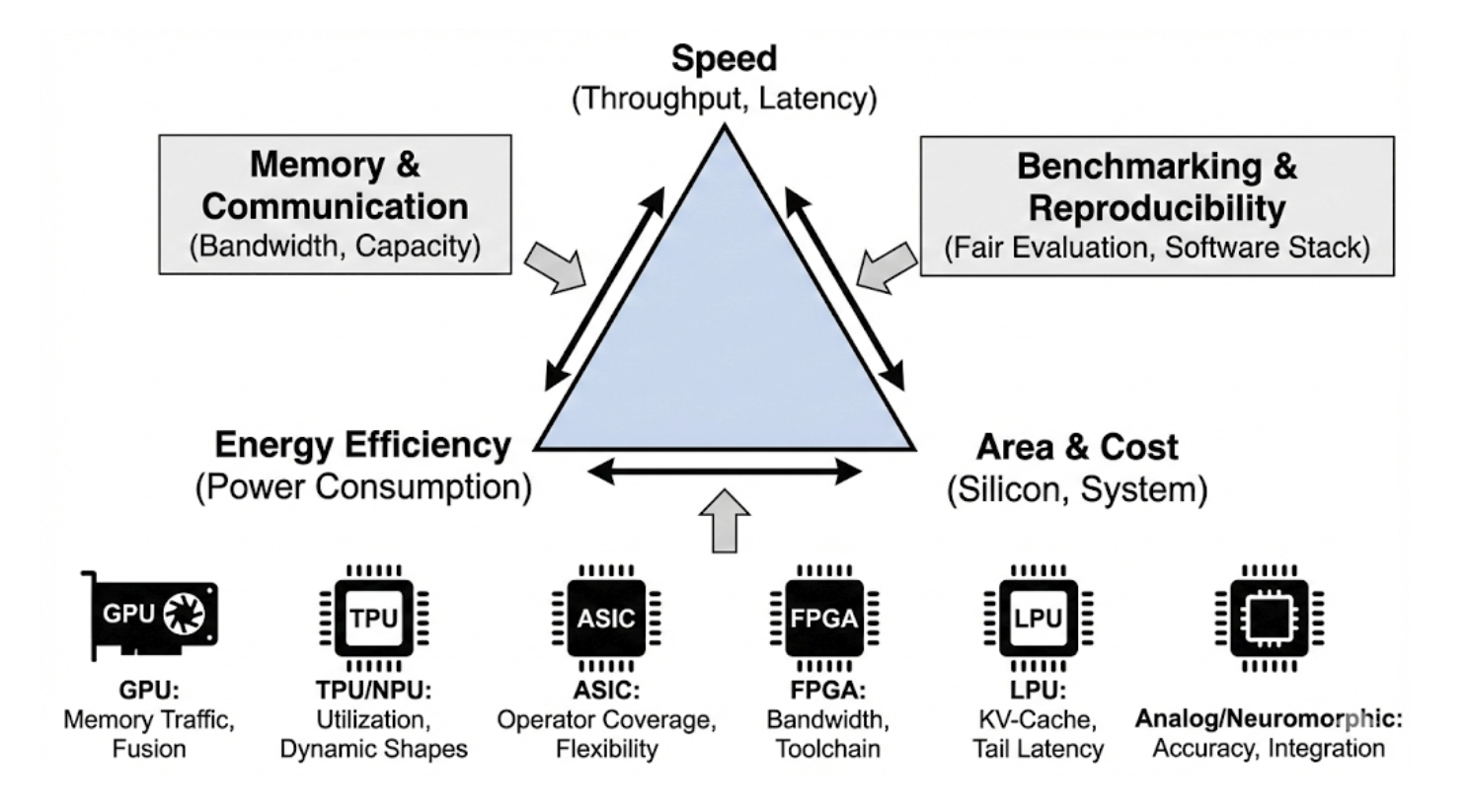}
\caption{Key limitations and trade-offs in hardware acceleration design, highlighting the multi-objective optimization problem involving performance (throughput/latency), energy efficiency, and deployment constraints such as area and cost.}
\label{fig:section2}
\end{figure}

Figure~\ref{fig:section2} summarizes key limitations that shape accelerator design choices across performance, energy, and deployability. Designing and deploying neural-network accelerators requires navigating a multi-objective optimization problem where improvements along one axis often degrade another. In practice, “speed” is not a single number: training emphasizes time-to-train and scalability, while inference emphasizes latency, tail latency under load, and throughput at a target quality. Across both settings, energy and resource budgets constrain what can be deployed, especially when models must run continuously or at the edge \cite{hennessy2019hwds}.

A recurring theme is that peak arithmetic throughput is rarely the limiting factor by itself. End-to-end behavior depends on the balance between compute, memory bandwidth, and communication, and it is often dominated by data movement across memory hierarchies and interconnects \cite{williams2009roofline,wulf1995memorywall}. This is amplified in modern workloads such as Transformers/LLMs, where attention and KV-cache introduce large memory footprints and bandwidth demands that scale with sequence length and concurrency \cite{vaswani2017attention,kwon2023vllm}.

Finally, accelerators are deployed as part of a software and systems stack. Kernel libraries, compilers, and runtimes determine which optimizations are realized in practice, and they influence not only performance but also portability and reproducibility \cite{chetluar2014cudnn,chen2018tvm,tillet2019triton}.

\subsection{Power and energy consumption}
Power consumption is a first-order constraint because it determines thermal limits, battery lifetime, and operating cost. A key observation is that energy is consumed not only by arithmetic, but also by moving data through memory hierarchies and interconnects; for many modern workloads, the energy per byte moved can dominate the energy per operation \cite{horowitz2014energy,williams2009roofline}. This motivates accelerator designs and mappings that reduce off-chip traffic through locality-aware dataflows and larger on-chip buffers \cite{chen2016eyeriss}.

\begin{figure}[thbp]
\centering
\includegraphics[width=0.9\linewidth]{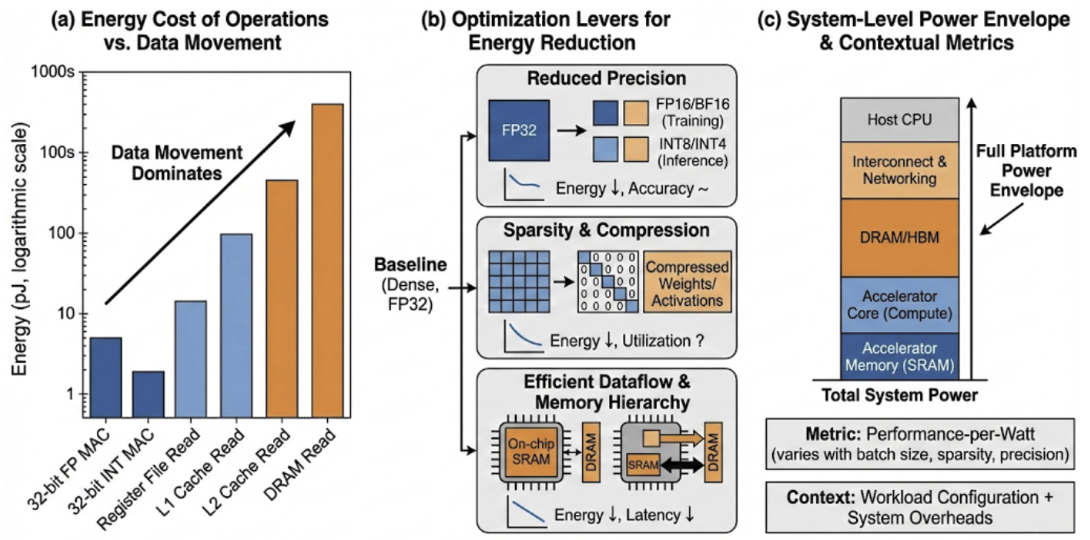}
\caption{Power consumption analysis highlighting (a) the dominance of data movement energy over arithmetic operations, (b) the impact of reduced precision on energy efficiency, and (c) the trade-offs involved in exploiting sparsity.}
\label{fig:section10}
\end{figure}

Figure~\ref{fig:section10} highlights the power-consumption perspective that motivates energy-aware mappings and reduced data movement. Precision and sparsity are primary levers for energy reduction. Mixed-precision training improves throughput and reduces memory traffic while maintaining convergence, typically using reduced-precision compute with higher-precision accumulation and stabilization techniques \cite{micikevicius2018mixedprecision}. For inference, quantization to integer arithmetic can significantly reduce energy and improve effective bandwidth, but it introduces calibration and accuracy trade-offs that depend on the model and deployment target \cite{jacob2018quantization}. Compression and sparsity reduce memory footprint and traffic, but unstructured sparsity can add indexing overhead and reduce utilization if the hardware/software stack is not sparsity-aware \cite{han2016deepcompression,gale2019state}.

Meaningful energy comparisons must therefore be contextual. Performance-per-watt numbers are sensitive to batch size, sequence length, precision, sparsity pattern, and system-level overheads (host CPU work, DRAM/HBM power, and network power). As a result, energy evaluation should report both workload configuration and the full platform power envelope rather than only accelerator core power.

\subsubsection{GPU-specific energy challenges}
On GPUs, energy is frequently dominated by memory traffic rather than arithmetic, especially when arithmetic intensity is low and intermediate tensors are repeatedly materialized \cite{williams2009roofline,horowitz2014energy}. Inference pipelines that lack operator fusion can spend a large fraction of energy moving activations between registers, caches, and off-chip memory, which motivates aggressive fusion in libraries and compiler stacks \cite{chetluar2014cudnn,chen2018tvm,tillet2019triton}. Mixed precision improves throughput-per-watt for training, but it can shift the bottleneck toward memory and communication when scaling across devices \cite{micikevicius2018mixedprecision,shoeybi2019megatronlm}. Sparsity can reduce memory traffic, yet unstructured sparsity may increase metadata and indexing overhead that erodes energy gains without sparsity-aware kernels \cite{gale2019state,han2016deepcompression}.

\subsubsection{TPU/NPU-specific energy challenges}
Tensor processors improve efficiency by specializing dense dataflows and maximizing on-chip reuse, but their energy advantage diminishes when workloads deviate from the assumed shapes or require frequent off-chip traffic \cite{jouppi2017tpu,jouppi2021tpuv4,williams2009roofline}. Attention-heavy workloads and long-context serving stress memory systems via KV-cache and intermediate activations, pushing energy back toward DRAM/HBM movement \cite{vaswani2017attention,kwon2023vllm,horowitz2014energy}. In edge NPUs, power is constrained by System-on-Chip (SoC)-level thermal limits and shared memory bandwidth; as a result, sustained performance-per-watt depends on compiler scheduling, operator coverage, and quantized datapaths \cite{jacob2018quantization,mazumder2021survey}. When sparsity or conditional execution is present (e.g., MoE), energy efficiency can suffer due to reduced reuse and load imbalance unless the stack explicitly supports these patterns \cite{shazeer2017outrageously,gale2019state}.

\subsubsection{ASIC-specific energy challenges}
ASICs can minimize energy by hardwiring data reuse, choosing an energy-efficient dataflow, and provisioning large on-chip Static Random Access Memory (SRAM) to reduce DRAM access \cite{chen2016eyeriss,chen2014dadiannao,horowitz2014energy}. However, energy efficiency can degrade when models include unsupported operators (e.g., attention variants or custom normalization) that force host fallback or additional data marshaling \cite{vaswani2017attention,chen2016eyeriss}. Compression and sparsity reduce weight/activation traffic, but the benefits depend on sparsity structure; irregular sparsity introduces indexing overhead and load imbalance that can negate energy savings \cite{han2016eie,han2016deepcompression,gale2019state}. Designs closer to sensors (e.g., vision-centric accelerators) reduce I/O energy, yet they are specialized and may not generalize to broader model families \cite{liu2015shidiannao}.

\subsubsection{FPGA-specific energy challenges}
FPGAs can be energy-efficient when a streaming pipeline eliminates intermediate DRAM writes and uses custom precision to reduce switching activity and memory traffic \cite{umuroglu2017finn,ma2018optimizing,horowitz2014energy}. In practice, performance-per-watt is sensitive to external memory bandwidth, on-chip Block RAM (BRAM)/Digital Signal Processor (DSP) utilization, and routing overhead; poor mapping can waste energy through underutilized pipelines and excess off-chip transfers \cite{nurvitadhi2017can,venieris2017fpgaconvnet}. Transformer/attention-style workloads further increase bandwidth pressure and can reduce the energy advantage if buffering and fusion are insufficient \cite{FPGATrans,vaswani2017attention}. In cloud FPGA deployments, host-device interfaces and multi-tenant overheads can dominate system energy unless amortized by sustained utilization and carefully engineered serving pipelines \cite{fowers2018brainwave,nurvitadhi2017can}.

\subsubsection{LPU/LLM-serving energy challenges}
For LLM serving, energy is tightly coupled to memory movement from KV-cache and to serving policy (batching, concurrency, and context length), not just to matmul efficiency \cite{kwon2023vllm,vaswani2017attention,horowitz2014energy}. IO-aware attention kernels reduce memory traffic and can improve energy per token, but they must be integrated with runtime memory management to realize system-level gains under concurrency \cite{dao2022flashattention,kwon2023vllm}. LPU-style designs can reduce control and dispatch overheads and target predictable execution, yet they still face the fundamental energy cost of sustaining high bandwidth per generated token \cite{groq2024lpu,williams2009roofline}. Heterogeneous serving (MoE, tool-calling) can further increase energy variability and complicate power provisioning due to conditional execution and load imbalance \cite{shazeer2017outrageously,gale2019state}.

\subsubsection{In-/near-memory and analog energy challenges}
Analog in-/near-memory approaches reduce data movement for matmul-heavy layers by performing multiply-accumulate in crossbars, but Analog-to-Digital Converter (ADC)/Digital-to-Analog Converter (DAC) overheads, calibration, and device non-idealities can offset energy benefits at the system level \cite{shafiee2016isaac,chi2016prime,anzaroot2019puma,horowitz2014energy}. Energy efficiency is sensitive to mapping: if only a fraction of the model maps to crossbars, the overall system can still be dominated by digital compute and data conversion costs \cite{anzaroot2019puma}. Attention-heavy models introduce softmax and KV-cache access patterns that are not naturally accelerated by crossbar matmul and can erode end-to-end savings \cite{vaswani2017attention,kwon2023vllm}. Consequently, practical designs often require heterogeneous integration and careful partitioning between analog and digital components \cite{hennessy2019hwds}.

\subsubsection{Neuromorphic energy challenges}
Neuromorphic systems can achieve extremely low energy when activity is sparse and event-driven computation matches the workload, but benefits are sensitive to encoding/decoding overheads and to whether sparsity is realized in practice \cite{davies2018loihi,merolla2014truenorth}. When spike rates increase, communication and routing energy rises, and the event-driven advantage diminishes \cite{davies2018loihi}. Mapping conventional dense networks to spiking representations can add conversion overhead and reduce effective sparsity, making comparisons with dense accelerators sensitive to modeling assumptions \cite{gale2019state,coleman2019dawnbench}. As a result, neuromorphic energy wins are most robust in always-on sensing and temporal workloads where sparse events are intrinsic to the data \cite{merolla2014truenorth}.

\subsection{Throughput, latency, and speed}
Throughput and latency requirements impose different architectural and scheduling pressures. Datacenter inference serving may prioritize tokens-per-second at high utilization while still meeting strict end-to-end latency targets, whereas interactive applications care about single-request latency and worst-case behavior (often measured as p95/p99 latency). Achieving high throughput typically benefits from large batches and deep pipelining, but batching can increase latency and memory footprint, particularly for Transformers where KV-cache grows with sequence length and active requests \cite{vaswani2017attention,kwon2023vllm}.

\begin{figure}[thbp]
\centering
\includegraphics[width=0.9\linewidth]{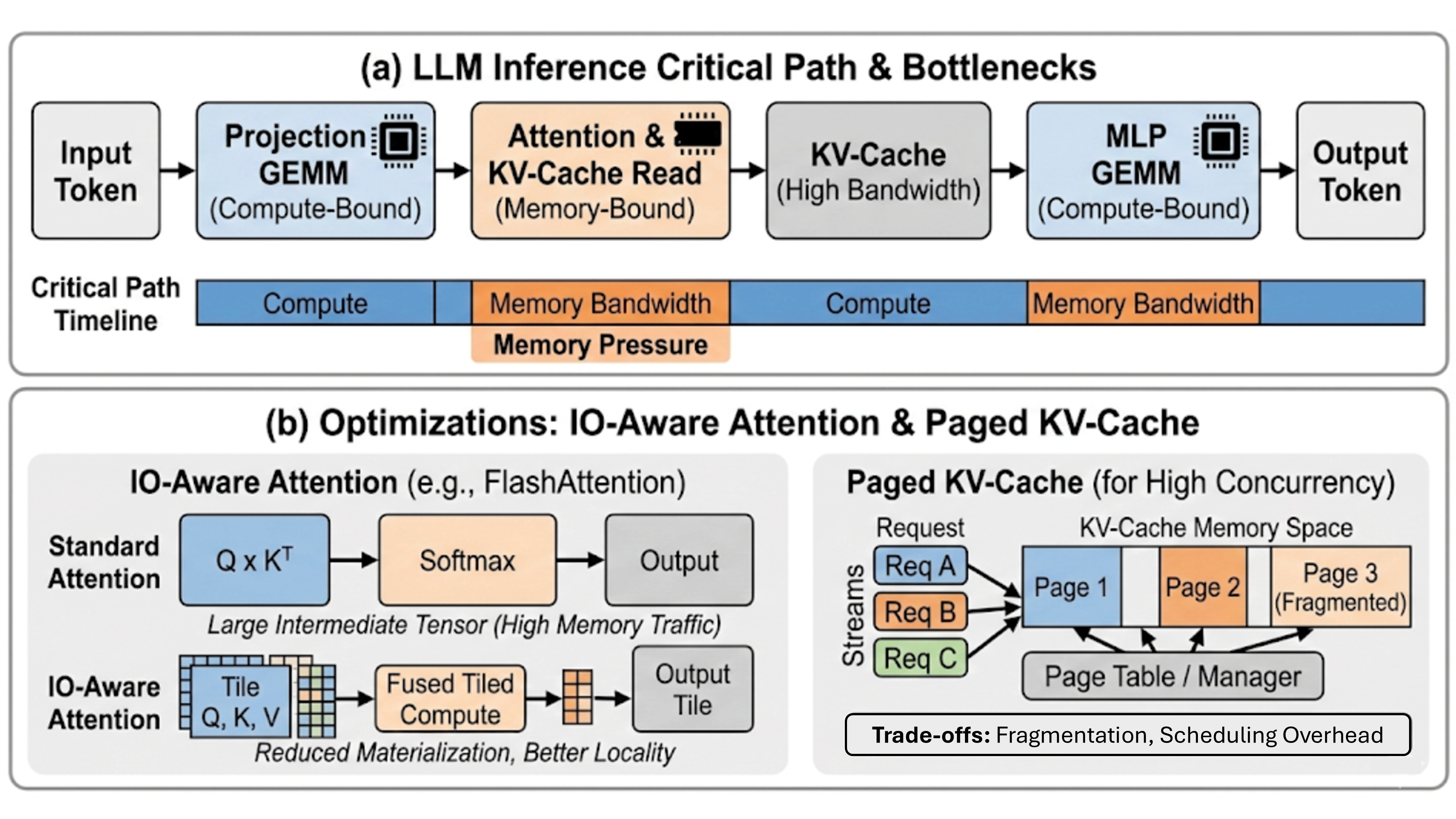}
\caption{LLM inference bottlenecks and optimization strategies, differentiating between the compute-bound prefill phase and the memory-bandwidth-bound decode phase, and illustrating techniques like KV-cache management, paging, and attention optimization.}
\label{fig:section9}
\end{figure}

For LLM inference, the critical path often alternates between compute-heavy projection/Multilayer Perceptron (MLP) General Matrix Multiplications (GEMMs) and bandwidth-heavy attention/KV-cache reads, as illustrated in Figure~\ref{fig:section9}. Input/Output (IO)-aware attention kernels reduce memory traffic by tiling attention and avoiding materialization of large intermediate tensors, improving both throughput and latency under memory pressure \cite{dao2022flashattention}. Runtime techniques such as paging KV-cache help maintain throughput at high concurrency, but they introduce their own trade-offs in fragmentation, scheduling overhead, and tail latency \cite{kwon2023vllm}.

At the software level, kernel fusion and code generation reduce launch overhead and improve locality, which matters especially for small GEMMs and latency-sensitive inference. Libraries and compiler stacks (e.g., vendor primitives, graph compilers, and kernel Domain-Specific Languages (DSLs)) play a central role in realizing these optimizations in practice \cite{chetluar2014cudnn,chen2018tvm,tillet2019triton}. Consequently, performance claims should specify not only the hardware but also the kernel/library and compilation path used.

\subsubsection{GPU-specific throughput/latency challenges}
GPUs excel at throughput via batching, but tail latency can suffer when batching increases queueing delay and when small kernels incur launch overhead or synchronization \cite{williams2009roofline,chetluar2014cudnn}. Latency is also sensitive to kernel fusion and code generation quality; insufficient fusion increases memory traffic and amplifies launch overhead, especially for small GEMMs and normalization/activation chains \cite{chen2018tvm,tillet2019triton}. For LLMs, efficient attention kernels reduce IO pressure, but predictable latency under concurrency requires careful KV-cache management and paging policies \cite{dao2022flashattention,kwon2023vllm}. Finally, sparsity and conditional execution can introduce latency variance due to load imbalance and irregular memory access \cite{gale2019state,shazeer2017outrageously}.

Achieving consistent p99 latency on GPUs is further complicated by the ``bulk synchronous'' programming model: a single long-running kernel can block other streams, making it hard to multiplex small, latency-critical tasks with background throughput-oriented work. Context switching on GPUs is relatively expensive, so preemption strategies for real-time serving are limited compared to CPUs. This forces serving systems to rely on cooperative scheduling (e.g., careful batching windows) or multi-instance GPU (MIG) partitioning to isolate workloads, though MIG partitions compute and memory statically rather than dynamically \cite{nvidia2023llm_inference_platforms,williams2009roofline}.

\subsubsection{TPU/NPU-specific throughput/latency challenges}
Tensor processors often achieve high throughput on static dense kernels, but latency can degrade when shapes are dynamic or when compilation and layout choices are not well matched to the input distribution \cite{jouppi2017tpu,jouppi2021tpuv4}. Attention-heavy inference introduces bandwidth-heavy phases and intermediate tensors that may not fit on-chip buffers, increasing latency variability unless the compiler and runtime manage tiling and memory explicitly \cite{vaswani2017attention,kwon2023vllm,dao2022flashattention}. Edge NPUs frequently prioritize deterministic latency, but they are constrained by shared memory bandwidth and quantized operator support; when an operator falls back to the CPU/GPU, latency can spike \cite{jacob2018quantization,mazumder2021survey}. As model architectures evolve, maintaining predictable latency becomes a joint problem of operator coverage, compiler scheduling, and system integration \cite{hennessy2019hwds}.

A key trade-off for TPU/NPU architectures is the ``compile-time vs. run-time'' balance. Relying on static compilation provides excellent determinism and dense packing for fixed graphs, but it can lead to severe latency penalties if dynamic shapes force recompilation or padding. For variable-length sequences (common in LLM serving), the system must either pad to the worst case (wasting throughput) or use specialized dynamic-shape compilers that may generate less optimal code. Managing this trade-off requires sophisticated runtime systems that can bucket requests by size or switch between pre-compiled binaries on the fly, adding complexity to the serving stack \cite{jouppi2021tpuv4,kwon2023vllm}.

\subsubsection{ASIC-specific throughput/latency challenges}
ASICs can deliver low, predictable latency when the operator set is fixed and dataflow is streamed, but they may struggle with emerging operators (e.g., attention variants) and dynamic control, which can force inefficient fallbacks or additional data marshaling \cite{chen2016eyeriss,vaswani2017attention}. Even for supported kernels, latency depends on whether activations and weights fit on-chip buffers; otherwise off-chip bandwidth dominates and can introduce jitter \cite{williams2009roofline,chen2016eyeriss}. Sparsity-aware ASICs can improve throughput, yet irregular sparsity can increase latency variability due to load imbalance and metadata handling \cite{han2016eie,gale2019state}. In practice, achieving low tail latency often requires constraining the model/operator set or providing sufficient programmability to handle evolving kernels \cite{hennessy2019hwds}.

Furthermore, the ``throughput at what cost'' question is acute for ASICs: maximizing peak Tera Operations Per Second (TOPS) often leads to large systolic arrays that have high startup latency and poor efficiency for small batches (batch-1 inference). To target low-latency serving, designers may split resources into multiple smaller cores or use finer-grained pipelining, but this complicates synchronization and on-chip interconnect design. If the workload is memory-bound (e.g., decode phase of LLMs), the ASIC’s peak compute advantage becomes irrelevant, and latency is governed solely by the available HBM bandwidth and the efficiency of the memory controller \cite{groq2024lpu,williams2009roofline}.

\subsubsection{FPGA-specific throughput/latency challenges}
FPGAs can provide strong latency determinism through streaming pipelines, but achieving high throughput while maintaining low latency depends on memory bandwidth, pipeline depth, and host I/O overhead \cite{nurvitadhi2017can,fowers2018brainwave}. Designs that require frequent external DRAM access can lose determinism and become bandwidth-limited, especially for large feature maps or long sequences \cite{williams2009roofline}. Transformer inference on FPGAs is particularly sensitive to buffering and fusion decisions because attention amplifies bandwidth pressure and increases intermediate tensor volume \cite{FPGATrans,vaswani2017attention}. In multi-tenant deployments, additional latency variability can come from batching and scheduling at the service layer rather than from the FPGA datapath itself \cite{fowers2018brainwave}.

However, achieving this determinism requires resolving the ``routing vs. logic'' frequency wall. Deeply pipelined designs can run at high clock speeds, but complex control logic or irregular memory access patterns (e.g., from sparse attention) can create routing congestion that lowers the achievable frequency, directly hurting throughput. Moreover, if the model does not fit entirely on-chip, the FPGA must buffer and swap weights/activations; if this swapping is not perfectly hidden by compute (double buffering), latency spikes occur. Modern FPGAs with HBM stacks help alleviate the bandwidth bottleneck, but maximizing their effective bandwidth requires wide data paths and careful memory controller tuning, which complicates the High-Level Synthesis (HLS)/Register Transfer Level (RTL) design effort \cite{nurvitadhi2017can,li2024hlstransform}.

\subsubsection{LPU/LLM-serving throughput/latency challenges}
LLM-serving accelerators must optimize for end-to-end token latency and throughput simultaneously under variable request patterns; paging and scheduling policies can dominate tail latency even when compute kernels are efficient \cite{kwon2023vllm}. Attention kernels and KV-cache access create a mix of compute- and bandwidth-bound phases that are difficult to keep balanced under fluctuating concurrency \cite{vaswani2017attention,dao2022flashattention}. LPU-style designs aim to reduce dispatch overhead and improve predictability, but they remain constrained by KV-cache bandwidth and context-length growth, which set a lower bound on token latency at high concurrency \cite{groq2024lpu,kwon2023vllm}. Conditional execution (MoE, tool-calling) can also increase tail latency by introducing load imbalance and unpredictable per-token work \cite{shazeer2017outrageously,gale2019state}.

Ideally, serving systems want \emph{throughput without latency degradation}, but batching inherently trades one for the other. Large batches improve GEMM efficiency and aggregate bandwidth utilization, but they increase the time each token spends waiting in queues or for other sequences to finish their step. This ``batching capability gap'' is where specialized architectures try to innovate: by managing control flow and dependency tracking in hardware (rather than GPU kernels + CPU driver), LPUs attempt to make fine-grained interleaved execution feasible, allowing high concurrency with lower queuing delays. Even so, the physical reality of moving KV-cache bytes means that scaling context length inevitably hurts latency unless memory bandwidth scales proportionally \cite{groq2024lpu,williams2009roofline}.

\subsubsection{In-/near-memory and analog throughput/latency challenges}
Analog accelerators can offer high throughput for large matmuls, but end-to-end latency can be limited by conversion overheads (ADC/DAC), calibration, and handling of non-matmul operators \cite{shafiee2016isaac,anzaroot2019puma}. Latency is sensitive to the size and batching of matmuls: small or irregular GEMMs underutilize crossbars and amplify peripheral overheads \cite{anzaroot2019puma}. Workloads with attention and dynamic control reduce the fraction of time spent in the matmul kernels where analog speedups apply and add additional latency for softmax and memory management \cite{vaswani2017attention,kwon2023vllm}. As a result, analog accelerators are often most compelling in pipelines where large dense layers dominate and where conversion can be amortized \cite{chi2016prime}.

A deeper latency challenge stems from the dataflow between analog and digital domains. If a network requires frequent normalization or activation functions that are best done digitally, data must be repeatedly converted and moved, killing the latency benefits of in-place compute. Pipelining these stages can hide throughput costs but increases pipeline depth (latency). Furthermore, writing weights to analog arrays (e.g., Resistive RAM (ReRAM) or Flash) is often slow and energy-intensive; this makes analog approaches less suitable for workloads requiring rapid model switching or dynamic LoRA adapters, restricting their ``speed'' advantage to static-weight inference scenarios \cite{shafiee2016isaac,kim2025hpim}.

\subsubsection{Neuromorphic throughput/latency challenges}
Neuromorphic devices can achieve low latency for event-driven sensing when spikes are sparse, but throughput and latency advantages diminish as activity increases or when dense preprocessing is needed \cite{davies2018loihi,merolla2014truenorth}. Latency can be dominated by encoding/decoding between frame-based signals and spikes, and by routing congestion when event rates increase \cite{davies2018loihi}. Mapping conventional models can add conversion overhead and reduce determinism, making end-to-end latency highly workload-dependent \cite{gale2019state}. Consequently, neuromorphic advantages are most visible in tasks where event-driven representations are natural and where computation is sparse by design \cite{merolla2014truenorth}.

Throughput in neuromorphic systems is also non-standard: it is often measured in ``events per second'' or ``inferences per second'' under a specific sparsity assumption. If the input data becomes dense (e.g., a complex visual scene), the event traffic can saturate the on-chip interconnect, causing spikes to be dropped or delayed, which degrades both accuracy and latency. This traffic-dependent latency profile makes it hard to guarantee real-time bounds compared to the predictable scan-line processing of a traditional accelerator. Thus, achieving high speed requires not just fast silicon but also algorithms that actively maintain sparsity and locality throughout the network execution \cite{davies2018loihi,coleman2019dawnbench}.

\subsection{Area and cost}
Area and cost constraints shape what can be built and deployed at scale. Silicon area (in \(\mathrm{mm}^2\)) is consumed by compute arrays, on-chip SRAM, Network-on-Chip (NoC) resources, and I/O Physical Layers (PHYs), and each choice impacts achievable frequency and yield. Increasing on-chip memory can reduce off-chip traffic and improve performance-per-watt, but it increases die area and may limit clock rate; conversely, relying on off-chip memory shifts the bottleneck toward bandwidth and increases energy \cite{williams2009roofline,horowitz2014energy}.

\begin{figure}[thbp]
\centering
\includegraphics[width=0.9\linewidth]{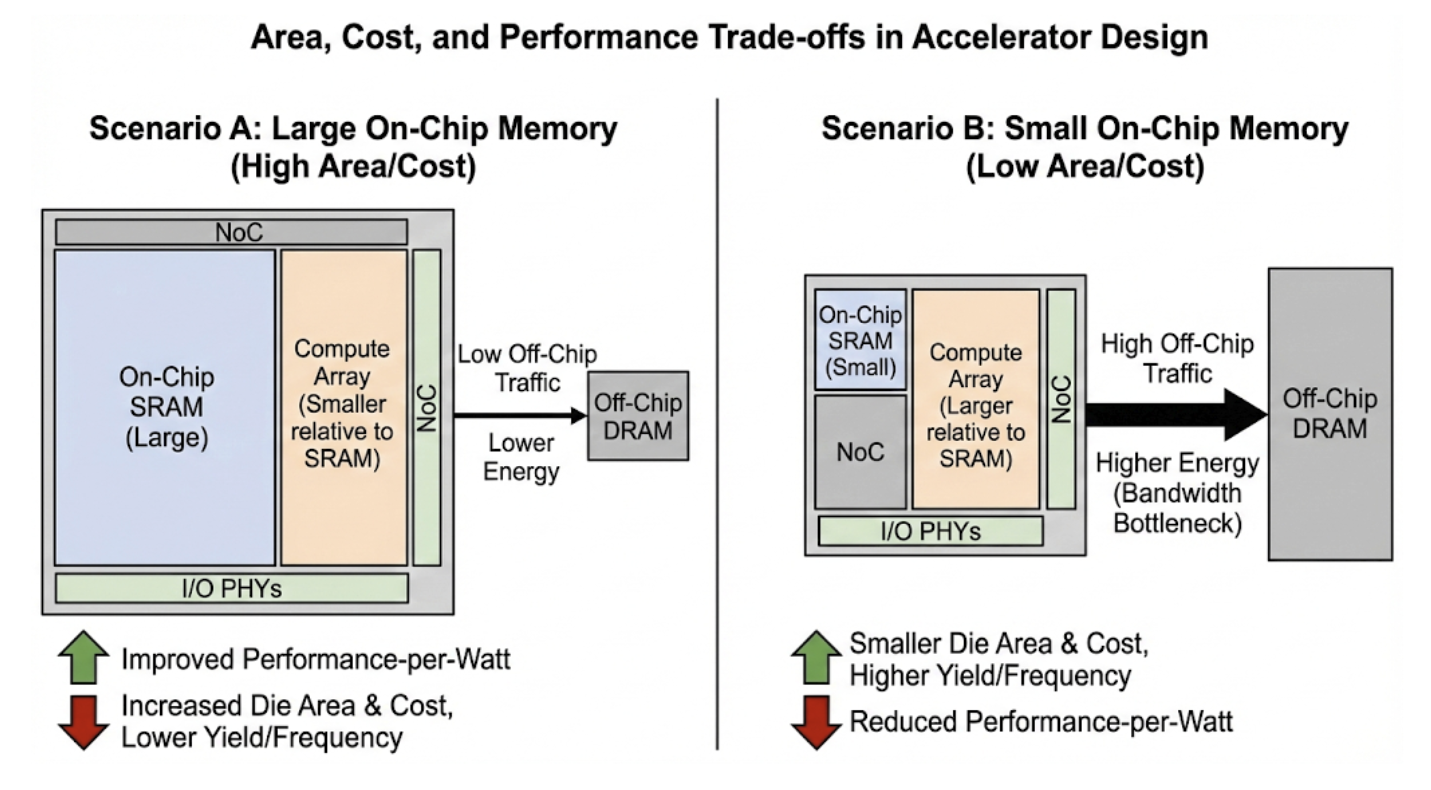}
\caption{The fundamental trade-off between performance and silicon area/cost, illustrating how increasing parallelism and memory capacity improves throughput but raises manufacturing and packaging costs.}
\label{fig:section11}
\end{figure}

At the system level, packaging technologies and memory stacks (e.g., HBM) can improve bandwidth but raise cost and power, which affects total cost of ownership and practical adoption; Figure~\ref{fig:section11} illustrates the resulting performance--area trade-off. The “right” design point differs across environments: edge SoCs optimize for unit cost and battery life, while datacenters optimize for throughput-per-dollar and energy efficiency under high utilization. Domain-specific training systems demonstrate that large investments in interconnect and memory can pay off when workloads and utilization are stable at scale \cite{jouppi2017tpu,jouppi2021tpuv4}.

Cost also includes engineering and opportunity cost. Highly specialized ASICs can be extremely efficient for a stable operator set, but they risk obsolescence as model architectures evolve. More programmable platforms (GPUs, FPGAs, and compiler-driven tensor processors) trade peak efficiency for adaptability, aligning with the broader trend toward hardware/software co-design \cite{hennessy2019hwds}.

\subsubsection{GPU-specific area/cost challenges}
GPUs amortize high die area and advanced packaging costs across a broad software ecosystem, but cost-efficiency depends on utilization; under low batch or latency-constrained serving, throughput-per-dollar can drop sharply \cite{williams2009roofline}. In many datacenters, GPU cost is coupled to the cost of power delivery and cooling, so improvements in performance-per-watt translate directly into operational savings \cite{horowitz2014energy}. For training clusters, scaling also requires expensive interconnect and system integration, making system-level cost a dominant factor beyond the accelerator chip itself \cite{hennessy2019hwds}. Software maturity further influences cost: improved kernel libraries and compilers can raise utilization without hardware changes \cite{chetluar2014cudnn,chen2018tvm}.

A major driver of GPU cost is the memory subsystem: HBM stacks and CoWoS (Chip-on-Wafer-on-Substrate) packaging are significant fractions of the BOM (Bill of Materials). For inference workloads that are memory-capacity bound (like hosting LLaMA-70B), users must purchase more GPUs just to fit the model weights and KV-cache, even if compute utilization is low. This ``capacity tax'' has spurred interest in memory-expansion technologies (e.g., CXL) and aggressive quantization (4-bit/3-bit), which effectively reduce the dollar-cost per parameter by fitting larger models onto fewer devices. However, these solutions trade off bandwidth or accuracy, reinforcing the link between algorithmic choices and deployment economics \cite{micikevicius2018mixedprecision,kwon2023vllm}.

\subsubsection{TPU/NPU-specific area/cost challenges}
TPU-style systems devote substantial area to systolic arrays and on-chip SRAM, and they rely on specialized interconnects and memory systems to sustain utilization at scale \cite{jouppi2017tpu,jouppi2021tpuv4}. These design choices can improve throughput-per-watt for stable dense workloads, but they require large capital investment in custom silicon and system infrastructure \cite{hennessy2019hwds}. Edge NPUs optimize for SoC integration and unit cost, but they face strict area budgets and must share packaging and DRAM cost with the rest of the device, often limiting on-chip memory and making compiler scheduling critical \cite{mazumder2021survey,jacob2018quantization}. Operator coverage also impacts cost: if important kernels fall back to CPU/GPU, the effective cost per inference rises due to duplicated resources \cite{vaswani2017attention}.

For datacenter TPUs, the ``pod'' architecture amortizes the cost of the high-speed torus interconnect across many chips, but this makes the system expensive to deploy in small increments. The fixed topology is highly efficient for large-scale training jobs that map well to the torus, but it can be rigid for fragmented inference workloads or for models that don’t naturally shard across the pod’s dimensions. This creates a utilization challenge: if a job doesn’t need a full pod slice, resources may be stranded. Thus, the total cost of ownership (TCO) depends on the scheduler's ability to bin-pack diverse jobs onto the available topology without fragmentation \cite{jouppi2021tpuv4}.

\subsubsection{ASIC-specific area/cost challenges}
ASIC inference engines can be very area-efficient for a narrow operator set, but adding programmability, larger SRAM, or support for emerging operators increases area and design complexity \cite{chen2016eyeriss,chen2014dadiannao}. Attention and LLM-serving workloads tend to demand more memory capacity and bandwidth, which can push ASIC designs toward larger SRAMs and more expensive packaging \cite{vaswani2017attention,kwon2023vllm}. Long design cycles and non-recurring engineering costs create risk when model architectures evolve rapidly, making ASIC viability tightly linked to workload stability and volume \cite{hennessy2019hwds}. Compression and sparsity can reduce required memory, but they introduce metadata and control complexity that can increase area and verification cost \cite{han2016eie,han2016deepcompression}.

Furthermore, the ``dark silicon'' problem manifests in ASICs when new model features emerge that the hardware wasn't designed for. If an accelerator devotes 30\% of its area to a specific sparsity engine or activation function that later falls out of favor (e.g., a move from ReLU to Swish/GELU), that area becomes wasted cost. This risk forces architects to over-provision programmable elements (vector processors or DSPs) alongside fixed-function units, diluting the area-efficiency advantage. As a result, the most successful commercial ASICs often pair efficient matrix cores with a fairly capable general-purpose vector Instruction Set Architecture (ISA) to future-proof the investment \cite{hennessy2019hwds,williams2009roofline}.

\subsubsection{FPGA-specific area/cost challenges}
FPGAs trade silicon efficiency for reconfigurability; Look-Up Table (LUT)/DSP/BRAM resources and routing overhead can limit peak throughput-per-area compared to ASICs \cite{nurvitadhi2017can}. Cost-efficiency depends on whether the deployment can keep the FPGA well utilized; otherwise fixed platform cost dominates. However, for deployments that value flexibility or low-volume customization, FPGAs can reduce overall engineering cost and time-to-deployment, particularly in cloud settings where the same fleet can be repurposed \cite{fowers2018brainwave}. Toolchain productivity is also a cost factor: design iteration time and the ability to map new operators can dominate total cost of ownership \cite{venieris2017fpgaconvnet}. Finally, partial reconfiguration and multi-tenant scheduling introduce additional complexity that can affect effective cost-per-inference \cite{fowers2018brainwave}.

High-end FPGAs with HBM and hardened NoCs are expensive, often rivaling GPUs in unit cost. To justify this premium, the design must deliver value that GPUs cannot—typically strictly deterministic latency or direct integration with network/storage (``smart Network Interface Card (NIC)'' or ``smart Solid State Drive (SSD)'' use cases). If the FPGA is used merely as a slower GPU for batch processing, the area/cost math rarely works out due to the overhead of programmable fabric. Therefore, cost-effective FPGA deployment often involves moving the compute \emph{to} the data (e.g., inside the network switch) to save system-level energy and bandwidth, rather than competing on raw FLOPs/\$ in a rack server \cite{fowers2018brainwave,nurvitadhi2017can}.

\subsubsection{LPU/LLM-serving area/cost challenges}
LPU-style accelerators must justify cost through high utilization in LLM serving, where demand can be bursty and multi-tenant. Because serving is often memory- and bandwidth-limited, cost-efficiency depends on balanced provisioning of compute and memory bandwidth rather than on peak matmul throughput alone \cite{kwon2023vllm,groq2024lpu,williams2009roofline}. Long-context and high-concurrency workloads increase KV-cache footprint, pushing designs toward more memory capacity per accelerator, which can raise cost \cite{vaswani2017attention,kwon2023vllm}. In addition, cost depends on the software ecosystem: serving runtimes, kernel libraries, and compatibility with popular model formats influence time-to-deployment and operational overhead \cite{kwon2023vllm,dao2022flashattention}. Finally, heterogeneity (MoE, retrieval) can reduce utilization and increase cost unless supported in scheduling and memory management \cite{shazeer2017outrageously}.

A specific cost driver for LPU serving is the need for SRAM-heavy architectures to hide latency. SRAM is orders of magnitude more expensive per bit than DRAM. While an SRAM-centric design (like Groq's) provides unmatched deterministic latency and throughput for small batches, scaling it to store the weights of a 70B+ parameter model requires daisy-chaining many chips, exploding the system cost. This creates a bifurcation in the market: SRAM-based LPUs for latency-critical, lower-parameter (or sharded) serving, and HBM-based GPUs/ASICs for cost-optimized, high-capacity serving. The ``right'' cost choice depends entirely on the user's willingness to pay for milliseconds of latency reduction \cite{groq2024lpu,williams2009roofline}.

\subsubsection{In-/near-memory and analog area/cost challenges}
In-/near-memory designs shift area into memory arrays and mixed-signal peripherals. While crossbars can be dense, ADC/DAC and peripheral circuitry can dominate area and cost, and device variability can increase calibration and testing complexity \cite{shafiee2016isaac,anzaroot2019puma}. Cost must also account for yield and reliability: analog non-idealities and endurance constraints can increase test time and reduce manufacturability \cite{chi2016prime}. Moreover, because not all operators map naturally to crossbars, practical systems may require additional digital logic, increasing area and integration complexity \cite{vaswani2017attention}. These factors mean that cost-per-inference depends on end-to-end mapping ratio and on the overheads of heterogeneous integration \cite{anzaroot2019puma}.

Fabrication cost is another hurdle: many dense non-volatile memory technologies (Resistive RAM (RRAM), Phase-Change Memory (PCM)) require specialized process steps that may not be available in standard logic finFET nodes, or they require integration via expensive 2.5D/3D stacking. This splits the manufacturing ecosystem and can delay access to the latest lithography nodes that benefit the digital control logic. Consequently, analog accelerators often lag behind digital counterparts in frequency and logic density, forcing them to compete purely on energy efficiency in niche markets (e.g., edge, IoT) rather than replacing general-purpose datacenter silicon \cite{hennessy2019hwds,chi2016prime}.

\subsubsection{Neuromorphic area/cost challenges}
Neuromorphic chips allocate area to distributed memory and communication fabrics to support event-driven execution. They can be cost-effective for always-on low-power sensing, but their value depends on workload fit and toolchain maturity rather than on conventional throughput-per-area metrics \cite{davies2018loihi,merolla2014truenorth}. If the application requires significant preprocessing or dense computation, the effective cost rises because the neuromorphic chip must be paired with conventional processors \cite{gale2019state}. In addition, software ecosystem and deployment tooling influence engineering cost, which is often the limiting factor for adopting non-standard execution models \cite{hennessy2019hwds}. As a result, neuromorphic cost-efficiency is best evaluated in complete systems and application pipelines, not in isolation \cite{coleman2019dawnbench}.

The area trade-off in neuromorphic designs is unique: they sacrifice dense arithmetic logic (large MAC arrays) to prioritize state storage (synapses) and fine-grained routing. For sparse workloads, this is area-efficient because inactive circuits consume little static power and no dynamic power. However, for dense workloads, the lack of time-multiplexed arithmetic units means the chip must physically instantiate more neurons/synapses to represent the model, potentially leading to a larger die size than a compact, reused systolic array. Thus, the ``cost'' of neuromorphic hardware is only justified when the sparsity factor is high enough to offset the lower area-density of computation \cite{davies2018loihi}.

\subsection{Performance limits from memory and communication}
End-to-end performance is often limited by memory capacity, bandwidth, and communication rather than peak compute. Training requires storing activations, gradients, and optimizer state; for very large models, optimizer state and activations can exceed device memory, motivating memory-reduction techniques and sharding strategies \cite{rajbhandari2020zero,micikevicius2018mixedprecision}. Inference for large language models requires managing KV-cache and intermediate tensors whose sizes depend on sequence length and batching, shifting the bottleneck toward memory systems even on compute-rich accelerators \cite{kwon2023vllm}.

\begin{figure}[thbp]
\centering
\includegraphics[width=0.9\linewidth]{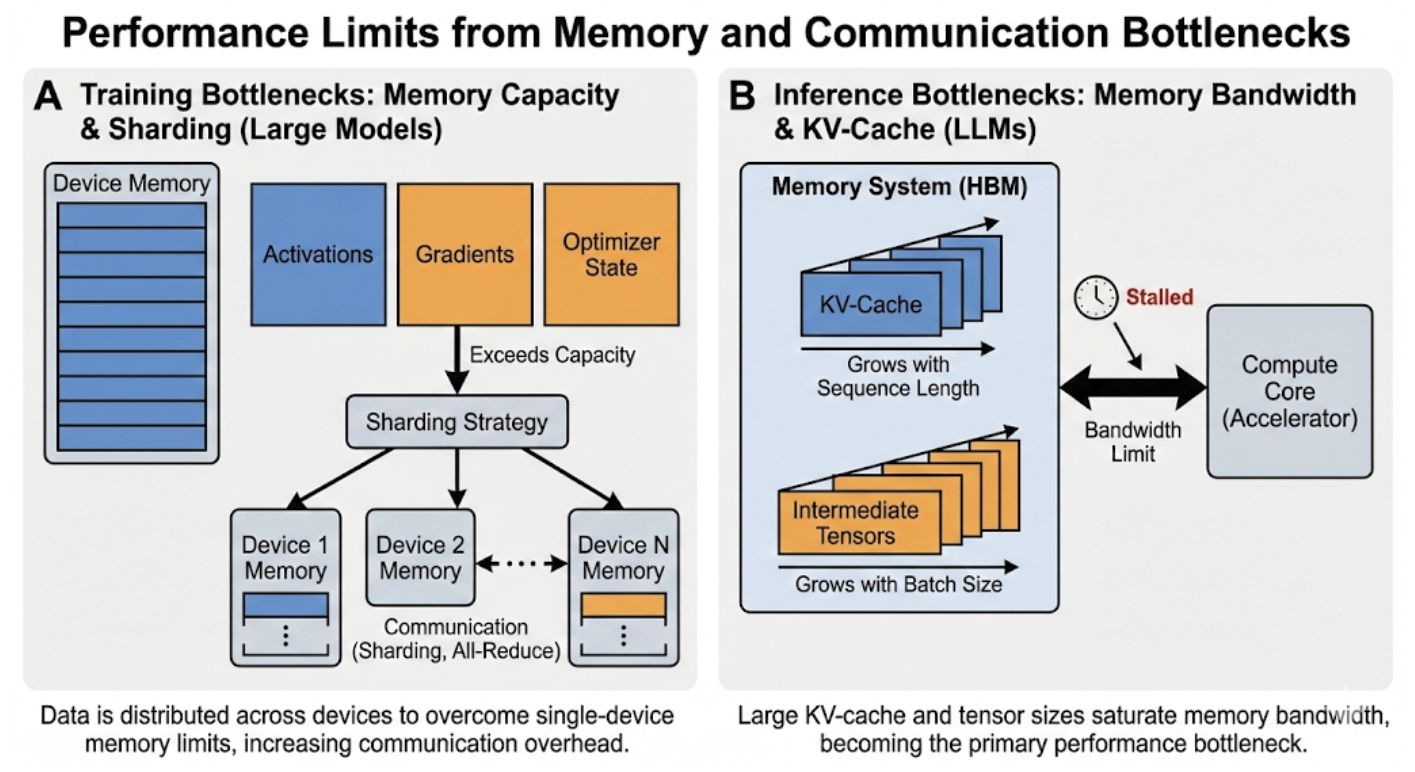}
\caption{Memory and communication bottlenecks in large-scale training and inference, emphasizing the impact of limited HBM bandwidth, interconnect latency, and the overhead of collective communication primitives.}
\label{fig:section12}
\end{figure}

Figure~\ref{fig:section12} summarizes common memory and communication bottlenecks that dominate large-model training and serving at scale. Distributed execution introduces collective communication and synchronization that can dominate iteration time. Data parallel training stresses all-reduce bandwidth, while tensor and pipeline parallelism introduce more frequent communication of activations and gradients; the optimal strategy depends on model size, sequence length, and interconnect characteristics \cite{shoeybi2019megatronlm}. Large-scale tensor-processor systems explicitly provision interconnect bandwidth and topology for these patterns, illustrating how system design shapes achievable scaling efficiency \cite{jouppi2021tpuv4}.

The practical implication is that reporting peak FLOPs or TOPS is insufficient. Accurate performance modeling and diagnosis require accounting for arithmetic intensity, memory traffic, and communication volume, along with the ability to overlap communication with compute \cite{williams2009roofline,wulf1995memorywall}.

\subsubsection{GPU-specific memory/communication challenges}
On GPUs, memory bandwidth and cache behavior frequently limit end-to-end performance, especially for attention and sparse workloads \cite{williams2009roofline,dao2022flashattention,gale2019state}. Long-context inference increases KV-cache footprint and can turn otherwise compute-heavy pipelines into bandwidth-bound workloads \cite{kwon2023vllm,vaswani2017attention}. Multi-GPU training adds all-reduce and synchronization costs; scaling efficiency depends on overlap and on the chosen parallelism strategy (data vs.\ tensor vs.\ pipeline) \cite{shoeybi2019megatronlm,rajbhandari2020zero}. Compiler and runtime choices influence effective communication by determining fusion, scheduling, and overlap behavior \cite{chen2018tvm,abadi2016tensorflow}.

Specifically, the ``memory wall'' on GPUs is exacerbated by the growing disparity between arithmetic throughput (scaling rapidly with tensor cores) and memory bandwidth (scaling more slowly). This puts pressure on software to maximize cache reuse through techniques like activation checkpointing (trading compute for memory) and aggressive kernel fusion (keeping data in registers). In distributed settings, communication bandwidth (NVLink/Infiniband) often determines the feasibility of splitting a single model across devices. If the interconnect is slow, tensor parallelism becomes prohibitively expensive due to frequent all-reduce operations, forcing users into less efficient pipeline parallelism or pure data parallelism with lower effective batch sizes \cite{shoeybi2019megatronlm}.

\subsubsection{TPU/NPU-specific memory/communication challenges}
TPU pods explicitly co-design interconnect and memory to match collective communication patterns of large-scale training, enabling higher scaling efficiency on stable workloads \cite{jouppi2021tpuv4,jouppi2017tpu}. Nevertheless, workloads with dynamic shapes, irregular sparsity, or long-context attention can stress memory capacity and reduce utilization when intermediate tensors exceed on-chip buffers \cite{kwon2023vllm,vaswani2017attention,gale2019state}. Quantized inference can reduce bandwidth demand, but only when the operator set is fully supported and conversions are minimized \cite{jacob2018quantization}. In edge NPUs, shared DRAM bandwidth and cache contention with other SoC components amplify memory bottlenecks, making placement and scheduling critical \cite{mazumder2021survey}.

A distinct challenge for TPUs is the management of ``HBM fragmentation'' and scratchpad allocation. The compiler must orchestrate data movement between HBM and on-chip memory with cycle-level precision. If a model’s activation working set slightly exceeds the SRAM capacity, the compiler may spill data to HBM, causing a sharp ``performance cliff.'' This binary behavior (fit vs. spill) makes performance less predictable than on cache-based architectures (GPUs), where performance degrades more gracefully. Consequently, developers often spend significant effort manual tuning batch sizes and partition strategies to stay on the ``fast path'' of the memory hierarchy \cite{jouppi2017tpu,williams2009roofline}.

\subsubsection{ASIC-specific memory/communication challenges}
ASIC accelerators often rely on large on-chip SRAM and carefully chosen dataflows to reduce DRAM traffic, but when model layers do not fit the assumed reuse patterns, off-chip bandwidth becomes the bottleneck \cite{chen2016eyeriss,williams2009roofline}. Attention and KV-cache behavior can further increase streaming bandwidth requirements and introduce irregular access patterns that are difficult to support with fixed dataflows \cite{vaswani2017attention,kwon2023vllm}. Compression and sparsity can reduce memory footprint, yet they introduce metadata and irregular accesses that must be handled efficiently to avoid shifting the bottleneck to indexing and control \cite{han2016eie,han2016deepcompression,gale2019state}. In multi-accelerator ASIC systems, interconnect and synchronization must be provisioned for the chosen parallelism strategy, otherwise scaling stalls on communication \cite{hennessy2019hwds}.

The ``bandwidth tax'' of programmability also affects ASICs. If an architecture relies on a host CPU to issue commands for every tile or layer, the PCIe bus or command-queue latency can become a bottleneck, especially for small-batch inference. To avoid this, advanced ASICs implement autonomous command processors or graph executors on-chip, allowing the device to run entire subgraphs without host intervention. However, this increases design complexity and requires a robust compiler to generate the command streams. Without this autonomy, the accelerator may sit idle waiting for the host to pointer-chase through the next set of descriptors, wasting the available DRAM bandwidth \cite{hennessy2019hwds}.

\subsubsection{FPGA-specific memory/communication challenges}
FPGA designs can minimize external traffic with streaming pipelines, but bandwidth to external DRAM and host-device I/O can dominate performance for large models or attention-heavy workloads \cite{nurvitadhi2017can,FPGATrans,vaswani2017attention}. For Transformers, buffering Query-Key-Value (QKV) projections and intermediate activations is challenging under limited on-chip BRAM, which can force frequent off-chip transfers and reduce throughput \cite{FPGATrans,williams2009roofline}. In cloud deployments, PCIe/host interfaces and multi-tenant scheduling can become communication bottlenecks, and the end-to-end system often determines observed performance \cite{fowers2018brainwave}. Toolchain decisions (HLS vs.\ RTL, memory partitioning) affect achievable bandwidth and hence the effective communication bottleneck \cite{venieris2017fpgaconvnet}.

Another constraint is the ``port limitation'' of on-chip memory. BRAMs typically have only two ports. If a compute kernel needs to read 16 operands per cycle to keep the pipeline full, the designer must bank and duplicate memory extensively or run the BRAMs at a higher clock multiple (if possible). This memory partitioning problem is NP-hard and is a frequent cause of routing congestion or lower-than-peak performance. HLS tools try to automate this, but for complex access patterns (like sliding windows with non-unit strides or sparse lookups), manual pragma insertion or RTL rewriting is often needed to saturate the external memory interface \cite{zhang2015optimizing,canis2013legup}.

\subsubsection{LPU/LLM-serving memory/communication challenges}
LLM serving is often limited by KV-cache capacity and bandwidth rather than compute, particularly at long context lengths and high concurrency \cite{kwon2023vllm,vaswani2017attention}. Paging and memory management policies are essential to scale concurrency, but they add overheads and can impact tail latency via fragmentation and scheduling decisions \cite{kwon2023vllm}. IO-aware attention reduces intermediate traffic but does not eliminate KV-cache movement, so bandwidth-per-token remains a central system constraint \cite{dao2022flashattention,horowitz2014energy}. LPU-style designs must therefore provision balanced memory bandwidth per token and minimize control overheads, while maintaining compatibility with serving runtimes and model formats \cite{groq2024lpu,kwon2023vllm}.

Communication between chips is critical for LPUs because a single chip rarely holds a full large model. The interconnect must support extremely low-latency, fine-grained tensor slicing to allow ``tensor-parallel'' execution across 16--64 chips without stall bubbles. Unlike GPU clusters that might tolerate microseconds of all-reduce latency, an LPU pipeline executing layer-by-layer requires nanosecond-scale synchronization to maintain its deterministic throughput guarantees. This drives LPU interconnects to be proprietary, statically routed, and integrated directly into the ISA, rather than relying on standard Ethernet or PCIe switches \cite{groq2024lpu,shazeer2017outrageously}.

\subsubsection{In-/near-memory and analog memory/communication challenges}
In-memory compute reduces the cost of moving weights and activations for matmul, but system-level communication remains for non-matmul operators and for moving results between analog and digital domains \cite{shafiee2016isaac,chi2016prime,anzaroot2019puma}. The communication bottleneck can reappear in the peripherals (ADC/DAC) and in interconnect between crossbar tiles and digital control, particularly when workloads are small or irregular \cite{anzaroot2019puma}. Attention workloads can remain bandwidth-bound due to KV-cache and softmax operations, which are not naturally accelerated by crossbar matmul \cite{vaswani2017attention,kwon2023vllm}. As a result, end-to-end speedups depend on mapping ratio and on how efficiently the heterogeneous system orchestrates data movement \cite{hennessy2019hwds}.

Additionally, the ``fan-out/fan-in'' communication within a crossbar array poses signal integrity challenges. As arrays grow larger to hold more weights, the analog signal path lengthens (increasing IR drop and capacitance), which limits the readout speed or precision. Digital communication between tiles (e.g., aggregating partial sums from multiple arrays) then becomes the new bottleneck. If the NoC connecting these tiles isn't provisioned with enough bandwidth, the fast analog cores will stall waiting to retire their results, negating the throughput benefit. Thus, PIM architectures effectively trade off-chip memory bandwidth problems for on-chip network design challenges \cite{shafiee2016isaac,kim2025hpim}.

\subsubsection{Neuromorphic memory/communication challenges}
Neuromorphic systems distribute memory and communication across cores; performance depends on sparse event traffic and locality \cite{davies2018loihi,merolla2014truenorth}. When spiking activity is dense, communication fabric contention can dominate, and the energy/latency benefits of event-driven execution diminish \cite{davies2018loihi}. End-to-end communication overhead also includes encoding/decoding between conventional sensors and spiking representations, which can outweigh benefits if sparsity is not intrinsic \cite{gale2019state}. These effects mean that neuromorphic communication challenges are best evaluated at the application level rather than by peak metrics alone \cite{coleman2019dawnbench}.

Scaling neuromorphic systems to large models introduces a ``synaptic storage'' challenge. While distributed SRAM is fast, it is not as dense as DRAM. Storing a billion-parameter model entirely in on-chip SRAM is cost-prohibitive. Some designs use multi-chip meshes, but this reintroduces inter-chip communication latency, which can disrupt the precise timing required for spike-based plasticity (Spike-Timing-Dependent Plasticity (STDP)). Therefore, efficient memory virtualization or hierarchical storage (local SRAM + backing DRAM) with smart prefetching is needed to support scale-out without killing the event-driven efficiency \cite{davies2018loihi}.

\subsection{Resource utilization and contention}
Resource consumption is broader than power and area: it includes compute utilization, memory footprint, bandwidth demand, and contention for shared resources. Real systems are constrained by multiple shared bottlenecks (Streaming Multiprocessor (SM) occupancy, cache capacity, memory controllers, and interconnect), so maximizing one resource in isolation can degrade overall throughput by increasing contention elsewhere \cite{williams2009roofline}.

\begin{figure}[thbp]
\centering
\includegraphics[width=0.9\linewidth]{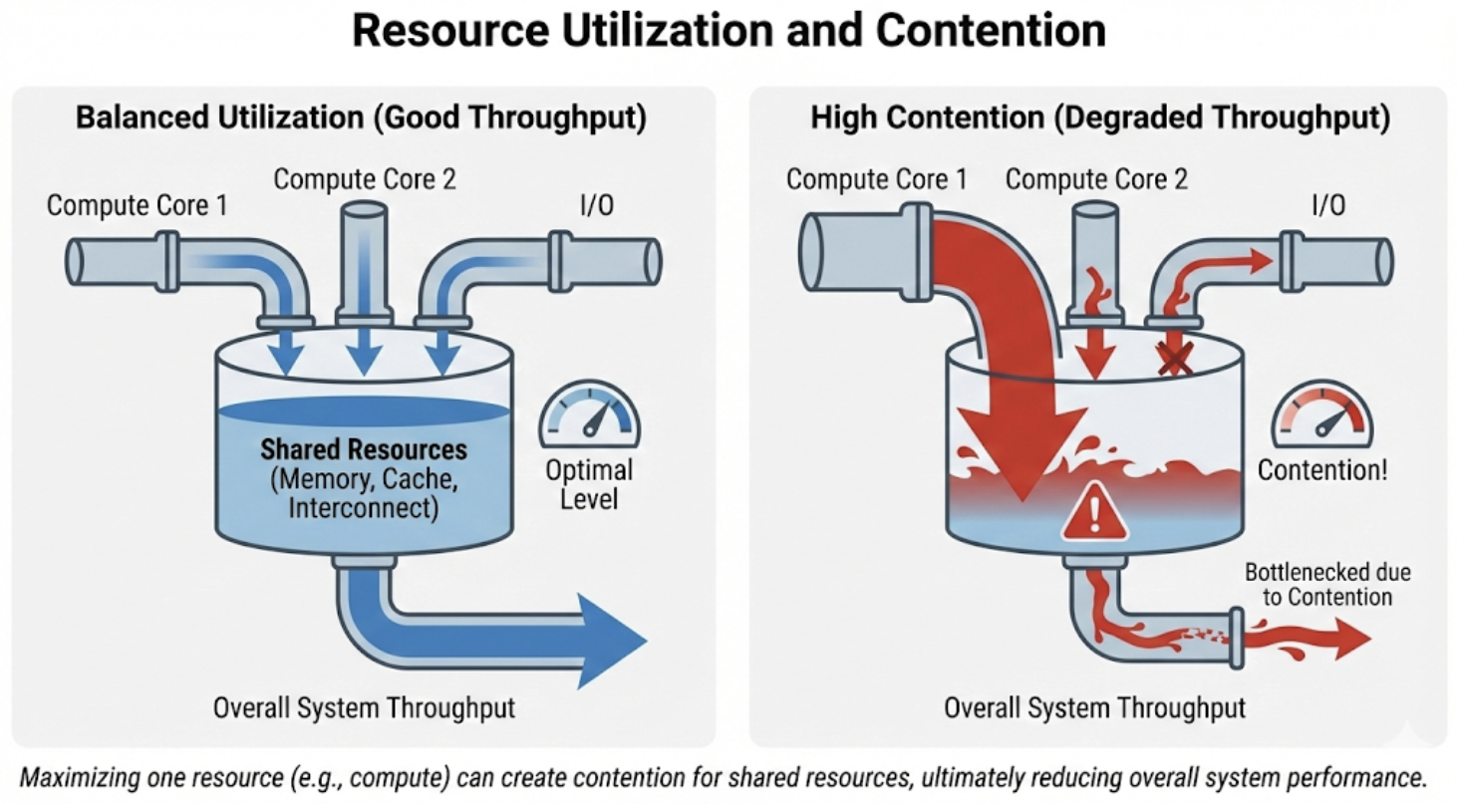}
\caption{Resource utilization challenges arising from irregular workloads, showing how sparsity, dynamic shapes, and conditional execution (e.g., MoE) can lead to load imbalance and underutilized compute units.}
\label{fig:section17}
\end{figure}

From Fig.\ref{fig:section17} shows that irregular workloads—arising from unstructured sparsity, dynamic shapes, mixture-of-experts routing, and conditional execution—can lead to load imbalance and poor utilization even on highly capable hardware \cite{gale2019state,shazeer2017outrageously}. Sparse accelerators and sparse kernels can mitigate this, but they require careful data structures and scheduling to avoid turning arithmetic savings into indexing overhead and random memory access \cite{parashar2017scnn,han2016eie}.

Support for these workloads often depends on compiler/runtime capabilities. Graph compilers and kernel DSLs can generate fused kernels and choose layouts that improve locality and reduce contention, but dynamic behavior (variable sequence length, adaptive computation, MoE routing) remains challenging for ahead-of-time optimization \cite{chen2018tvm,tillet2019triton}. As models become more heterogeneous, maintaining high utilization becomes increasingly a systems problem rather than a kernel-level problem.

\subsubsection{GPU-specific utilization challenges}
GPU utilization can collapse on small-batch or latency-constrained serving due to insufficient parallelism and kernel launch overheads, and it can degrade under unstructured sparsity due to load imbalance \cite{gale2019state,parashar2017scnn}. Utilization is also sensitive to memory behavior: cache misses and bandwidth saturation can stall SMs even when compute resources are plentiful \cite{williams2009roofline}. Compiler-driven fusion and kernel generation help mitigate overheads by reducing intermediate writes and launch count, but dynamic shapes and control flow remain challenging for ahead-of-time optimization \cite{chen2018tvm,tillet2019triton}. For LLMs, paging and KV-cache layout decisions in the runtime can determine whether hardware utilization is sustained under concurrency \cite{kwon2023vllm}.

Another utilization pitfall is the ``tail effect'' in distributed training. If one GPU in a 1000-GPU cluster is slow (due to thermal throttling, Error Correction Code (ECC) errors, or straggler tasks), the entire synchronous training step waits, effectively reducing the cluster-wide utilization to zero during the wait time. Fault tolerance and straggler mitigation strategies are thus essential for utilization at scale. Similarly, in pipeline parallelism, ``pipeline bubbles'' (idle time during fill/drain phases) reduce effective FLOPs. Minimizing these bubbles requires complex interleaved schedules (e.g., 1F1B) that increase memory pressure, creating a tension between utilization and memory capacity \cite{shoeybi2019megatronlm,rajbhandari2020zero}.

\subsubsection{TPU/NPU-specific utilization challenges}
TPUs/NPUs achieve high utilization on static dense kernels, but utilization can suffer when models include irregular operators, dynamic control flow, or sparse routing that breaks the accelerator’s preferred dataflow \cite{jouppi2017tpu,shazeer2017outrageously}. Attention kernels with long contexts can also reduce utilization if intermediate tensors do not fit on-chip buffers and the execution becomes bandwidth-bound \cite{vaswani2017attention,kwon2023vllm}. Compiler scheduling and operator coverage are therefore critical determinants of realized performance, both for datacenter TPUs and for edge NPUs \cite{jouppi2021tpuv4,mazumder2021survey}. Quantization improves throughput when supported end-to-end, but mixed precision and conversion overhead can reduce utilization if not carefully managed \cite{jacob2018quantization}.

The rigidity of systolic arrays poses a utilization risk for non-GEMM operations. While matmuls run at near-peak efficiency, element-wise ops (activations, normalization) or reductions often run on separate vector units with much lower throughput. If a model is heavy on LayerNorm or Softmax (relative to matmul), these vector units become the bottleneck, leaving the massive systolic arrays idle. This ``Amdahl’s Law'' effect means that as matrix engines get faster, the utilization bottleneck shifts aggressively to the vector/transcendental units and data-reshaping logic \cite{jouppi2017tpu,williams2009roofline}.

\subsubsection{ASIC-specific utilization challenges}
ASICs can be highly utilized when the workload matches the designed dataflow, but utilization suffers when operator mix shifts (e.g., attention variants) or when data-dependent sparsity introduces irregularity \cite{chen2016eyeriss,vaswani2017attention}. Off-chip bandwidth limits can also reduce utilization when the intended reuse pattern breaks down, forcing MAC arrays to idle while waiting for data \cite{williams2009roofline,chen2016eyeriss}. Sparsity-aware engines can help, but they require careful scheduling and metadata handling to avoid idle compute and to prevent indexing overhead from dominating \cite{han2016eie,gale2019state}. Maintaining utilization over time often requires either a sufficiently expressive programming model or frequent redesign, reflecting the ASIC flexibility trade-off \cite{hennessy2019hwds}.

A subtle utilization killer for ASICs is the ``batching mismatch.'' If an ASIC is designed with a very wide vector width (e.g., 512-byte SIMD) to maximize peak TOPS, it requires a minimum batch size to fill those vectors. In real-time serving where batch size might be 1, the hardware effectively runs at a fraction of its capacity (e.g., 1/32th utilization). Techniques like ``batch-1 optimization'' or systolic arrays that stream weights (weight-stationary) can help, but they often require different on-chip buffering strategies than training-optimized designs. Thus, an ASIC built for training may show abysmal utilization in inference, and vice-versa \cite{hennessy2019hwds}.

\subsubsection{FPGA-specific utilization challenges}
FPGA utilization depends on balancing pipelines, memory ports, and limited DSP/BRAM resources. Designs can become bandwidth-bound or routing-limited, and achieving high utilization across diverse models is difficult without reconfiguration or overprovisioning \cite{nurvitadhi2017can,venieris2017fpgaconvnet}. For attention/Transformer workloads, limited on-chip buffering can reduce utilization by forcing frequent external memory access \cite{FPGATrans,vaswani2017attention}. In cloud inference services, utilization is also shaped by batching and multi-tenant scheduling, which may be dominated by service-level constraints rather than by the FPGA datapath \cite{fowers2018brainwave}. Toolchain maturity affects achievable utilization because memory partitioning and pipelining decisions are often compiler/HLS-limited \cite{venieris2017fpgaconvnet}.

Furthermore, ``logic utilization'' on FPGAs has a practical ceiling (often around 70-80\%) before routing congestion makes timing closure impossible. A design that theoretically uses 95\% of the DSPs might fail to route or run at a very low clock frequency, effectively reducing throughput. This routing overhead means that the ``usable peak'' is significantly lower than the ``datasheet peak.'' Moreover, the time spent reconfiguring the FPGA (programming the bitstream) is dead time; in multi-tenant clouds, this reconfiguration latency discourages fine-grained time-sharing, leading to static partitioning that may leave regions underutilized \cite{fowers2018brainwave,williams2009roofline}.

\subsubsection{LPU/LLM-serving utilization challenges}
Serving workloads are bursty and heterogeneous (prompt lengths, concurrency), so utilization depends on dynamic batching and memory management rather than on steady-state kernel throughput \cite{kwon2023vllm}. KV-cache placement and paging can determine whether the accelerator sustains high utilization or becomes bandwidth-bound due to scattered memory access \cite{kwon2023vllm,dao2022flashattention}. Conditional execution (MoE, tool-calling) introduces additional load-balance challenges that can reduce utilization even on specialized serving hardware \cite{shazeer2017outrageously,gale2019state}. Achieving consistent utilization therefore requires co-design of the serving runtime, model architecture, and hardware scheduling policy \cite{hennessy2019hwds}.

Specifically, the ``prefill-decode imbalance'' creates utilization gaps. During prefill, compute units are saturated; during decode, they are starved by memory bandwidth. In a naive system, the compute logic sits 90\% idle during the long decode phase. Mixed-phase scheduling (running prefill for request A while request B is in decode) can recover some utilization, but it requires sophisticated resource isolation to prevent the prefill burst from destroying the decode latency Service Level Objective (SLO). LPU architectures often include specialized scheduling hardware to manage this mixing without Operating System (OS) overhead, aiming to keep the matrix units fed even when request phases diverge \cite{groq2024lpu,kwon2023vllm}.

\subsubsection{In-/near-memory and analog utilization challenges}
Analog accelerators can be underutilized when the workload includes many small matmuls, frequent conversions, or substantial non-matmul compute. Achieving high utilization often requires batching and mapping that maximizes crossbar occupancy, which can conflict with low-latency serving \cite{shafiee2016isaac,anzaroot2019puma}. Utilization also depends on whether weights remain resident in the arrays; if frequent remapping is needed, setup and calibration overhead can dominate \cite{chi2016prime}. Attention-heavy models reduce the fraction of time spent in large dense matmuls and therefore reduce analog utilization unless the system is redesigned around these operators \cite{vaswani2017attention,kwon2023vllm}. As a result, utilization is tightly coupled to model structure and to the partitioning between analog and digital components \cite{hennessy2019hwds}.

The ``mapping fragmentation'' problem is severe for crossbars. If a layer’s weight matrix doesn't perfectly align with the physical crossbar size (e.g., a 100x100 matrix on a 128x128 array), the unused cells (rows/columns) are wasted area and power. Unlike digital memories that can be packed, analog weights are spatially fixed. This leads to low effective utilization for layers with odd shapes or for the tail ends of large matrices. Solutions involve virtualizing the arrays or using complex interconnects to route inputs to sub-tiles, but these add overhead. Consequently, utilization is highest for uniform, large-model layers and drops for irregular, optimized architectures (like MobileNets) \cite{shafiee2016isaac,chi2016prime}.

\subsubsection{Neuromorphic utilization challenges}
Neuromorphic utilization depends on spike sparsity and event locality. When activity is dense, communication contention increases and the advantage of event-driven computation diminishes \cite{davies2018loihi,merolla2014truenorth}. Utilization can also be limited by the mapping of the network to the hardware fabric; poor mapping can concentrate spikes and create hotspots even when global activity is low \cite{davies2018loihi}. Converting conventional models to spiking forms can reduce effective sparsity and therefore reduce utilization benefits \cite{gale2019state}. Consequently, utilization must be evaluated jointly with the encoding, network design, and application workload \cite{coleman2019dawnbench}.

Moreover, ``temporal utilization'' in neuromorphic chips can be low if the input events are bursty. The hardware must be provisioned for the peak event rate to avoid dropping spikes, but during quiet intervals, the dedicated routing and update logic sits idle. Unlike clock-gated digital logic where idle means zero dynamic power, neuromorphic circuits often have leakage or bias currents that persist. Thus, the effective utilization (useful work per unit time / provisioned capacity) can be quite low for sporadic signals, challenging the TCO justification unless the standby power is exceptionally well managed \cite{davies2018loihi}.

\subsection{Benchmarking and reproducibility}
\begin{figure}[thbp]
\centering
\includegraphics[width=0.9\linewidth]{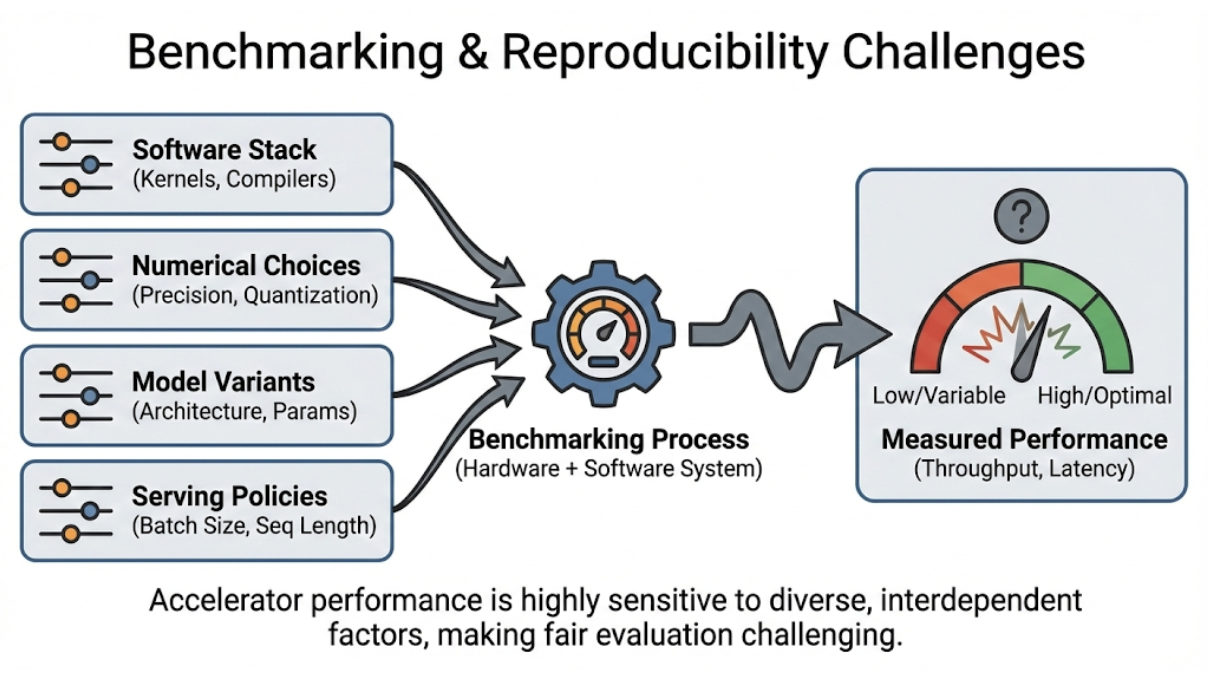}
\caption{Benchmarking and reproducibility challenges in hardware acceleration, underscoring the sensitivity of results to software versions, compiler settings, model variants, and measurement methodology.}
\label{fig:section18}
\end{figure}

Fair evaluation remains challenging(Fig.\ref{fig:section18}). Accelerator results are sensitive to software stack maturity (kernels, compilers, graph optimizers), numerical choices (precision and quantization), model variants, and serving policies. Even small changes in batch size, sequence length, or kernel fusion can change whether a workload is compute-bound or memory-bound, leading to large swings in measured performance \cite{williams2009roofline,kwon2023vllm}.

Benchmark suites such as MLPerf aim to standardize tasks, accuracy targets, and reporting rules for both training and inference, improving comparability across vendors and platforms \cite{mattson2020mlperf}. However, representative evaluation still requires careful choice of workload mix: production systems often run diverse models, experience non-stationary request distributions, and care about tail latency and cost, not only peak throughput. End-to-end benchmarks and competitions (e.g., time-to-train style evaluations) complement kernel microbenchmarks by capturing the interaction between the model, the input pipeline, the optimizer, and the distributed system \cite{coleman2019dawnbench}.

Reproducible comparisons therefore require clear reporting of hardware configuration, software versions, workloads, accuracy targets, compilation settings, and measurement methodology. For inference, reporting should include batch/concurrency, sequence lengths, and latency percentiles; for training, reporting should include scaling strategy, communication overlap, and convergence criteria. Without this context, results can be misleading even when individual measurements are correct.

\subsubsection{GPU benchmarking challenges}
GPU results are highly dependent on kernel/library versions and compiler paths (e.g., vendor libraries vs. compiler-generated kernels), and on serving policies such as batching and KV-cache management \cite{chetluar2014cudnn,chen2018tvm,tillet2019triton,kwon2023vllm}. IO-aware attention kernels can dramatically change the memory/computation balance, so benchmarks should explicitly report attention implementations and sequence-length distributions \cite{dao2022flashattention,vaswani2017attention}. MLPerf provides standardized reporting, but reproducing results still requires careful control of software versions, compilation flags, and measurement methodology \cite{mattson2020mlperf}. For training, reporting should include scaling strategy and memory optimizations (e.g., ZeRO) that change communication and memory footprints \cite{rajbhandari2020zero,shoeybi2019megatronlm}.

Furthermore, benchmarking often ignores the ``warm-up'' phase, Just-In-Time (JIT) compilation time, and memory fragmentation effects, which are critical in production. A benchmark that measures steady-state throughput after 1000 iterations hides the initial 30-second compilation delay or the gradual performance degradation from heap fragmentation. For dynamic serving, measuring \emph{sustainable} throughput (throughput at a fixed latency SLO) is more honest than peak throughput, but harder to standardize. Benchmarks should therefore report the ``latency-throughput curve'' rather than a single point, capturing the behavior under load \cite{mattson2020mlperf,kwon2023vllm}.

\subsubsection{TPU/NPU benchmarking challenges}
For TPUs/NPUs, performance depends on compilation, layout, and sharding decisions that may be opaque or highly tuned for specific workloads \cite{jouppi2017tpu,jouppi2021tpuv4}. Benchmarking should report compiler settings and input distributions (shapes, sequence lengths), especially for attention-heavy models where KV-cache and intermediate tensor sizes drive memory behavior \cite{vaswani2017attention,kwon2023vllm}. In edge NPUs, benchmarks should also report operator coverage and fallback behavior, since CPU/GPU fallback can dominate latency and energy \cite{mazumder2021survey,jacob2018quantization}. MLPerf-style rules help, but reproducibility still depends on consistent compiler versions and system configuration \cite{mattson2020mlperf}.

The closed nature of some NPU software stacks adds a ``black box'' variability. An OS update or driver patch might change the NPU frequency governor or memory allocation policy, altering results significantly. Unlike open GPU kernels (e.g., in Triton or CUDA), NPU behavior is often hidden behind firmware blobs. This makes attribution difficult: did performance improve because of the hardware or because the compiler recognized a specific subgraph pattern? Fair benchmarking requires freezing not just the model but the entire firmware/driver image \cite{mazumder2021survey}.

\subsubsection{ASIC benchmarking challenges}
ASIC benchmarks must clearly specify operator coverage, precision, and how unsupported operations are handled (host fallback vs. accelerator implementation) \cite{chen2016eyeriss,chen2014dadiannao}. Because ASIC efficiency is tightly tied to the chosen dataflow and memory hierarchy, reporting should include on-chip SRAM size, off-chip bandwidth, and any assumed reuse or tiling strategy \cite{chen2016eyeriss,williams2009roofline}. Sparsity and compression results should report sparsity pattern, metadata overhead, and utilization to avoid misleading comparisons \cite{han2016eie,han2016deepcompression,gale2019state}. For modern attention-based workloads, benchmarks should clarify how softmax/KV-cache operations are handled, since these can dominate end-to-end behavior \cite{vaswani2017attention,kwon2023vllm}.

Additionally, ASIC power measurements are often reported at the ``chip'' level, excluding DRAM or host power. This can be misleading for memory-bound workloads where DRAM power is 30-50\% of the total. A ``system-level'' benchmark (wall power) is necessary to compare an ASIC against a GPU fairly. Also, many research ASICs report simulation results rather than silicon measurements; simulation assumptions (e.g., zero DRAM latency, perfect clock gating) should be scrutinized. Benchmarks should explicitly state whether numbers are measured on silicon, emulated on FPGA, or estimated via architectural simulators \cite{hennessy2019hwds,williams2009roofline}.

\subsubsection{FPGA benchmarking challenges}
FPGA results are sensitive to toolchain choices (HLS vs. RTL), frequency targets, and memory configuration, and they often depend on end-to-end pipeline integration (host I/O, streaming, reconfiguration) \cite{nurvitadhi2017can,fowers2018brainwave}. Reporting should include resource utilization (DSP/BRAM/LUT), achieved clock, and I/O constraints, because these determine whether the design is compute- or bandwidth-limited \cite{venieris2017fpgaconvnet,williams2009roofline}. Transformer benchmarking should include sequence lengths and attention implementation details, since buffering and external memory access dominate performance \cite{FPGATrans,vaswani2017attention}. In cloud settings, multi-tenant effects and batching policies can change both throughput and tail latency and should be reported explicitly \cite{fowers2018brainwave}.

A common pitfall is benchmarking only the ``kernel execution time'' on the FPGA, ignoring the PCIe transfer time to move data to/from the device. For small batches, this transfer dominates. Benchmarks must report end-to-end latency including all data movement. Also, precision matters: comparing an 4-bit Integer (INT4) FPGA implementation against an 16-bit Floating Point (FP16) GPU implementation is an apples-to-oranges comparison unless accuracy metrics are also provided. The ``Pareto frontier'' of Accuracy vs. Throughput/Watt is the only robust way to compare flexible precision FPGAs against fixed-precision logic \cite{nurvitadhi2017can,umuroglu2017finn}.

\subsubsection{LPU/LLM-serving benchmarking challenges}
LLM serving benchmarks must report latency percentiles, concurrency, prompt/response lengths, and cache policies because these dominate performance and cost in practice \cite{kwon2023vllm}. “Tokens per second” without context can hide tail-latency regressions, memory-capacity limits, and instability under bursty load \cite{mattson2020mlperf}. Benchmarks should also specify attention kernels and paging strategy, since IO-aware attention and KV-cache management can fundamentally change memory behavior \cite{dao2022flashattention,kwon2023vllm}. When MoE or conditional execution is present, benchmarks should report load balance and routing behavior because these affect utilization and tail latency \cite{shazeer2017outrageously,gale2019state}.

Furthermore, the choice of ``random'' vs. ``real'' text inputs affects benchmarking for LLMs due to tokenization and sparsity. If a benchmark uses random noise, the tokenizer might produce a different number of tokens than for English text, skewing throughput numbers. If the model uses content-dependent sparsity (MoE or early exit), random data might trigger worst-case paths. Therefore, standard datasets (like ShareGPT or Alpaca) are preferred over synthetic tensors to capture realistic execution divergence and cache hit rates \cite{mattson2020mlperf,kwon2023vllm}.

\subsubsection{In-/near-memory and analog benchmarking challenges}
Analog and in-memory compute results must report accuracy under non-idealities, calibration procedure, ADC/DAC overheads, and the fraction of the model mapped to the accelerator \cite{shafiee2016isaac,chi2016prime,anzaroot2019puma}. End-to-end comparisons should include the cost of non-matmul operators and data conversions, because these can dominate when only part of the graph is mapped to crossbars \cite{vaswani2017attention,williams2009roofline}. Results should also clarify whether weights are assumed stationary and how often remapping occurs, as this affects both throughput and energy \cite{anzaroot2019puma}. Finally, for attention-heavy workloads, benchmarks should state how softmax and KV-cache are handled, since these are not naturally accelerated by crossbar matmul \cite{kwon2023vllm}.

The definition of ``operation'' in analog is tricky. Is a noisy, low-precision MAC equivalent to a deterministic digital MAC? Benchmarks should use ``effective'' throughput at a target accuracy (e.g., ImageNet Top-1) rather than raw TOPS. If the analog chip requires a larger model or retraining to regain accuracy lost to noise, that overhead must be penalized. A metric like ``ISO-accuracy energy efficiency'' is more meaningful than ``peak efficiency at unknown accuracy'' \cite{chi2016prime,jacob2018quantization}.

\subsubsection{Neuromorphic benchmarking challenges}
Neuromorphic benchmarks must report spike encoding, activity sparsity, and application-level accuracy, since performance and energy benefits depend strongly on event sparsity and representation \cite{davies2018loihi,merolla2014truenorth}. Benchmarks should also report event rates and communication patterns, because routing congestion can dominate latency and energy when activity increases \cite{davies2018loihi}. Comparing against dense accelerators requires careful definition of equivalent workloads and latency/accuracy targets, and should account for preprocessing and encoding overheads \cite{coleman2019dawnbench,gale2019state}. As with analog accelerators, end-to-end evaluation is essential: reporting only chip-level energy without the full pipeline can be misleading \cite{mattson2020mlperf}.

Finally, the ``baseline'' problem is pervasive: neuromorphic papers often compare against unoptimized CPU/GPU code. A fair comparison requires comparing against a highly optimized sparse-dense matrix engine or a dedicated low-power edge NPU running a quantized version of the same model. Benchmarking suites like ``NeuroBench'' attempt to standardize this by defining tasks where temporal dynamics are essential, ensuring that the event-driven nature is actually exercising a relevant capability rather than just running MNIST inefficiently \cite{davies2018loihi,coleman2019dawnbench}.

\section{Hardware Accelerator Architectures for Neural Networks}

\begin{figure}[thbp]
\centering
\includegraphics[width=0.9\linewidth]{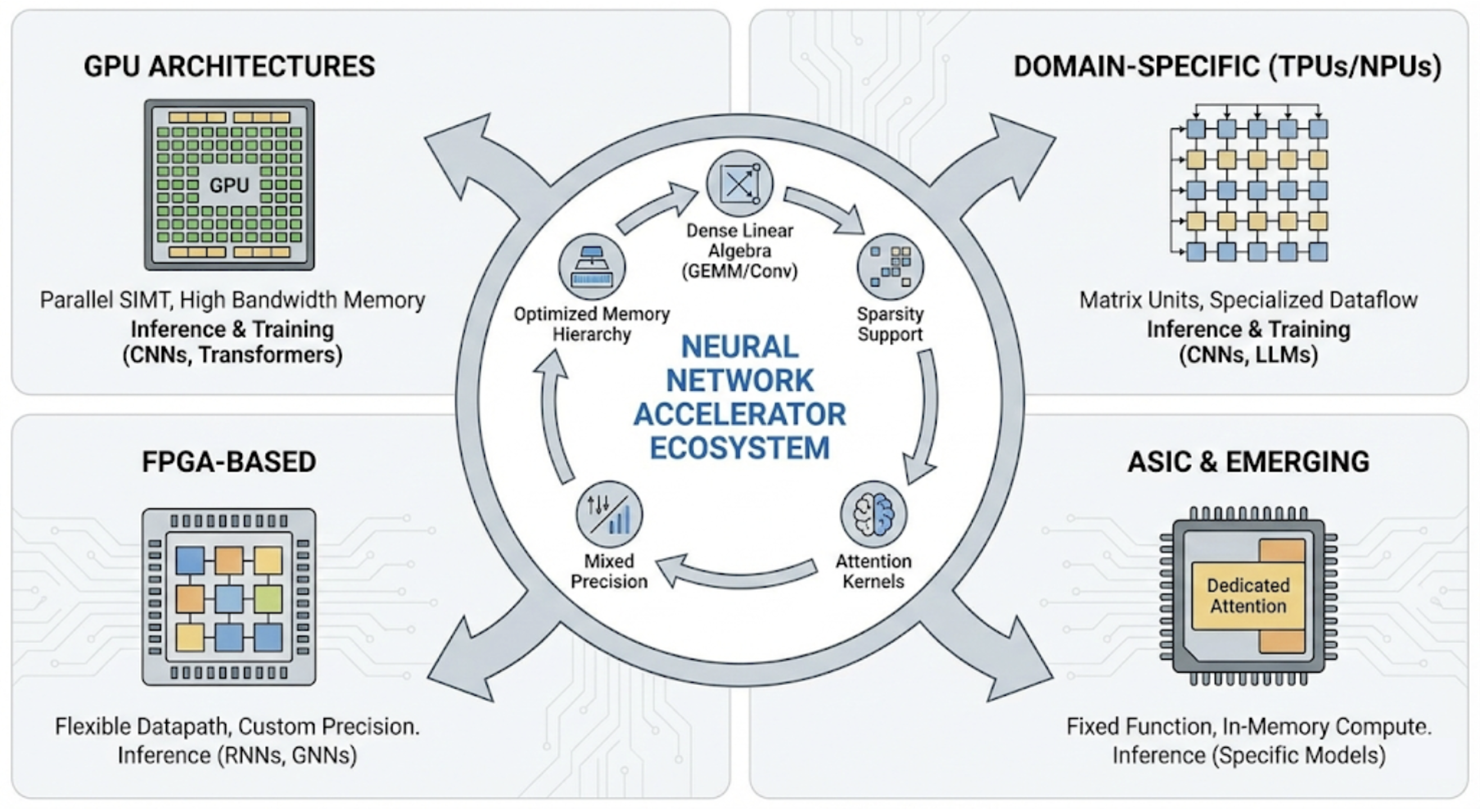}
\caption{Overview of the neural-network acceleration pipeline, mapping the flow from high-level model definitions through graph compilation and optimization to execution on specialized hardware architectures.}
\label{fig:section4}
\end{figure}

Neural-network acceleration has evolved into a heterogeneous ecosystem where different platforms optimize different points in the design space. At a high level, accelerators implement dense linear algebra efficiently (GEMM/conv), reduce data movement through carefully designed memory hierarchies and dataflows, and increasingly incorporate support for sparsity, mixed precision, and attention-centric kernels. This section surveys major architectural families and discusses how they are used for inference and training across common model classes (CNNs, RNNs, GNNs, and Transformers/LLMs) \cite{williams2009roofline,vaswani2017attention,jouppi2017tpu}. Figure~\ref{fig:section4} provides an overview of the neural-network acceleration pipeline and where architectural choices impact performance.

\subsection{GPU architectures}
GPUs remain the dominant platform in many production and research settings due to their programmability, mature software ecosystem, and strong support for both training and inference. Architecturally, modern GPUs combine wide Single Instruction, Multiple Threads (SIMT) execution, deep multithreading to hide latency, and high-throughput memory systems (large on-chip caches and register files, plus high-bandwidth off-chip memory), together with specialized matrix-multiply units (tensor cores) that accelerate dense linear algebra at reduced precision \cite{williams2009roofline,micikevicius2018mixedprecision}. In practice, GPU performance is inseparable from the software stack: operator libraries (e.g., highly tuned convolution and GEMM kernels), graph compilers, and kernel DSLs determine whether a model saturates compute or becomes bottlenecked by memory bandwidth, kernel launch overheads, or synchronization \cite{chetluar2014cudnn,chen2018tvm,tillet2019triton}.

GPUs are used across model families. CNNs map well to convolution/GEMM kernels with high reuse, while RNNs and sequence models expose different bottlenecks such as limited parallelism and recurrent dependencies. Transformers and LLMs place sustained pressure on memory bandwidth and capacity through attention and KV-cache, and they are sensitive to kernel fusion and memory-aware scheduling \cite{vaswani2017attention,kwon2023vllm}. As models adopt structured sparsity and mixture-of-experts routing, GPUs increasingly rely on sparse-friendly primitives and compiler support to preserve utilization under irregular access patterns \cite{gale2019state,shazeer2017outrageously,parashar2017scnn}.

\subsubsection{Inference on GPUs}
For inference, GPUs deliver high throughput by batching and by mapping convolutions and GEMMs onto highly optimized kernels. CNN inference benefits from implicit GEMM lowering, data layout transformations, and fusion of activation and normalization operations to reduce memory traffic. For Transformers and LLMs, inference efficiency hinges on fast attention and softmax kernels, fused MLP blocks, and careful KV-cache management; at high concurrency, KV-cache can dominate memory footprint and memory bandwidth \cite{vaswani2017attention,kwon2023vllm,dao2022flashattention}. Kernel DSLs and compiler stacks are increasingly used to generate fused kernels and to adapt to rapidly changing operator mixes (e.g., different attention variants), which helps avoid fragmentation across hand-written kernels \cite{tillet2019triton,chen2018tvm}.

Quantization is widely used to reduce bandwidth and improve throughput, particularly for edge-like deployment constraints and cost-efficient serving. Integer arithmetic and mixed-precision inference can be effective, but end-to-end gains depend on calibration, accuracy targets, and whether kernels can exploit the reduced precision without falling back to inefficient conversion paths \cite{jacob2018quantization}. Sparsity and compression can further reduce memory traffic, yet unstructured sparsity can harm utilization without dedicated sparse kernels and scheduling that preserves coalesced memory access and load balance \cite{gale2019state,han2016deepcompression}.

\subsubsection{Training on GPUs}
Training emphasizes throughput, numerical stability, and scalability. Mixed-precision training has become a standard approach to improve performance while maintaining convergence, typically using 16-bit Floating Point (FP16)/Brain Floating Point (BF16) compute with 32-bit Floating Point (FP32) accumulation and loss-scaling or similar techniques \cite{micikevicius2018mixedprecision}. Memory footprint is a dominant constraint: activations, gradients, and optimizer states can exceed device memory for large models, motivating activation checkpointing, optimizer-state partitioning, and distributed parallelism strategies (data, tensor, pipeline, and expert parallelism) \cite{rajbhandari2020zero,shoeybi2019megatronlm,shazeer2017outrageously}. These strategies trade off communication volume, synchronization, and pipeline bubbles against memory savings and compute efficiency.

At scale, training performance is often limited by communication and synchronization overheads (collectives for data parallelism; all-reduce and all-gather for tensor parallelism), making overlap strategies and interconnect bandwidth critical to end-to-end time-to-train \cite{williams2009roofline}. As workloads evolve, compiler and runtime support becomes increasingly important for selecting numerically stable kernels, fusing memory-bound operations, and scheduling communication to hide latency behind compute \cite{chen2018tvm,abadi2016tensorflow}.

\subsection{Domain-specific tensor processors (TPUs and NPUs)}
Domain-specific accelerators pursue higher performance-per-watt by specializing the datapath and dataflow for dense linear algebra. A common pattern is a systolic-array-like matrix engine (or multiple such engines) coupled with on-chip SRAM and a compiler stack that performs graph compilation, operator fusion, and layout planning to match hardware constraints \cite{jouppi2017tpu,kung1982systolic}. Compared to GPUs, these processors often make stronger assumptions about operand shapes and memory access patterns, which enables higher utilization and better energy efficiency on the targeted kernels. In edge settings, NPUs are frequently integrated into SoCs, where they must operate under strict power, thermal, and latency envelopes and share memory bandwidth with CPUs/GPUs and other accelerators \cite{mazumder2021survey}.

\begin{figure}[thbp]
\centering
\includegraphics[width=0.9\linewidth]{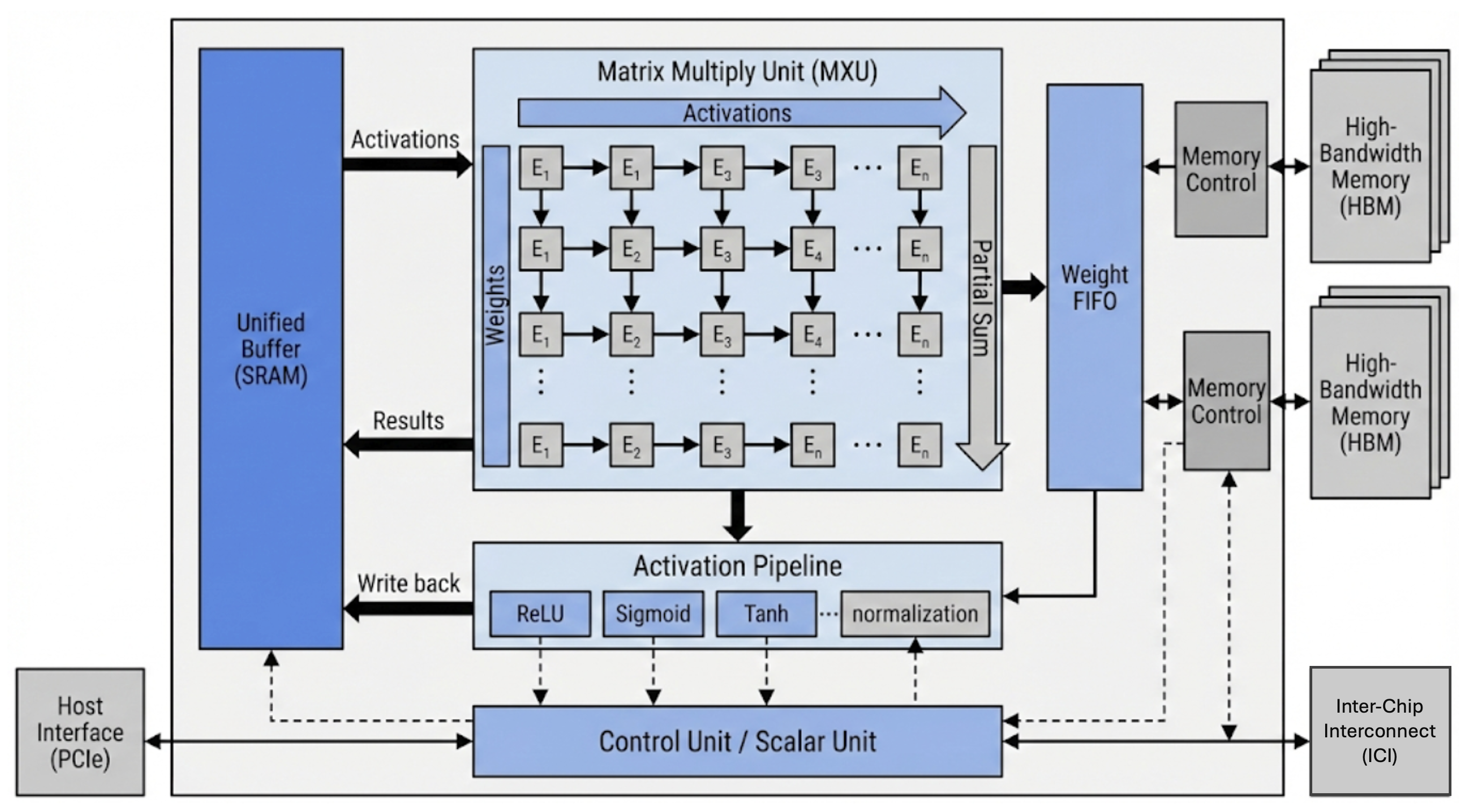}
\caption{Representative TPU architecture featuring a large systolic array for dense matrix multiplication, coupled with unified on-chip buffers and a specialized control unit to maximize data reuse.}
\label{fig:section8}
\end{figure}

\subsubsection{Inference on TPUs/NPUs}
Figure~\ref{fig:section8} illustrates a TPU-style tensor-processor organization used to motivate compilation, tiling, and on-chip reuse. The “tensor processor” category includes both data center-class training systems (e.g., TPU pods with specialized interconnect) and embedded inference engines. Across both, the core architectural challenge is balancing compute with on-chip reuse: increasing MAC throughput is only beneficial if activations and weights can be delivered at sufficient bandwidth without thrashing the memory hierarchy. This is especially visible for attention-heavy models, where KV-cache and intermediate activations introduce large, streaming memory demands that may not match the accelerator’s preferred dense reuse patterns \cite{vaswani2017attention,kwon2023vllm}. Inference on tensor processors benefits from predictable dense kernels and static shapes, enabling aggressive compilation, operator fusion, and on-chip reuse. Many deployments target integer arithmetic (8-bit Integer (INT8) and below) to reduce energy and improve throughput, and hardware often includes dedicated datapaths for quantized GEMM/conv and activation functions to avoid expensive conversions \cite{jacob2018quantization}. For CNNs, static-shape inference allows compilers to pre-plan tiling and buffering to maximize weight/activation reuse; for RNNs, throughput is more sensitive to sequence length and batch size due to limited parallelism and recurrent dependencies.

Transformer inference introduces additional constraints because attention and KV-cache behavior can pressure memory. Even with efficient matmul engines, end-to-end latency can be dominated by memory movement and softmax-like operators, motivating compiler/runtime techniques that manage cache growth, memory fragmentation, and kernel fusion for attention blocks \cite{vaswani2017attention,kwon2023vllm,dao2022flashattention}. When workloads include structured sparsity or MoE routing, utilization depends on whether the platform supports sparse-friendly data layouts and load-balanced scheduling \cite{gale2019state,shazeer2017outrageously}.

\subsubsection{Training on TPUs/NPUs}
Training on TPUs has been demonstrated at datacenter scale using large systolic arrays, high-bandwidth memory systems, and scalable interconnects designed for collective communication \cite{jouppi2017tpu,jouppi2021tpuv4}. These systems rely on compilation to map graphs onto hardware efficiently, including partitioning, operator fusion, and layout transforms that maximize systolic-array utilization while minimizing off-chip traffic. Mixed precision is central for achieving high utilization and managing memory footprint, often in conjunction with numerically stable accumulation and loss-scaling techniques \cite{micikevicius2018mixedprecision}.

Distributed training on tensor processors introduces system-level trade-offs between parallelism strategy, communication volume, and memory footprint. Data parallelism stresses all-reduce bandwidth, while tensor/pipeline parallelism introduces more frequent synchronization and activation transfers; MoE adds routing and load-balance issues that can reduce utilization if not well matched to the interconnect and compiler \cite{shazeer2017outrageously,shoeybi2019megatronlm}. Overall, achieving high performance-per-watt requires co-design between the model (operator mix and shapes), the compiler (fusion, sharding, scheduling), and the hardware (on-chip memory and interconnect).

\subsection{ASIC inference accelerators}
ASIC inference accelerators focus on maximizing throughput-per-area and throughput-per-watt for fixed or semi-fixed operator sets. Many designs exploit dataflow specialization (e.g., weight-stationary, output-stationary, or row-stationary mappings), large on-chip SRAM to reduce DRAM traffic, and support for reduced precision, pruning, and compression \cite{chen2016eyeriss,chen2014dadiannao,han2016deepcompression}. Unlike GPUs, ASICs can hardwire control and buffering patterns to reduce overhead, but this comes at the cost of reduced flexibility when operators and data layouts evolve. Consequently, ASIC designs often target stable kernels (conv/GEMM) while providing limited programmability for control and scheduling \cite{hennessy2019hwds}.

Architecturally, ASIC accelerators typically include arrays of MAC units, multi-bank on-chip SRAM scratchpads, and an on-chip network designed to sustain the chosen dataflow. Some families emphasize bringing computation closer to sensors or memory to reduce I/O energy, while others target datacenter inference with high throughput at modest precision. As models evolve toward attention-heavy and mixture-based architectures, ASICs increasingly require either richer programmability or carefully chosen operator coverage to avoid falling back to a host CPU/GPU for unsupported kernels \cite{vaswani2017attention,shazeer2017outrageously}.

\subsubsection{Inference on ASICs}
CNN inference has been a primary driver of ASIC accelerators due to its structured computation and high reuse. Eyeriss introduced a spatial architecture and dataflow taxonomy aimed at minimizing energy by reducing data movement through careful mapping and local reuse \cite{chen2016eyeriss}. Complementary designs such as the DianNao family explore microarchitectural specialization for neural operators, including support for common activation functions and efficient handling of weight storage and reuse \cite{chen2014dadiannao,liu2015shidiannao}. Beyond dense compute, engines such as EIE demonstrate how exploiting compression and sparsity can reduce memory bandwidth demand and improve energy efficiency by skipping zeros and storing compressed weights \cite{han2016eie,han2016deepcompression}.

Sparsity-aware ASICs can achieve large gains when sparsity is structured or when the architecture can avoid the overhead of irregular indexing. However, unstructured sparsity introduces load imbalance, irregular memory access, and reduced reuse, which must be handled by sparse data structures, scheduling, and often dedicated hardware support \cite{gale2019state,parashar2017scnn}. For attention-based models, inference ASICs must also address memory-intensive KV-cache behavior and the growing prominence of non-matmul components (softmax, normalization, sampling), which complicates a purely matmul-centric datapath \cite{vaswani2017attention,kwon2023vllm}.

\begin{figure}[thbp]
\centering
\includegraphics[width=0.9\linewidth]{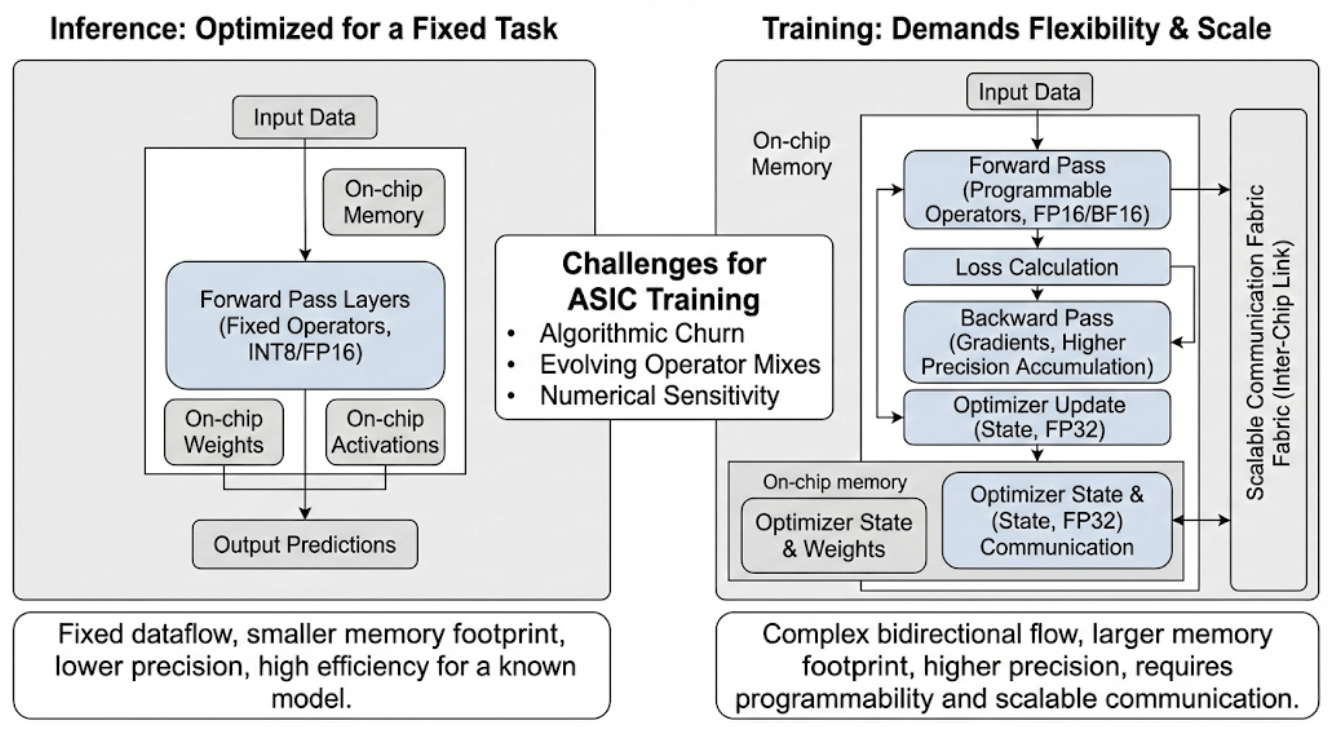}
\caption{Challenges in designing ASICs for training, highlighting the need for higher precision, backward-pass support, and flexibility to handle evolving optimization algorithms compared to inference-only designs.}
\label{fig:section19}
\end{figure}

\subsubsection{Training on ASICs}
Training on fixed-function ASICs is harder due to frequent algorithmic changes, evolving operator mixes, and the need for flexibility in numerical formats and optimizers. Compared to inference, training requires backward operators, optimizer updates, and larger memory footprints for activations and optimizer state; it is also more sensitive to numerical accuracy and stability, which limits the most aggressive quantization options \cite{micikevicius2018mixedprecision}. As a result, training-capable ASICs typically provide a broader set of supported operations, higher precision accumulation, and a scalable communication substrate.

Nevertheless, domain-specific training systems show that carefully chosen primitives (matmul, collective communication, and a sufficiently expressive compiler/runtime) can support a wide range of training workloads while delivering strong performance-per-watt \cite{jouppi2021tpuv4,hennessy2019hwds}. The key is to co-design hardware capabilities with compiler abstractions so that new models can be expressed as compositions of supported primitives, and to ensure the memory system and interconnect are provisioned for the communication patterns of large-scale training.

\subsection{FPGA-based accelerators}
FPGAs offer reconfigurability and the ability to tailor datapaths, memory access, and I/O interfaces to a workload, making them attractive for edge inference, low-latency pipelines, and rapid architectural experimentation. FPGA accelerators commonly use deeply pipelined streaming architectures, custom precision, and specialized memory layouts, and can integrate tightly with sensors and networking to reduce end-to-end latency \cite{nurvitadhi2017can,umuroglu2017finn,venieris2017fpgaconvnet}. Unlike GPUs, FPGAs allow designers to specialize the datapath for a fixed model (or model family) and to exploit determinism in the workload, which can yield strong energy efficiency when memory bandwidth is sufficient and the operator set is well covered.

The FPGA design space also includes cloud-scale deployments where reconfigurable logic is used as a shared inference service. In these settings, throughput and tail latency depend not only on compute pipelines but also on host-device interfaces, batching policies, and the ability to switch between models with minimal reconfiguration overhead \cite{fowers2018brainwave,nurvitadhi2017can}. Toolchains and abstraction layers (HLS, OpenCL, compiler-based mapping) play an outsized role in productivity and performance portability, influencing how quickly designs can adapt to new model operators and sparsity patterns \cite{venieris2017fpgaconvnet}.

\begin{figure}[thbp]
\centering
\includegraphics[width=0.9\linewidth]{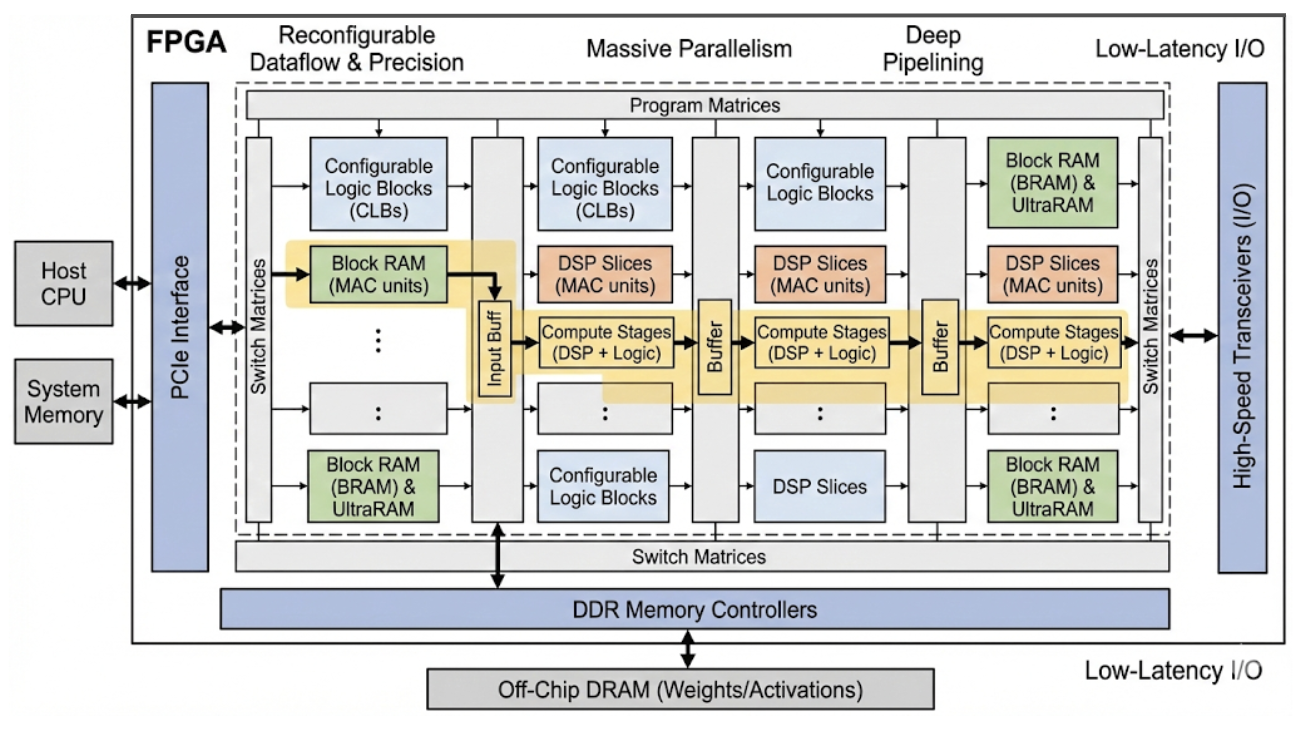}
\caption{FPGA-based acceleration architecture illustrating a streaming datapath approach with custom buffering and I/O integration, designed to minimize off-chip memory access for deterministic inference.}
\label{fig:section7}
\end{figure}

\subsubsection{Inference on FPGAs}
Figure~\ref{fig:section7} sketches a representative FPGA-based acceleration setup used to discuss streaming datapaths, buffering, and I/O constraints. Inference on FPGAs has a long history of efficient CNN and Binarized Neural Network (BNN) implementations using quantization, folding, and pipeline parallelism \cite{umuroglu2017finn,ma2018optimizing}. Designs typically trade flexibility for efficiency by fixing layer parameters and building a streaming pipeline that reuses on-chip buffers and minimizes off-chip traffic. When the model changes frequently or includes operators outside the accelerator’s coverage, the benefits can diminish due to host fallback and data marshaling overhead \cite{nurvitadhi2017can,venieris2017fpgaconvnet}.

Recent work also targets Transformer inference by accelerating attention, softmax, and feed-forward layers with structured sparsity patterns and custom compute engines \cite{FPGATrans}. Attention-heavy workloads stress memory bandwidth and require efficient buffering strategies for QKV projections and intermediate activations; they also benefit from kernel fusion to reduce intermediate writes. Compared to GPUs, FPGAs can deliver strong latency determinism and energy efficiency, but performance depends heavily on external memory bandwidth, operator coverage, and toolchain quality (HLS scheduling, memory partitioning, and DSP utilization) \cite{nurvitadhi2017can}. Recent work has also demonstrated the efficacy of FPGAs for accelerating Digital Twin learning and model recovery in mission-critical edge applications \cite{xu2025accelerated,xu2025fast,xu2025hardware}.

\subsubsection{Training on FPGAs}
Full training on FPGAs is less common due to limited on-chip memory, lower peak throughput than high-end GPUs, and challenges in supporting rapidly evolving training pipelines. Training requires storing intermediate activations and gradients, supporting backpropagation for a wide operator set, and performing optimizer updates, all of which can stress both FPGA resources and memory bandwidth. Mixed precision can help, but numerical stability and the complexity of implementing training kernels in reconfigurable logic remain barriers \cite{micikevicius2018mixedprecision}.

Hybrid approaches that offload selected kernels (e.g., matmul) or specialize for specific models and fixed training regimes can still be effective, particularly when the end-to-end pipeline benefits from tight I/O integration or when energy constraints dominate \cite{nurvitadhi2017can}. Another pragmatic approach is to use FPGAs for inference and for hardware-in-the-loop workloads, while reserving training for GPUs/TPUs, which allows the FPGA design to focus on deterministic low-latency execution.

\subsection{LLM-serving accelerators (LPUs and related designs)}
Recently, specialized accelerators have emerged targeting the inference-serving regime of large language models. These systems often emphasize predictable execution, low-latency token generation, and high throughput under real-time constraints, with architecture and software co-designed around attention-centric workloads and memory management \cite{groq2024lpu,kwon2023vllm}. Recent surveys highlight that achieving ``faster and lighter'' LLMs requires a combination of algorithmic compression (quantization, pruning) and system-level optimization, reinforcing that hardware cannot be designed in isolation from the serving stack \cite{chavan2024faster}. This deployment regime differs from classical batch inference: it combines strict latency targets with fluctuating concurrency, diverse prompt lengths, and complex scheduling policies for multi-tenant serving. As a result, accelerator design must address not only peak GEMM throughput but also memory capacity, memory bandwidth, and the overheads of dynamic batching and KV-cache allocation \cite{vaswani2017attention,kwon2023vllm}.

From an architectural perspective, LLM serving is dominated by repeated execution of relatively small GEMMs (per token) and bandwidth-heavy memory accesses for KV-cache. This encourages designs that reduce dispatch overhead, favor predictable pipelines, and provide sufficient memory bandwidth per unit of compute. Software remains central: runtime systems that manage KV-cache paging, kernel fusion, and scheduling are often as important as the hardware datapath \cite{kwon2023vllm,dao2022flashattention}.

\subsubsection{Inference on LPUs}

\begin{figure}[thbp]
\centering
\includegraphics[width=0.9\linewidth]{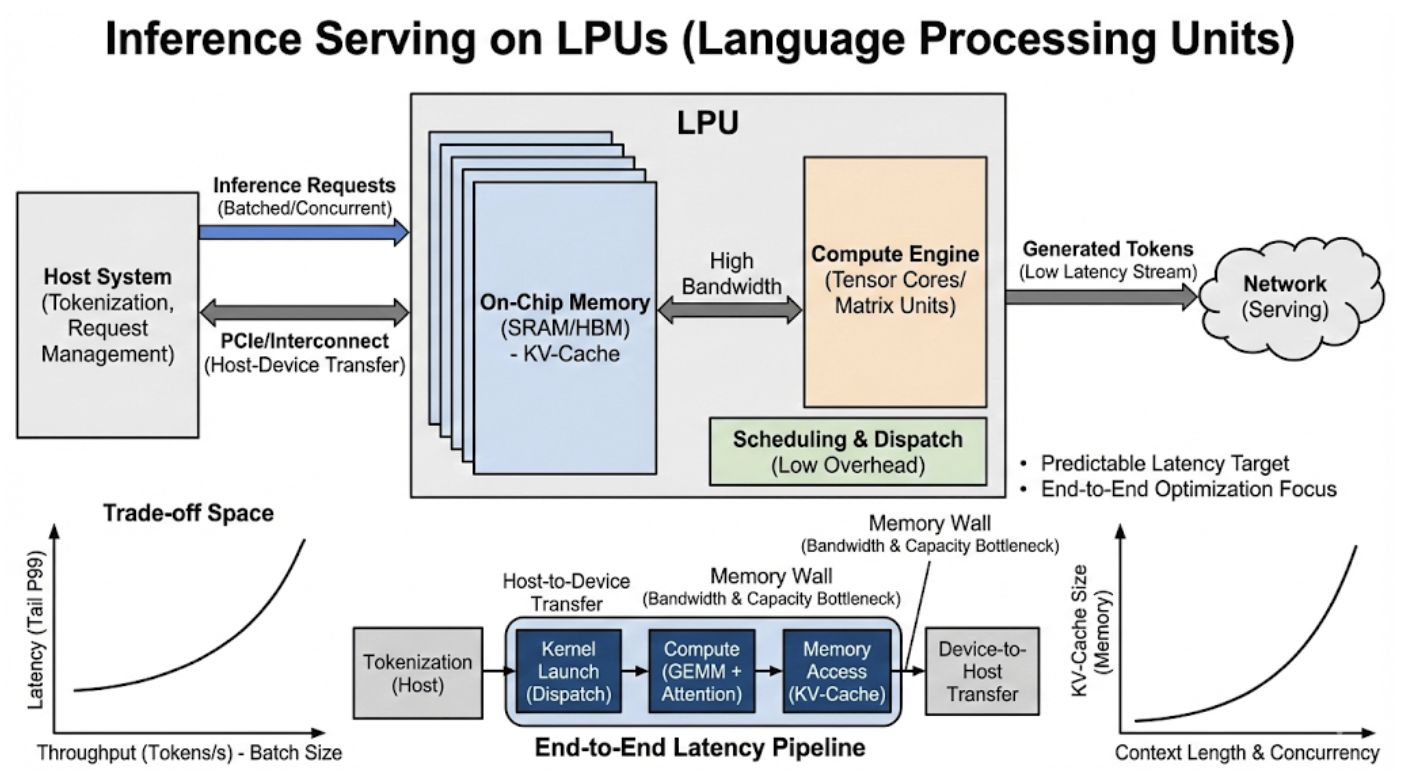}
\caption{Inference on LPU-style accelerators, focusing on deterministic low-latency token generation for LLMs by managing control flow in hardware and optimizing memory access patterns for the decode phase.}
\label{fig:section6}
\end{figure}

Inference serving is frequently bottlenecked by memory capacity and bandwidth due to KV-cache growth, especially for long contexts and high concurrency \cite{kwon2023vllm}, as contextualized in Figure~\ref{fig:section6}. LPU-style designs aim to reduce scheduling overheads and improve predictability, but they still must manage fundamental trade-offs between batching (throughput) and latency, as well as between compute provisioning and memory-system design \cite{groq2024lpu}. Systems-level factors such as tokenization overhead, host-device transfer, and kernel launch granularity can materially affect tail latency, motivating end-to-end pipeline design rather than isolated kernel optimization \cite{williams2009roofline}.

Another emerging factor is model heterogeneity: mixtures of experts, retrieval-augmented generation, and tool-calling introduce conditional execution and variable compute per token, which can reduce predictability and complicate scheduling. Serving accelerators must therefore either constrain the model interface (to preserve determinism) or provide sufficient flexibility in control and memory management to handle dynamic behaviors without collapsing utilization \cite{shazeer2017outrageously,gale2019state}.

\subsubsection{Training implications}
Most LPU-style systems primarily target inference rather than training. However, their emphasis on deterministic execution, low-overhead dispatch, and efficient attention kernels can inform training-system design (e.g., operator implementations and memory management), especially as training increasingly incorporates long-context sequences and MoE-style routing \cite{shazeer2017outrageously}. In particular, techniques that reduce attention memory traffic or that improve cache locality can benefit both inference and training, though training places additional demands on precision and backward-pass support \cite{micikevicius2018mixedprecision}.

More broadly, LLM serving highlights that memory is a first-class resource alongside FLOPs. Training systems already face similar pressures through activation storage and optimizer state, so ideas developed for inference memory management (paging, fragmentation control, layout-aware scheduling) may translate into improved training efficiency when combined with standard memory-reduction techniques \cite{rajbhandari2020zero}.

\subsection{In-/near-memory and analog accelerators}
To address the energy cost of data movement, in-/near-memory computing places computation closer to storage, often using crossbar arrays for analog matrix-vector multiplication. Such designs can offer high energy efficiency for dense linear algebra by exploiting physical laws for multiply-accumulate, but they face challenges in precision, device variability, ADC/DAC overheads, endurance, and system integration \cite{shafiee2016isaac,chi2016prime,anzaroot2019puma}. In addition, mapping modern networks is not only about matmul: normalization, attention/softmax, activation functions, and data-dependent control can erode the end-to-end benefit if they require frequent conversion between analog and digital domains \cite{vaswani2017attention}.

A key architectural question is where computation should occur relative to memory. Near-memory approaches may attach compute engines to DRAM/HBM stacks or to ReRAM banks, while in-memory approaches place compute directly inside the memory arrays. Both approaches attempt to reduce the energy per byte moved, but they trade this against increased complexity in data placement, programming models, and accuracy management \cite{chi2016prime,anzaroot2019puma}.

\subsubsection{Inference}
Analog in-memory accelerators are well matched to inference settings where weights are fixed and calibration can be amortized. Architectures such as ISAAC, PRIME, and PUMA demonstrate high throughput for convolution and fully connected layers when mapping is feasible and when the analog compute can be kept busy with sufficiently large matmuls \cite{shafiee2016isaac,chi2016prime,anzaroot2019puma}. Inference pipelines can also tolerate certain approximations if they preserve task-level accuracy, motivating quantization and algorithm-hardware co-design that accounts for analog noise and non-idealities \cite{jacob2018quantization}.

However, end-to-end inference increasingly includes components that are not straightforward dense matmuls. Attention introduces softmax and KV-cache-like access patterns, and modern LLM serving often becomes memory bound even on highly optimized digital accelerators \cite{vaswani2017attention,kwon2023vllm}. These trends suggest that in-memory/analog accelerators must either broaden operator support or be integrated into heterogeneous systems where they accelerate the dominant matmul portions while other processors handle control-heavy components.

\subsubsection{Training}
Training with in-memory/analog hardware is more challenging because weight updates, gradient noise, and optimizer dynamics increase sensitivity to numerical error and device drift. The backward pass introduces additional operators and accumulation patterns that are less tolerant to analog noise, and frequent weight updates stress device endurance and calibration overhead. Research explores mixed-signal training strategies and algorithm-hardware co-design, but broad applicability across model classes and training regimes remains an open problem \cite{hennessy2019hwds}.

One promising direction is hybrid training where analog crossbars accelerate the dominant GEMM/conv kernels while digital logic performs sensitive reductions and optimizer updates. Another is to use analog compute primarily for inference or fine-tuning regimes with constrained updates, while using conventional accelerators for full pretraining \cite{micikevicius2018mixedprecision}.

\subsection{Neuromorphic accelerators}
Neuromorphic processors explore event-driven computation and local learning rules, targeting sparse spiking workloads and ultra-low-power operation. Architecturally, they emphasize distributed local memory, spike-based communication, and asynchronous execution, which can be advantageous when activity is sparse and when always-on sensing requires strict energy budgets \cite{davies2018loihi,merolla2014truenorth}. However, these systems depart significantly from the dense linear algebra abstraction that dominates mainstream deep learning, so mapping conventional models typically requires conversion to spiking representations or training with spiking dynamics.

\begin{figure}[thbp]
\centering
\includegraphics[width=0.9\linewidth]{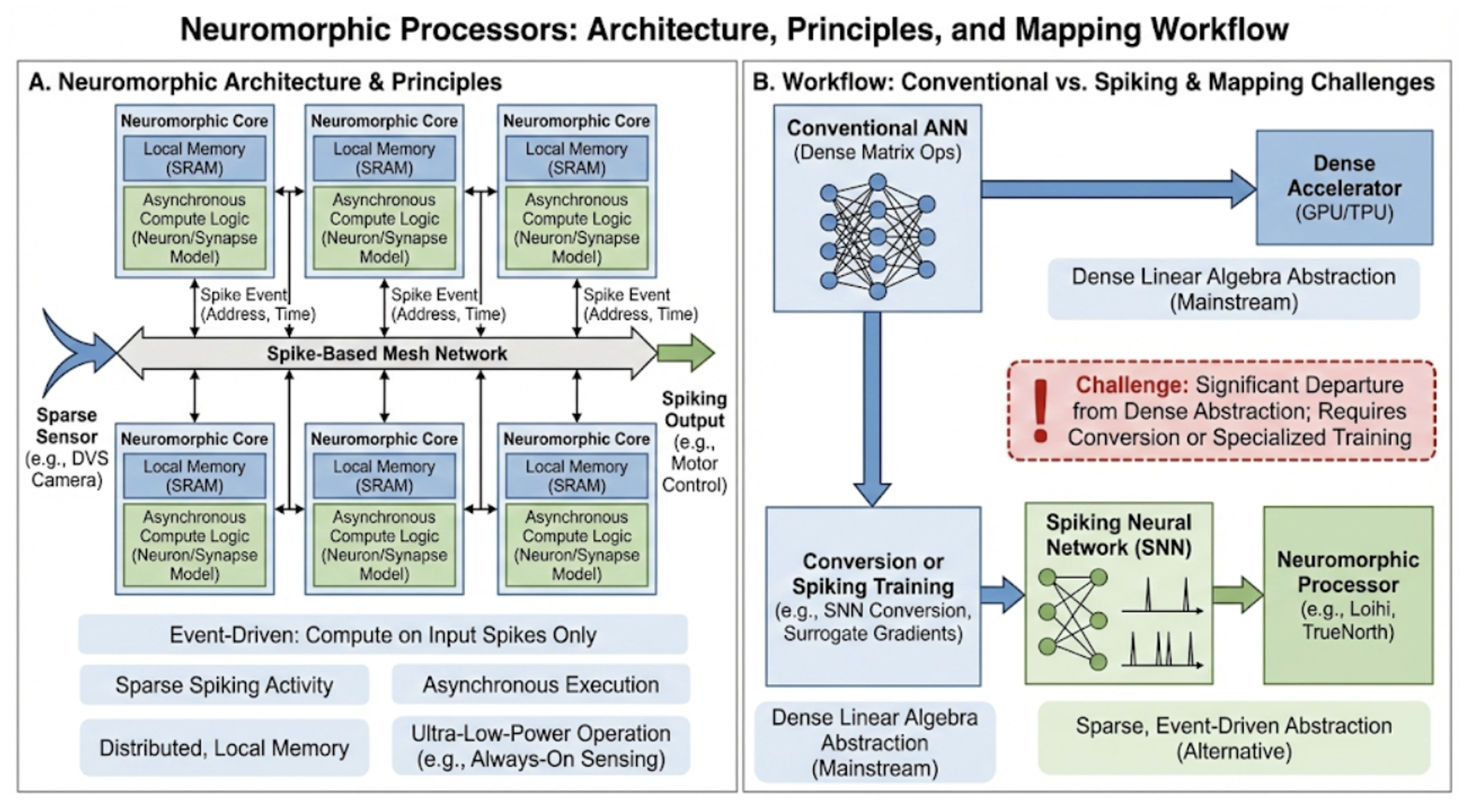}
\caption{Neuromorphic accelerator architecture emphasizing event-driven processing, distributed local memory, and asynchronous communication to achieve high energy efficiency for sparse workloads.}
\label{fig:section13}
\end{figure}

Neuromorphic accelerators are therefore best viewed as a complementary platform rather than a drop-in replacement for GPUs/TPUs. Figure~\ref{fig:section13} depicts a representative neuromorphic organization used to motivate event-driven execution, local memory, and spike-based communication. They can excel on certain temporal or sensory workloads, but their utility for large-scale dense models is limited by representation mismatch, toolchain maturity, and the cost of interfacing between event-driven and frame-based computation \cite{davies2018loihi}.

\subsubsection{Inference}
Event-driven inference can be efficient when activity is sparse and latency constraints are tight, as computation and communication occur only on spikes. This can make neuromorphic systems attractive for always-on perception, keyword spotting, or certain control workloads where sparse events capture the essential dynamics. Mapping conventional deep networks to spiking representations typically requires conversion or training with spiking dynamics, and performance depends strongly on sparsity and the cost of encoding/decoding between representations \cite{davies2018loihi,merolla2014truenorth}.

Compared to dense accelerators, the primary performance metric is often energy per inference rather than peak throughput. As with other sparse approaches, benefits depend on whether sparsity is structured and predictable enough for the hardware to exploit without introducing large overheads for routing and buffering \cite{gale2019state}.

\subsubsection{Training}
On-chip learning mechanisms exist for some neuromorphic systems, but training large-scale spiking models with competitive accuracy and efficiency remains an active research area. Learning rules may be local and biologically inspired, or they may approximate gradient-based updates, but bridging these approaches to mainstream training pipelines remains challenging. In many pipelines, neuromorphic devices are used primarily for inference, while training is performed offline on conventional accelerators \cite{davies2018loihi}.

A practical implication is that neuromorphic systems often participate in heterogeneous workflows: GPUs/TPUs perform training and model development, and neuromorphic devices execute specialized low-power inference when the application and representation are a good match. Improving toolchains for conversion, calibration, and deployment is therefore central to broader adoption \cite{davies2018loihi}.

\section{Accelerator Approaches}

\begin{figure}[thbp]
\centering
\includegraphics[width=0.9\linewidth]{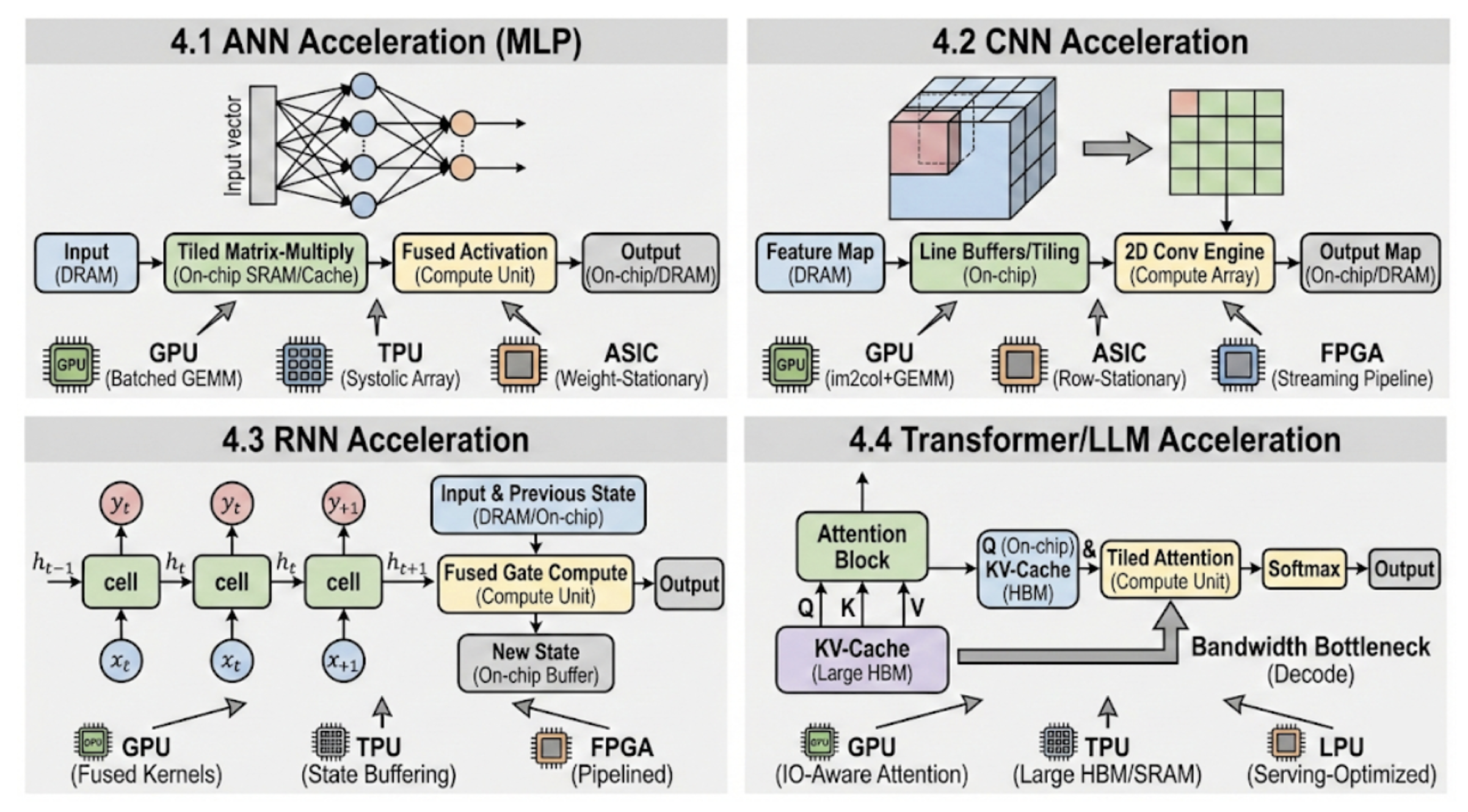}
\caption{Taxonomy of accelerator architectures tailored for different neural network types (ANNs, CNNs, RNNs, Transformers), showing how workload characteristics drive specific hardware design choices.}
\label{fig:section3}
\end{figure}

Several machine learning approaches, such as artificial neural networks (ANNs), convolutional neural networks (CNNs), and recurrent neural networks (RNNs), are implemented on hardware. Figure~\ref{fig:section3} provides an organizing view of accelerator architectures across these workload families. This section discusses common accelerator approaches for each category, emphasizing how workload structure drives compute mapping, dataflow, and memory-system design \cite{williams2009roofline,hennessy2019hwds}.

\subsection{ANN acceleration (MLP and fully connected networks)}
Classical feed-forward ANNs (e.g., multilayer perceptrons) are dominated by dense matrix multiplications (GEMMs) and elementwise activations. As a result, the primary accelerator approach is to maximize throughput on dense linear algebra while minimizing memory traffic:
\begin{itemize}
  \item \textbf{Matrix-engine mapping}: map each layer to GEMM on tensor cores / systolic arrays / MAC arrays, typically using mixed precision (FP16/BF16 for training; INT8 and below for inference when accuracy permits) \cite{micikevicius2018mixedprecision,jacob2018quantization,jouppi2017tpu}.
  \item \textbf{Tiling and reuse}: block weights and activations so they can be reused from on-chip storage (registers/SRAM) rather than repeatedly fetched from DRAM/HBM \cite{williams2009roofline}.
  \item \textbf{Fusion}: fuse activation and normalization chains with GEMM outputs to avoid materializing intermediate tensors in off-chip memory \cite{chen2018tvm,tillet2019triton}.
\end{itemize}

\begin{figure}[thbp]
\centering
\includegraphics[width=0.9\linewidth]{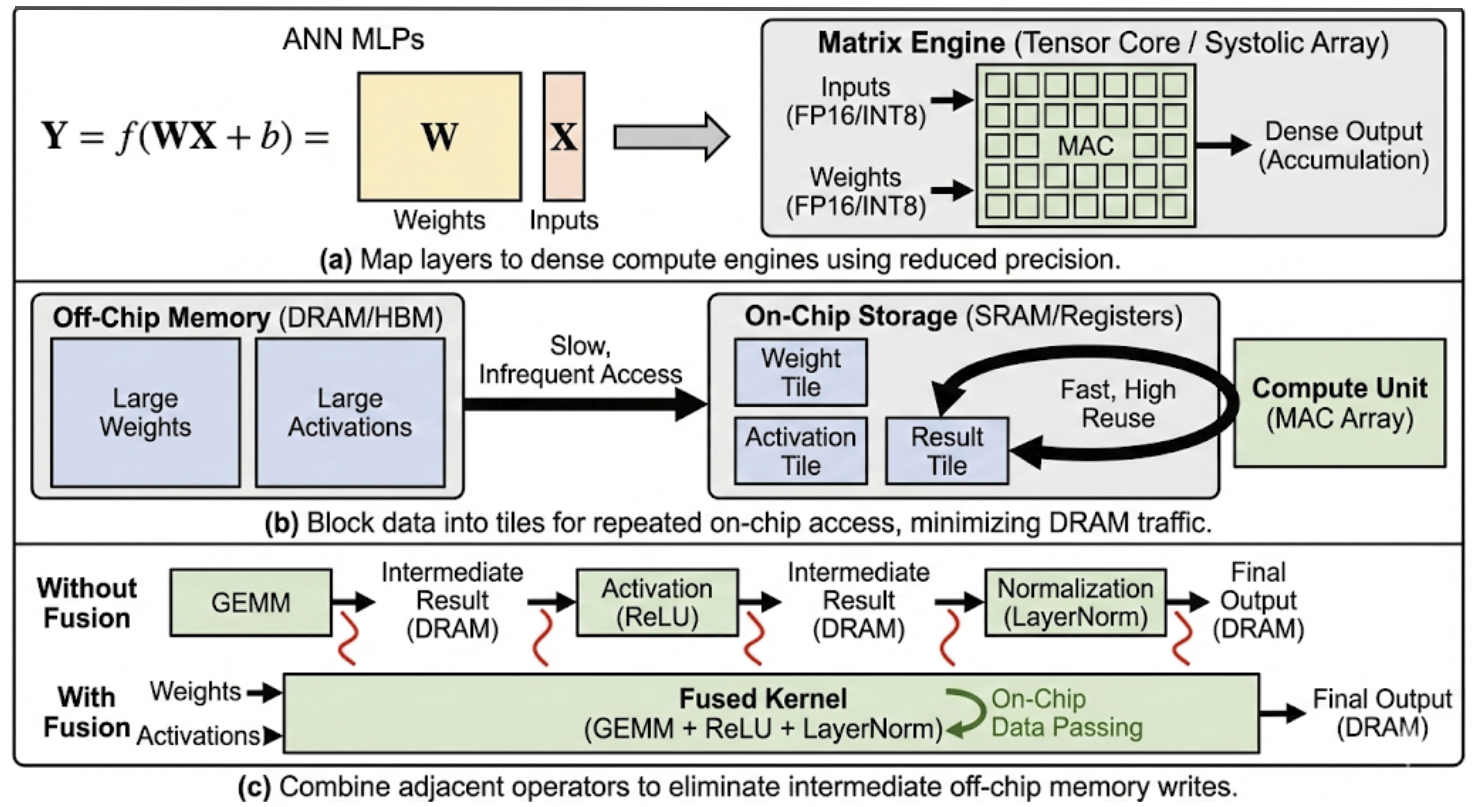}
\caption{Standard Artificial Neural Network (ANN) structure used as a reference for acceleration discussions, featuring dense fully connected layers that benefit from matrix-multiplication optimization.}
\label{fig:section15}
\end{figure}

When models are small or batch sizes are low, launch/dispatch overhead and memory latency can dominate; Figure~\ref{fig:section15} provides a reference ANN structure for the optimization discussion. Thus, practical ANN accelerators rely heavily on compiler/runtime fusion and layout selection to sustain utilization \cite{chen2018tvm,tillet2019triton}.

\subsubsection{ANNs on GPUs}
On GPUs, MLP layers are typically executed as a sequence of GEMMs where the software stack selects kernels, layouts, and epilogues (bias, activation, dropout) to reduce global-memory traffic and kernel launch overhead \cite{chetluar2014cudnn,tillet2019triton}. A common approach is to fuse as much of the per-layer ``tail'' as possible into the GEMM epilogue so that intermediate tensors remain in registers/shared memory rather than being written out and reloaded \cite{tillet2019triton,chen2018tvm}. For end-to-end networks, graph compilers further fuse adjacent pointwise operators and choose tensor layouts that improve locality and coalesced access, which can matter as much as peak GEMM throughput \cite{chen2018tvm,williams2009roofline}.

Training commonly uses mixed precision (FP16/BF16 compute with higher-precision accumulation) to increase effective throughput while maintaining numerical stability \cite{micikevicius2018mixedprecision}. Inference acceleration often focuses on quantized GEMMs (e.g., INT8) to improve throughput-per-watt and reduce memory bandwidth demand, provided the accuracy target allows \cite{jacob2018quantization}. Quantization is most effective when it is applied end-to-end so that repeated conversions do not reintroduce memory traffic and overhead \cite{jacob2018quantization}.

The dominant bottlenecks depend on regime. At large batch sizes and large hidden dimensions, the workload can approach compute-bound behavior; at small batch sizes, MLP inference becomes launch- and latency-dominated, making fusion and kernel granularity critical \cite{williams2009roofline,tillet2019triton}. Even when compute is ample, many MLP blocks are effectively bandwidth-bound because weights and activations must stream through the memory hierarchy every layer; reducing materialization (fusion), improving reuse (tiling), and selecting favorable layouts are therefore the primary ``accelerator approach'' on GPUs \cite{williams2009roofline,chen2018tvm}.

\subsubsection{ANNs on TPUs/NPUs}
TPU/NPU-style tensor processors accelerate ANNs by mapping GEMMs onto systolic arrays (or similar dense matrix engines) with compiler-managed tiling and on-chip SRAM reuse \cite{jouppi2017tpu,kung1982systolic}. The key approach is to treat the matrix engine as the ``throughput core'' and use the compiler to stage operands through on-chip buffers so the array receives a steady stream of tiles. This requires selecting blocking factors that match both SRAM capacity and the array geometry, and choosing layouts that minimize data reshaping between layers \cite{jouppi2017tpu,williams2009roofline}.

Because many MLP layers are structurally similar, these platforms benefit from static-shape compilation: the compiler can pre-plan data movement, fuse epilogues, and schedule transfers to overlap memory with compute. When shapes are dynamic or when layers are too small, utilization drops because the array cannot be fully occupied, and overheads (padding, layout transforms) become visible in end-to-end latency \cite{williams2009roofline}. In training settings, mixed precision plays a similar role as on GPUs: reduced-precision compute with higher-precision accumulation increases throughput while keeping convergence stable \cite{micikevicius2018mixedprecision}.

In edge NPUs integrated in SoCs, quantized inference (INT8 and below) is common because it reduces both compute energy and memory bandwidth. In practice, realized speedups depend heavily on \emph{operator coverage} and avoiding costly format conversions or CPU fallbacks in the middle of the graph \cite{jacob2018quantization,mazumder2021survey}. The ``accelerator approach'' here is therefore as much about compilation and operator availability as it is about raw TOPS: keep the computation inside the NPU with a consistent layout/precision so that data movement does not dominate \cite{williams2009roofline}.

\subsubsection{ANNs on ASIC accelerators}
ASIC inference engines accelerate fully connected layers using fixed or semi-programmable MAC arrays coupled with multi-bank on-chip SRAM scratchpads to keep weights and activations close to compute \cite{chen2014dadiannao,chen2016eyeriss}. The ``accelerator approach'' is to hardwire a small set of dense primitives (GEMM and simple elementwise ops) and then organize the memory system so that most accesses hit local SRAM rather than DRAM. Compared to GPUs, ASICs can reduce control overhead (scheduling, instruction fetch, dispatch) by constraining programmability and by specializing the datapath for the dominant operations \cite{hennessy2019hwds}.

Dataflow choice is central. Weight-stationary mappings are natural for MLP layers because weights are reused across many activations within a batch; output-stationary mappings can reduce partial-sum movement when accumulation dominates. The architecture provisions buffering and an on-chip interconnect to sustain the chosen reuse pattern and to stream tiles through the MAC array efficiently \cite{chen2016eyeriss,williams2009roofline}. When batch sizes are small, the ability to keep weights resident and to minimize DRAM traffic is still valuable, but the design must also avoid excessive underutilization due to limited parallelism.

Quantization and compression reduce footprint and bandwidth, but practical gains depend on whether the design supports efficient quantized datapaths and avoids repeated conversion or unpacking overhead \cite{jacob2018quantization,han2016deepcompression}. Unstructured sparsity can further reduce arithmetic, but it introduces indexing and load-balance overhead; benefits are largest when sparsity structure matches the hardware’s scheduling and dataflow \cite{gale2019state}. Overall, ASICs are most compelling when the operator set and deployment workload are stable enough that specialization remains useful over the product lifetime \cite{hennessy2019hwds}.

\subsubsection{ANNs on FPGAs}
On FPGAs, ANN acceleration often uses spatial pipelines: a GEMM is implemented as a tiled matrix-multiply engine with customized precision, and data are streamed through compute stages to avoid intermediate DRAM writes \cite{nurvitadhi2017can,venieris2017fpgaconvnet}. Rather than launching kernels, the design builds a fixed datapath where each cycle advances partial sums through a pipeline, enabling deterministic latency when the workload shape is fixed. Tiling controls how much parallelism is unrolled in hardware and how operands are buffered on-chip \cite{williams2009roofline}.

The approach emphasizes (i) matching the compute pipeline to available DSP resources, (ii) partitioning on-chip BRAM to supply sufficient parallel memory ports, and (iii) selecting quantized formats that reduce bandwidth while meeting accuracy constraints \cite{umuroglu2017finn,jacob2018quantization}. Quantized and binary arithmetic can map efficiently onto FPGA resources, but end-to-end efficiency depends on keeping data in a consistent format and avoiding expensive packing/unpacking steps at stage boundaries \cite{umuroglu2017finn,jacob2018quantization}.

Performance is frequently bounded by external memory bandwidth and by the overhead of moving tensors across the host--FPGA interface in system deployments \cite{nurvitadhi2017can}. Consequently, many effective FPGA ANN designs try to keep weights resident on-chip (when feasible), stream activations, and fuse adjacent elementwise operations into the datapath so that the pipeline remains compute-fed rather than I/O-limited \cite{williams2009roofline,venieris2017fpgaconvnet}.

\subsection{CNN acceleration}
CNNs are characterized by structured spatial reuse: the same filters are applied across many positions, and adjacent outputs share overlapping input windows. Hardware acceleration approaches therefore focus on exploiting this regularity:
\begin{itemize}
  \item \textbf{Convolution lowering and optimized kernels}: map convolution to direct conv kernels or GEMM-like forms, using carefully chosen data layouts and fusion of pointwise operators to reduce bandwidth \cite{chetluar2014cudnn,chen2018tvm}.
  \item \textbf{Specialized dataflows}: designs often adopt weight-/output-/row-stationary dataflows to maximize reuse and reduce DRAM energy, supported by multi-bank on-chip buffers \cite{chen2016eyeriss}.
  \item \textbf{Quantization and compression}: CNN inference commonly adopts INT8 (and lower) arithmetic; sparsity/compression reduce memory footprint but require kernels that preserve regular access and load balance \cite{jacob2018quantization,han2016deepcompression,gale2019state}.
\end{itemize}

\begin{figure}[thbp]
\centering
\includegraphics[width=0.9\linewidth]{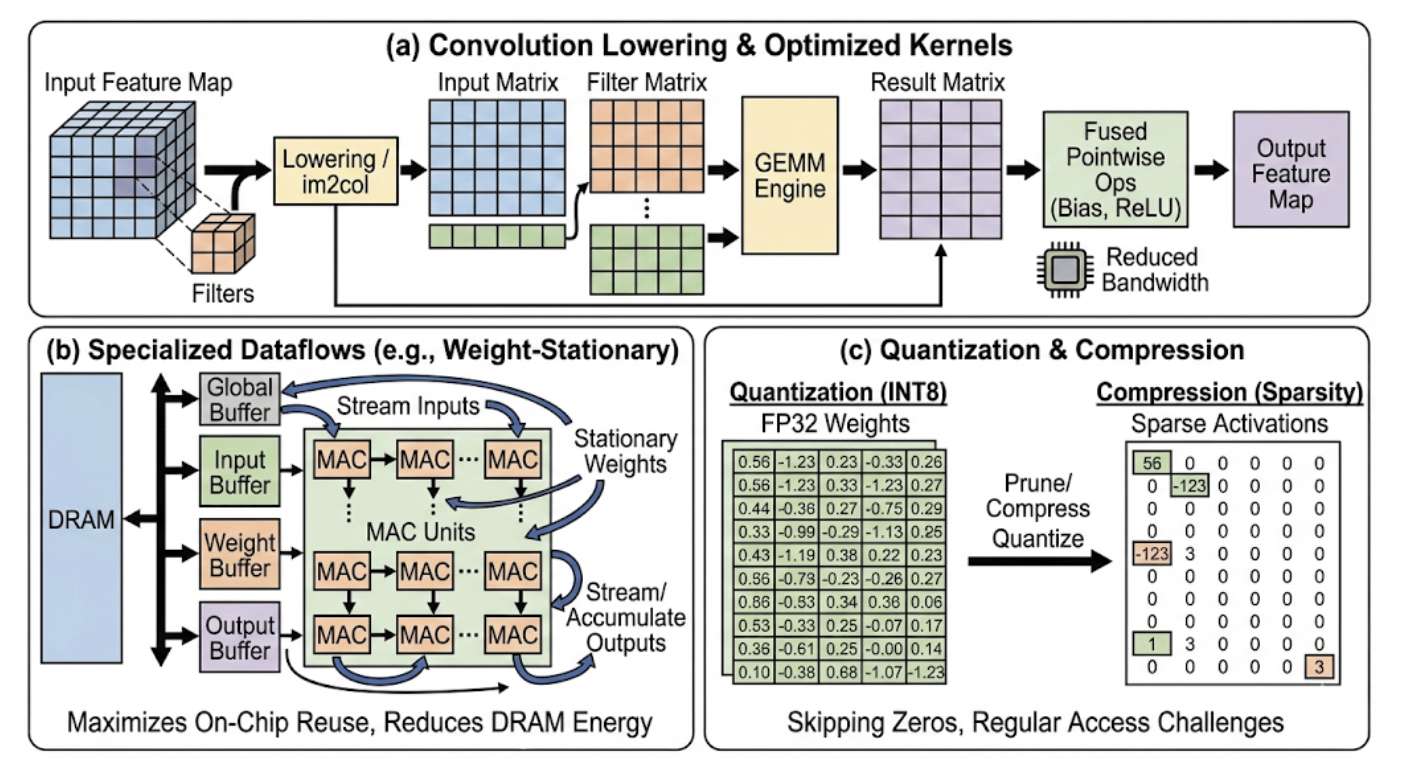}
\caption{Convolutional Neural Network (CNN) structure highlighting spatial locality and weight reuse, which are exploited by hardware through specialized dataflows and buffering strategies.}
\label{fig:section14}
\end{figure}

Compared to generic dense layers, CNN accelerators can achieve high energy efficiency because reuse is predictable and amenable to spatial architectures and streaming pipelines (e.g., ASIC and FPGA designs); Figure~\ref{fig:section14} summarizes a representative CNN structure \cite{chen2016eyeriss,umuroglu2017finn,venieris2017fpgaconvnet}.

\subsubsection{CNNs on GPUs}
GPUs accelerate CNNs primarily through highly tuned convolution kernels and implicit GEMM mappings, where the software stack selects layouts (e.g., NCHW variants) and fuses pointwise ops to reduce memory traffic \cite{chetluar2014cudnn,chen2018tvm}. In practice, convolution performance is shaped by data layout, tiling strategy, and how well the kernel uses on-chip memory (registers/shared memory/L2) to reuse input patches and filters. Vendor libraries provide many specialized kernels (for different strides, filter sizes, depthwise vs.\ standard conv), and compilers help select and fuse them into efficient end-to-end graphs \cite{chetluar2014cudnn,chen2018tvm}.

Mixed precision is common for training to increase throughput and reduce activation bandwidth, while inference frequently uses quantized convolution/GEMM for efficiency \cite{micikevicius2018mixedprecision,jacob2018quantization}. Quantized inference is most effective when adjacent layers can stay in the same quantized format and when activation/weight scales are handled without repeated conversion, otherwise the conversion overhead and extra memory traffic erode gains \cite{jacob2018quantization}. Sparsity and pruning can reduce arithmetic, but they require sparse-aware kernels; unstructured sparsity can introduce irregular access patterns and load imbalance that reduce utilization \cite{gale2019state}.

CNN performance is often limited by memory movement when kernels are not fused or when feature maps are small (reducing arithmetic intensity), motivating graph-level optimization, operator fusion, and careful kernel selection \cite{williams2009roofline,tillet2019triton}. For latency-sensitive inference, another limiting factor is kernel launch and synchronization overhead across many small layers; fusing conv+activation+normalization chains and using compiler-generated kernels helps amortize this overhead and reduce intermediate materialization \cite{chen2018tvm,tillet2019triton}.

\subsubsection{CNNs on TPUs/NPUs}
On TPUs/NPUs, convolutions are mapped to systolic arrays or dedicated convolution engines with compiler-selected tiling to maximize on-chip reuse \cite{jouppi2017tpu}. The compiler chooses how to block input feature maps, filters, and output tiles so that partial sums and reused data remain on-chip across many MAC operations. This is particularly effective for standard CNN layers with regular shapes, where blocking and layout decisions can be planned ahead of time and reused across many inferences \cite{williams2009roofline}.

Because CNN shapes are typically static in inference, compilers can pre-plan blocking and buffering to keep the array utilized and minimize off-chip traffic. For training, reduced precision compute (FP16/BF16) increases throughput in a similar manner as GPUs, but large activation tensors and gradient/optimizer state still stress memory capacity and bandwidth, making tiling and reuse central to efficiency \cite{micikevicius2018mixedprecision,williams2009roofline}. When models include less common operators or dynamic behavior, compilation may require additional layout transforms, which can increase memory traffic and reduce end-to-end performance.

Edge NPUs often target INT8 pipelines end-to-end; the key practical requirement is \emph{operator coverage} (conv, depthwise conv, pooling, activations) so that the graph does not fall back to CPU/GPU and incur extra copies and latency \cite{mazumder2021survey,jacob2018quantization}. Realized performance therefore depends not only on the conv engine but also on memory-system constraints (shared DRAM bandwidth on an SoC) and on whether quantization is applied consistently without repeated conversions \cite{jacob2018quantization}. When shapes deviate from expected regimes (very small feature maps, unusual channel counts) or when layers are too small, utilization can drop even if peak TOPS are high \cite{williams2009roofline}.

\subsubsection{CNNs on ASIC accelerators}
CNN ASICs exploit convolution’s regular reuse with spatial architectures and carefully chosen dataflows. Eyeriss demonstrated how row-stationary mapping and hierarchical buffering reduce energy by minimizing data movement, and similar ideas appear in many CNN inference engines \cite{chen2016eyeriss}. The accelerator approach is to bind the convolution loop nest to a fixed dataflow that maximizes local reuse: keep either weights, partial sums, or input tiles stationary in local buffers while streaming the remaining operands through the MAC array \cite{chen2016eyeriss,williams2009roofline}.

The hardware typically includes a MAC array, local buffers for weights/activations/partials, and an on-chip interconnect to sustain the chosen reuse pattern. DMA engines and buffering orchestrate streaming from DRAM, with the goal of turning expensive off-chip access into long, predictable bursts that can be amortized across many MACs. Because CNN layers vary in shape, practical ASICs may support multiple dataflow modes or configurable tiling so that both early (large feature map) and late (small feature map) layers run efficiently \cite{hennessy2019hwds}.

Quantization (INT8 and below) is widely used to reduce memory footprint and energy, while pruning/compression can help when sparsity is structured enough to be exploited without large indexing overhead \cite{jacob2018quantization,han2016deepcompression,gale2019state}. Unstructured sparsity can reduce arithmetic but can also introduce irregular memory access and load imbalance; the energy/performance benefit therefore depends on having sparse-aware scheduling and data formats that keep accesses regular enough for the memory system \cite{gale2019state}. Overall, CNN ASICs aim to trade flexibility for efficiency by specializing for the stable, regular operators that dominate CNN inference \cite{hennessy2019hwds}.

\subsubsection{CNNs on FPGAs}
FPGA CNN accelerators often use streaming dataflows where feature maps flow through a pipeline of convolution/activation stages, avoiding intermediate off-chip traffic and providing deterministic latency \cite{venieris2017fpgaconvnet,yan2024fpga}. A common pattern is to use line buffers and sliding-window datapaths that reuse input pixels across neighboring convolution windows, matching convolution’s spatial locality. The pipeline can be ``folded'' to trade throughput for area by time-multiplexing a smaller compute engine across channels or layers, while still maintaining a streaming schedule \cite{venieris2017fpgaconvnet,williams2009roofline}.

Quantized and even binary networks can map efficiently to FPGA LUT/DSP resources, enabling high throughput-per-watt when bandwidth is sufficient \cite{umuroglu2017finn}. In these designs, precision is a first-class architectural knob: fixed-point widths are chosen to reduce BRAM and external bandwidth while still meeting accuracy constraints, and keeping the entire pipeline in a consistent quantized format avoids conversion overhead \cite{umuroglu2017finn,jacob2018quantization}.

Practically, the design must balance DSP utilization, BRAM partitioning, and external memory bandwidth; otherwise the pipeline stalls on memory rather than compute \cite{nurvitadhi2017can,williams2009roofline}. System-level bottlenecks such as host I/O (PCIe), buffering, and reconfiguration overhead also affect end-to-end performance in deployment settings, motivating designs that stream directly from input sources and fuse as many layers as feasible into a single datapath \cite{nurvitadhi2017can,venieris2017fpgaconvnet}.

\subsection{RNN acceleration}
RNNs introduce temporal dependencies: hidden-state updates create a sequential critical path across timesteps. Acceleration approaches therefore combine dense-math specialization with techniques to reduce the overheads of recurrent execution:
\begin{itemize}
  \item \textbf{Gate fusion}: fuse multiple gate computations (e.g., in Long Short-Term Memory (LSTM)/Gated Recurrent Unit (GRU)-style cells) into fewer kernels to reduce memory traffic and kernel-launch overhead, improving latency at small batch sizes \cite{chetluar2014cudnn,tillet2019triton}.
  \item \textbf{Batching and sequence packing}: increase parallelism by batching sequences and packing variable-length inputs when latency constraints allow, trading off throughput against responsiveness \cite{williams2009roofline}.
  \item \textbf{On-chip state buffering}: keep recurrent state and frequently reused parameters in on-chip SRAM/registers where possible to reduce round-trips to DRAM \cite{williams2009roofline}.
\end{itemize}

\begin{figure}[thbp]
\centering
\includegraphics[width=0.9\linewidth]{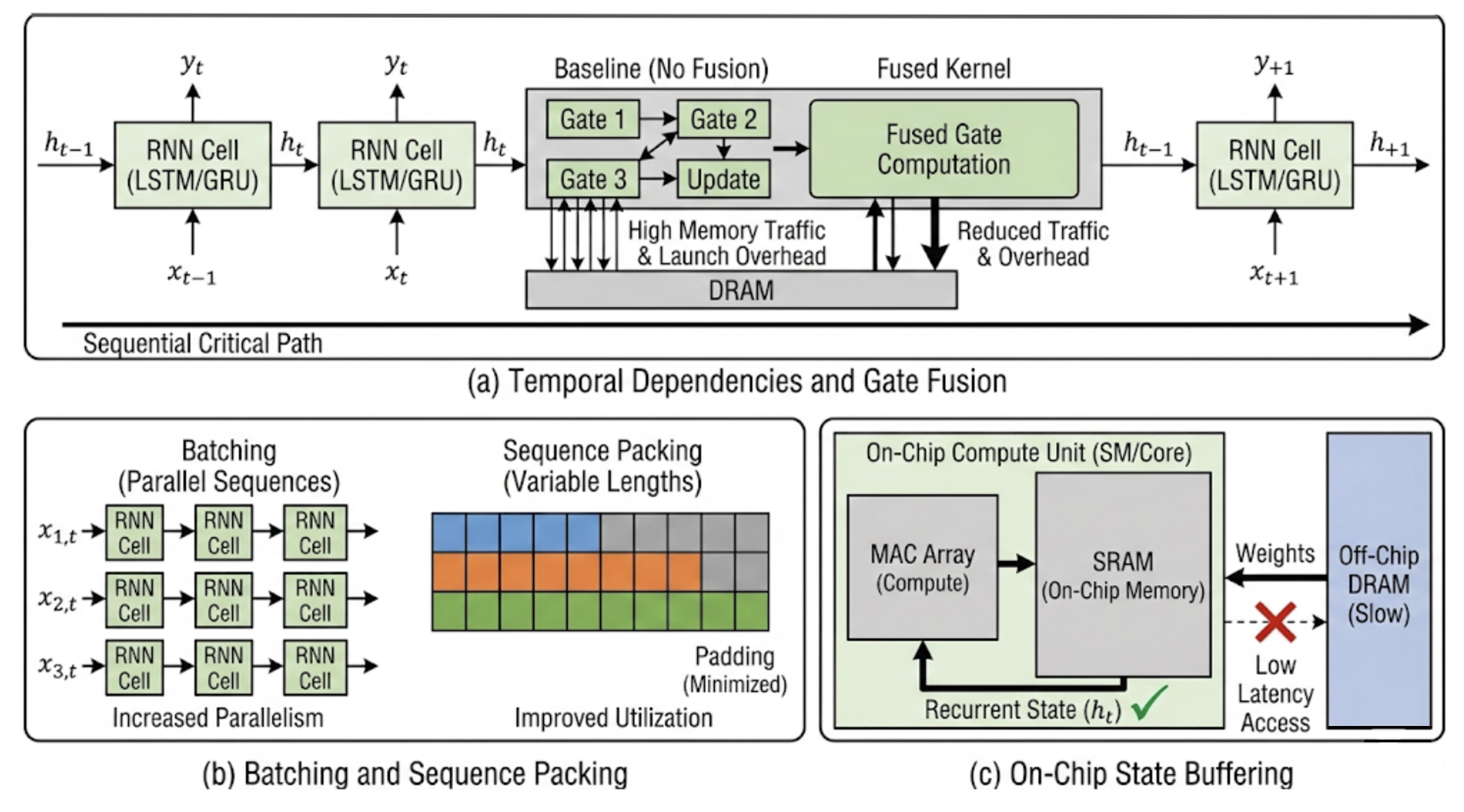}
\caption{Recurrent Neural Network (RNN) structure showing temporal dependencies and feedback loops, which present challenges for parallelization and require specialized state-buffering techniques.}
\label{fig:section16}
\end{figure}

In practice, RNN acceleration is often limited less by peak MAC throughput than by the ability of the software stack to schedule recurrent kernels efficiently and to minimize overhead between timesteps; Figure~\ref{fig:section16} provides the reference RNN structure for this discussion \cite{chen2018tvm}.

\subsubsection{RNNs on GPUs}
GPU RNN acceleration relies on fusing the per-timestep gate computations into a small number of kernels and using persistent/fused implementations that reduce kernel launches and global-memory traffic \cite{chetluar2014cudnn,tillet2019triton}. The main idea is to treat each timestep as a compact block (input projection + recurrent projection + nonlinearities) and to execute that block with minimal intermediate materialization, keeping gate activations and partial sums in registers/shared memory when possible \cite{tillet2019triton,chen2018tvm}. This is especially important for latency-sensitive inference, where each extra kernel launch or synchronization step adds overhead that cannot be amortized across many parallel tokens \cite{williams2009roofline}.

Mixed precision improves throughput for training, but sequential dependencies limit parallelism across timesteps, so utilization can be sensitive to batch size, hidden dimension, and sequence length \cite{micikevicius2018mixedprecision}. For high throughput, systems often rely on batching and sequence packing to increase parallel work per timestep, while for low latency they prioritize persistent kernels and fusion to minimize overhead between timesteps \cite{williams2009roofline}. Compiler stacks and kernel DSLs are increasingly used to generate fused recurrent kernels and to tailor layouts to specific hidden sizes, which helps reduce launch overhead and improve locality \cite{chen2018tvm,tillet2019triton}.

Practical bottlenecks include launch overhead (small batches), memory bandwidth (materialized intermediates), and synchronization between timesteps. When the hidden state cannot remain in fast memory, repeated reads/writes of state and gate tensors can become bandwidth-bound, making fusion and on-chip buffering more important than peak FLOPs \cite{williams2009roofline,chen2018tvm}. As a result, GPU ``acceleration'' for RNNs often looks like a compilation and kernel-engineering problem rather than a new hardware primitive: reduce the number of kernels, reduce memory traffic, and keep the recurrent loop tight \cite{tillet2019triton}.

\subsubsection{RNNs on TPUs/NPUs}
On tensor processors, RNNs are accelerated by mapping each timestep’s dense transforms onto systolic arrays, while the compiler/runtime manages loop structure, layout, and buffering to reuse weights and state \cite{jouppi2017tpu}. The primary approach is to keep the per-timestep GEMMs large enough (via batching or packing) to efficiently use the array, and to stage recurrent state through on-chip SRAM so that the loop does not become dominated by off-chip traffic \cite{williams2009roofline}. Because the computation repeats over many timesteps, keeping weights resident and minimizing layout changes across iterations can be critical for latency and energy.

Compared to CNNs, recurrent control flow can be harder to optimize ahead-of-time, and utilization depends on keeping the matrix engine busy despite timestep-level dependencies. When the compiler can unroll or partially unroll loops with static sequence lengths, it can fuse parts of the recurrent body and schedule data movement to overlap with compute; when sequence lengths are dynamic, additional bookkeeping and padding can reduce efficiency \cite{williams2009roofline}. Mixed precision provides throughput benefits for training, but the sequential structure still limits parallelism across time \cite{micikevicius2018mixedprecision}.

In edge NPUs, support depends on whether the device/compiler provides fused recurrent primitives; otherwise, repeated small operators may fall back and lose efficiency \cite{mazumder2021survey}. In these settings, end-to-end performance depends less on peak array throughput and more on staying within the NPU execution envelope (supported ops, supported layouts) so that the recurrent loop does not repeatedly cross the CPU/NPU boundary \cite{mazumder2021survey,williams2009roofline}.

\subsubsection{RNNs on ASIC accelerators}
Dedicated RNN ASIC acceleration is less common than CNN-focused designs, but the same principles apply: map gate GEMMs to a MAC array with on-chip SRAM buffers to keep recurrent state local and reduce DRAM traffic \cite{chen2014dadiannao}. The accelerator approach is to hardwire efficient dense transforms and to treat the recurrent state as a first-class on-chip resident tensor, minimizing off-chip reads/writes each timestep. With sufficient buffering, weights for the recurrent matrices can remain on-chip and be reused across timesteps, while activations stream through the datapath in a predictable order \cite{williams2009roofline}.

The main challenge is balancing efficiency with flexibility, since RNN variants and sequence lengths can vary and the sequential nature reduces opportunities for large-batch amortization \cite{hennessy2019hwds}. If the ASIC is specialized too narrowly (specific hidden sizes, specific cell types), it risks poor coverage as models evolve; if it is made more programmable, it loses some of the efficiency advantage that motivates ASIC design in the first place \cite{hennessy2019hwds}. Additionally, supporting training requires backward operators and higher precision accumulation, which further increases complexity compared to inference-oriented designs \cite{micikevicius2018mixedprecision}.

When such accelerators are deployed, they are most effective in stable inference workloads (fixed cell type, fixed dimensions) where deterministic execution and low DRAM traffic dominate performance and energy. In these regimes, specialized buffering, streaming schedules, and reduced-precision datapaths can yield strong efficiency, provided the system integration does not reintroduce overhead through host-device transfers \cite{williams2009roofline,hennessy2019hwds}.

\subsubsection{RNNs on FPGAs}
FPGAs can accelerate RNNs using deeply pipelined datapaths that stream inputs through fused gate computations, offering deterministic latency when dimensions are fixed \cite{nurvitadhi2017can}. Similar to FPGA CNNs, the design emphasizes a spatial schedule: compute for one timestep advances through a pipeline, while recurrent state is buffered and fed back with minimal control overhead. When hidden sizes are fixed, designers can unroll inner products and gates into a balanced pipeline that uses DSP blocks efficiently and sustains a steady initiation interval \cite{williams2009roofline}.

The approach benefits from keeping weights/state on-chip and using custom precision to reduce bandwidth and DSP usage. Fixed-point formats are often chosen to fit BRAM and reduce external bandwidth, and keeping the entire recurrent loop in a consistent format avoids conversion overhead \cite{jacob2018quantization}. Inference is generally a better fit than training because weight updates and backward-pass accumulation are harder to implement efficiently in a fixed pipeline \cite{micikevicius2018mixedprecision}.

However, long sequences can stress on-chip buffering, and performance is often bounded by external memory bandwidth or host I/O if the pipeline cannot keep data resident \cite{williams2009roofline,venieris2017fpgaconvnet}. In system deployments, additional overhead can come from host-device transfers and from how sequences are batched/packed; therefore, practical FPGA RNN acceleration often co-designs the data ingress/egress pipeline with the compute datapath so that the recurrent loop remains streaming rather than I/O-bound \cite{nurvitadhi2017can,williams2009roofline}.

\subsection{Transformer and LLM acceleration}
Transformers are now a dominant neural architecture for language, vision, and multimodal workloads, with large language models (LLMs) typically implemented as decoder-only Transformers \cite{vaswani2017attention}. While the compute core still consists of dense GEMMs (QKV projections and MLP blocks), end-to-end efficiency is increasingly dominated by \emph{attention} and by memory-system behavior (especially KV-cache during autoregressive inference) \cite{vaswani2017attention,kwon2023vllm,williams2009roofline}.

\subsubsection{Different Transformer and LLM model variants}
From an accelerator perspective, Transformer variants differ mainly in their attention pattern, sequence lengths, and whether execution is dense or conditional. These choices directly determine arithmetic intensity, memory footprint, and the amount of irregular control (routing, masking) that the hardware/software stack must handle efficiently \cite{vaswani2017attention,williams2009roofline}.
\begin{itemize}
  \item \textbf{Encoder-only Transformers}: used for representation learning (e.g., classification/embedding). They process the full sequence in parallel, so throughput is driven by batched matmuls and attention over the full context (e.g., Bidirectional Encoder Representations from Transformers (BERT)-style encoders) \cite{vaswani2017attention,devlin2019bert}.
  \item \textbf{Encoder--decoder Transformers}: used for sequence-to-sequence tasks. They combine self-attention with cross-attention, increasing memory traffic and operator count due to additional attention blocks and intermediate activations (e.g., T5-like models) \cite{vaswani2017attention,raffel2020t5}.
  \item \textbf{Decoder-only LLMs}: used for autoregressive generation (e.g., Generative Pre-trained Transformer (GPT)-style and LLaMA-style). Inference alternates between a \emph{prefill} phase (processing the prompt) and a \emph{decode} phase (generating tokens). Decode is often bandwidth-limited because each token requires reading/writing KV-cache and executing relatively small GEMMs \cite{brown2020gpt3,touvron2023llama,kwon2023vllm,williams2009roofline}.
  \item \textbf{Long-context models}: increase sequence length, amplifying KV-cache footprint and attention IO. This shifts bottlenecks from compute toward memory capacity/bandwidth and makes tiling and paging policies critical \cite{kwon2023vllm}. Architecturally, long-context variants may use sparse attention (e.g., Longformer/BigBird), recurrence (Transformer-XL), or linearized attention (Performer) to reduce the quadratic cost of attention, changing the balance between compute and memory access patterns \cite{beltagy2020longformer,zaheer2020bigbird,dai2019transformerxl,choromanski2021performer}.
  \item \textbf{Mixture-of-Experts (MoE) Transformers}: route tokens to a subset of experts, reducing arithmetic but introducing conditional execution, routing overhead, and load imbalance; benefits depend on runtime scheduling and interconnect (e.g., Switch Transformer) \cite{shazeer2017outrageously,fedus2022switchtransformer,gale2019state}.
\end{itemize}
These variants also differ in their \emph{serving behavior}. Encoder-only and encoder--decoder models are often used for batched inference with full-sequence computation, whereas decoder-only LLMs are typically deployed in interactive settings where decode speed (tokens/s) and tail latency matter most \cite{kwon2023vllm}. As a result, accelerator strategies that work well for training or offline inference (large GEMMs, large batches) can underperform for real-time decode unless kernels and runtimes are explicitly optimized for small, repeated per-token work \cite{williams2009roofline}.

\subsubsection{Key kernels and bottlenecks: attention, KV-cache, and fusion}

\begin{figure}[thbp]
\centering
\includegraphics[width=0.9\linewidth]{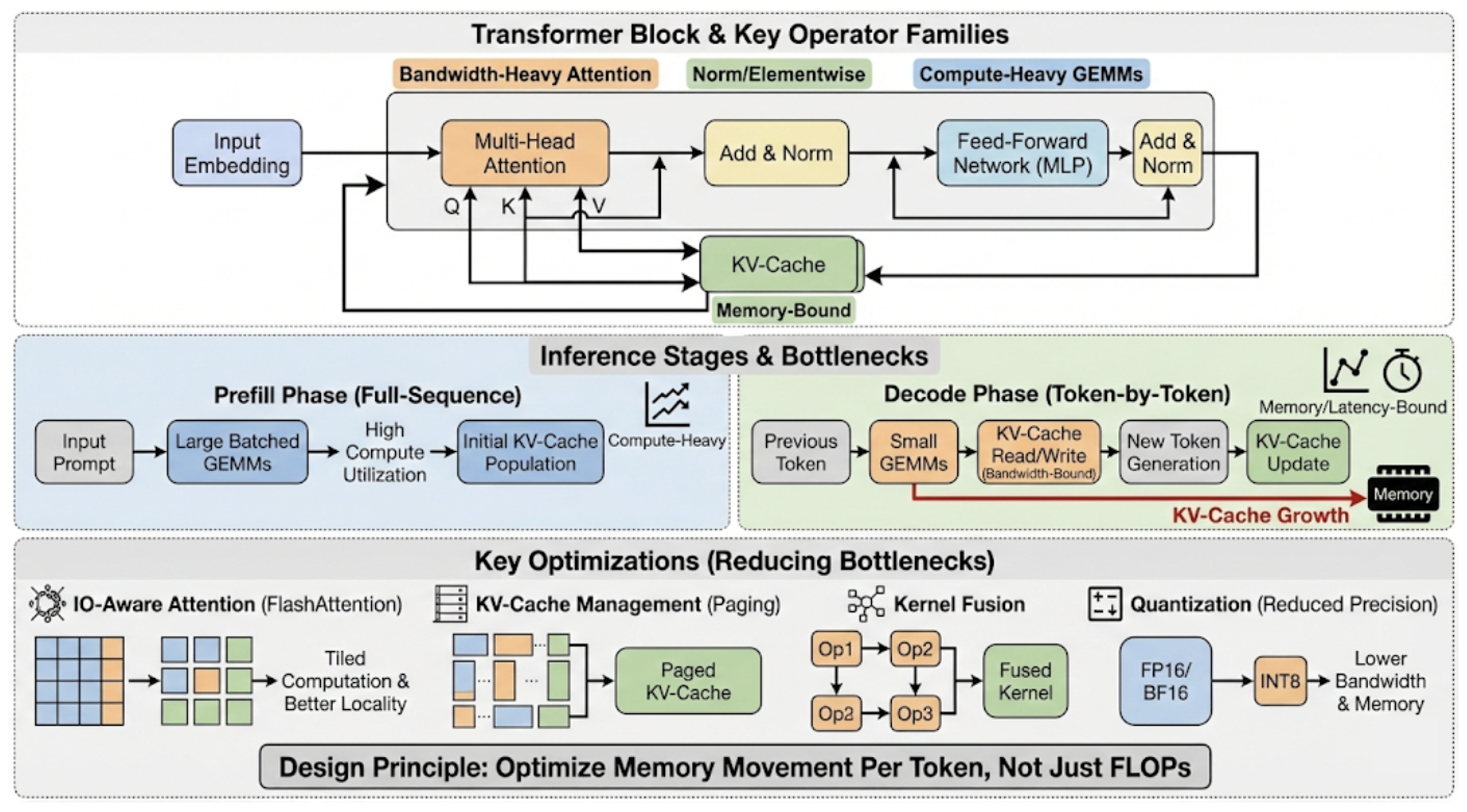}
\caption{Critical kernels and system bottlenecks in LLM acceleration, detailing the interplay between dense GEMM projections, bandwidth-intensive attention mechanisms, and the impact of sequence length on memory capacity.}
\label{fig:section5}
\end{figure}

Transformer blocks are a mix of compute-heavy GEMMs (projections and MLP) and bandwidth-heavy attention, as summarized in Figure~\ref{fig:section5}. A useful lens is to separate \emph{prefill} (full-sequence computation) from \emph{decode} (token-by-token generation) for decoder-only LLM inference. Prefill can be GEMM-heavy and benefits from batching, while decode often becomes memory- and latency-bound due to KV-cache access and small GEMM shapes \cite{kwon2023vllm,williams2009roofline}.

Key operator families include: (i) dense GEMMs for QKV projections and MLP blocks, (ii) attention score computation + softmax + weighted value aggregation, and (iii) normalization and elementwise chains (layer norm, activations). Many of these operators are individually optimized, but end-to-end performance depends on how well the stack \emph{reduces intermediate tensor materialization} and \emph{keeps data resident} in fast memory across the block \cite{chen2018tvm,tillet2019triton}.

During inference serving, KV-cache growth makes memory a first-class resource: bandwidth per generated token and KV-cache capacity often set the performance envelope, especially at long contexts and high concurrency \cite{kwon2023vllm,williams2009roofline}. IO-aware attention implementations tile computation to reduce intermediate materialization and improve locality (e.g., FlashAttention-style tiling), while runtime techniques such as paging KV-cache are critical to scaling concurrency under fixed device memory \cite{dao2022flashattention,kwon2023vllm}. These techniques illustrate a key design principle for LLM accelerators: optimize \emph{memory movement per token}, not just FLOPs.

Fusion and compilation further matter because Transformer blocks contain many short operator chains around the GEMMs. Fused kernels reduce launch overhead and global-memory traffic across the attention--MLP pipeline, and layout/epilogue selection often determines whether the workload is compute-bound or bandwidth-bound \cite{tillet2019triton,chen2018tvm,williams2009roofline}. Precision choices matter throughout: mixed precision is standard for training, while quantized inference reduces bandwidth and improves throughput-per-watt when the stack supports it end-to-end \cite{micikevicius2018mixedprecision,jacob2018quantization}. Sparsity can reduce arithmetic, but it introduces irregular access patterns and scheduling challenges unless sparsity structure is supported by kernels and data formats \cite{gale2019state}.

Recent acceleration strategies for Transformer/LLM inference increasingly target \emph{decode efficiency} and \emph{memory traffic per token}. Examples include faster IO-aware attention implementations beyond the original FlashAttention design \cite{dao2022flashattention,dao2023flashattention2}, attention variants that reduce KV-cache bandwidth (e.g., grouped-query attention, which reduces the number of KV heads relative to Q heads) \cite{ainslie2023gqa}, and serving-time techniques such as speculative decoding that trade extra parallel work for lower end-to-end latency \cite{leviathan2023speculative}. Post-training quantization has also evolved into practical, high-accuracy pipelines for LLMs---including SmoothQuant, GPTQ, and activation-aware weight quantization (AWQ)---which reduce bandwidth and enable lower-precision kernels when supported by hardware and libraries \cite{xiao2022smoothquant,frantar2022gptq,lin2023awq,jacob2018quantization,gong2024survey}. More radically, the "1-bit LLM" paradigm (e.g., BitNet b1.58) proposes ternary weights \(\{-1, 0, 1\}\) to drastically reduce memory footprint and replace expensive FP16 multiplications with integer addition, promising significant energy savings if hardware support matures \cite{ma2024era}.

\subsubsection{Transformers/LLMs on GPUs}
On GPUs, the dominant approach is to maximize tensor-core utilization for GEMMs while using specialized attention kernels and aggressive fusion to reduce memory traffic \cite{tillet2019triton,dao2022flashattention}. For training and offline inference, large batched GEMMs in projections and MLPs can reach high utilization; for interactive LLM serving, performance hinges on decode efficiency where GEMMs are smaller and the runtime must orchestrate many concurrent requests \cite{kwon2023vllm,williams2009roofline}.

Practical LLM serving on GPUs emphasizes (i) efficient decode for small GEMMs (kernel selection and fusion), (ii) KV-cache layout and paging to scale concurrency under limited device memory, and (iii) dynamic batching/scheduling to balance throughput and tail latency \cite{kwon2023vllm}. Attention kernels are particularly important: IO-aware tiling reduces intermediate tensor size and improves bandwidth efficiency, which can directly improve tokens/s when the workload is KV-cache and bandwidth dominated \cite{dao2022flashattention}. Graph compilers and kernel DSLs are used to fuse epilogues and pointwise chains around GEMMs to reduce launch overhead and global memory traffic \cite{chen2018tvm,tillet2019triton}.

Mixed precision (FP16/BF16) is widely used for training; for inference, quantization can be effective but requires kernel/library support and careful handling of conversions to avoid eroding gains \cite{micikevicius2018mixedprecision,jacob2018quantization}. In addition, memory hierarchy effects matter: even with fast GEMMs, repeated reads of KV-cache and activations can saturate bandwidth and stall compute, making layout and caching policy central to performance \cite{williams2009roofline,kwon2023vllm}.

When MoE is used, GPU efficiency depends on routing implementation, expert parallelism, and load balance; otherwise, conditional execution can reduce utilization and increase tail latency even when compute capacity is high \cite{shazeer2017outrageously,gale2019state}. Overall, GPUs remain the default platform because the software ecosystem (libraries, compilers, serving runtimes) evolves rapidly with model variants and can incorporate new attention kernels and scheduling policies without changing hardware \cite{chen2018tvm,tillet2019triton}. Continued architectural scaling is evident in recent generations such as NVIDIA's Blackwell platform, which integrates narrower precision support (FP4/FP6) and massive chip-to-chip bandwidth to support trillion-parameter training and inference \cite{nvidia_blackwell_2024}. Similarly, AMD's Instinct MI300X series targets the same high-bandwidth, high-capacity regime by integrating CPU and GPU cores with large shared HBM pools to reduce host-device transfer overheads \cite{amd_mi300x_2024}.

\subsubsection{Transformers/LLMs on TPUs/NPUs}
On TPUs/NPUs, dense GEMMs map naturally to systolic arrays, and compilation plans tiling/layout to keep the array utilized while managing on-chip SRAM reuse \cite{jouppi2017tpu,kung1982systolic}. A central accelerator approach is therefore \emph{compile-time dataflow planning}: choose blocking that maximizes operand reuse, schedule transfers to and from SRAM, and select layouts that reduce reshaping between layers \cite{jouppi2017tpu,williams2009roofline}.

For large-scale training, TPU-class systems demonstrate that system co-design (HBM bandwidth, on-chip SRAM, and high-bandwidth interconnect) matters as much as per-chip compute. Scaling LLM training requires sharding and collective communication that interact with compilation decisions, and overall efficiency depends on overlap between compute and communication as well as on memory footprint (activations, optimizer state) \cite{jouppi2021tpuv4,micikevicius2018mixedprecision}. Recent iterations, such as Google's TPU v5p, further optimize this balance by increasing inter-chip interconnect bandwidth and HBM capacity specifically to handle the communication-intensive training of large generative models \cite{google_tpu_v5p_2024}. Reduced precision (BF16/FP16) is widely used to increase throughput and manage memory traffic \cite{micikevicius2018mixedprecision}.

The main challenge for inference and serving is that attention and KV-cache can become bandwidth- and capacity-limited at long contexts and high concurrency, reducing utilization even when peak compute is high \cite{kwon2023vllm,williams2009roofline}. Similar to GPUs, IO-aware attention and careful KV-cache management are required to avoid turning the execution into a sequence of bandwidth-bound memory operations \cite{dao2022flashattention,kwon2023vllm}.

In edge NPUs, feasibility depends on operator coverage (attention, softmax, layer norm) and on keeping the graph inside the accelerator without CPU fallbacks, because cross-device fallbacks introduce extra copies and latency \cite{mazumder2021survey}. Quantized inference is common for power/thermal reasons, but end-to-end gains depend on consistent quantized execution and minimizing conversions \cite{jacob2018quantization}.

\subsubsection{Transformers/LLMs on ASICs and LLM-serving accelerators}
ASIC approaches aim to optimize the stable core (dense GEMM + attention primitives) with efficient dataflows and large on-chip buffers, but end-to-end wins depend on provisioning memory bandwidth/capacity for KV-cache and intermediate tensors \cite{chen2016eyeriss,kwon2023vllm}. From a design standpoint, the challenge is that Transformer inference mixes a matmul-centric datapath with bandwidth-heavy attention and cache access; thus, an ASIC must balance MAC throughput with memory-system throughput, rather than maximizing one in isolation \cite{williams2009roofline}.

A common approach is to implement a high-throughput matrix engine (MAC array or systolic array) and pair it with multi-level buffering and an on-chip interconnect that sustains a chosen dataflow. Large on-chip SRAM helps reduce DRAM traffic for weights/activations that can be tiled, but KV-cache for long-context serving often exceeds on-chip capacity, forcing reliance on high external bandwidth and careful cache layout \cite{kwon2023vllm}. Attention and softmax-like kernels may require dedicated support or efficient microcode because they can become performance-critical in decode regimes \cite{vaswani2017attention,dao2022flashattention}.

Dedicated LLM-serving accelerators (e.g., LPU-style designs) emphasize predictable low-overhead execution for token generation by reducing dispatch complexity and targeting deterministic pipelines \cite{groq2024lpu}. However, they remain constrained by the fundamental bandwidth-per-token cost of reading/writing KV-cache, which sets a system-level lower bound on achievable latency and energy per token at high concurrency \cite{kwon2023vllm,williams2009roofline}. As a result, practical designs focus on holistic system efficiency: memory provisioning, cache management, and minimizing control overheads that add to tail latency \cite{kwon2023vllm}.

For MoE-like workloads, system support for routing and load balance is essential; otherwise, conditional execution shifts the bottleneck to scheduling and communication \cite{shazeer2017outrageously,gale2019state}. Compared with GPUs, ASICs can offer strong efficiency for stable operator sets, but they risk underperforming when models evolve rapidly (new attention variants, new normalization/activation patterns) unless sufficient programmability is provided \cite{hennessy2019hwds}.

\subsubsection{Transformers/LLMs on FPGAs}
\begin{figure}[thbp]
\centering
\includegraphics[width=0.9\linewidth]{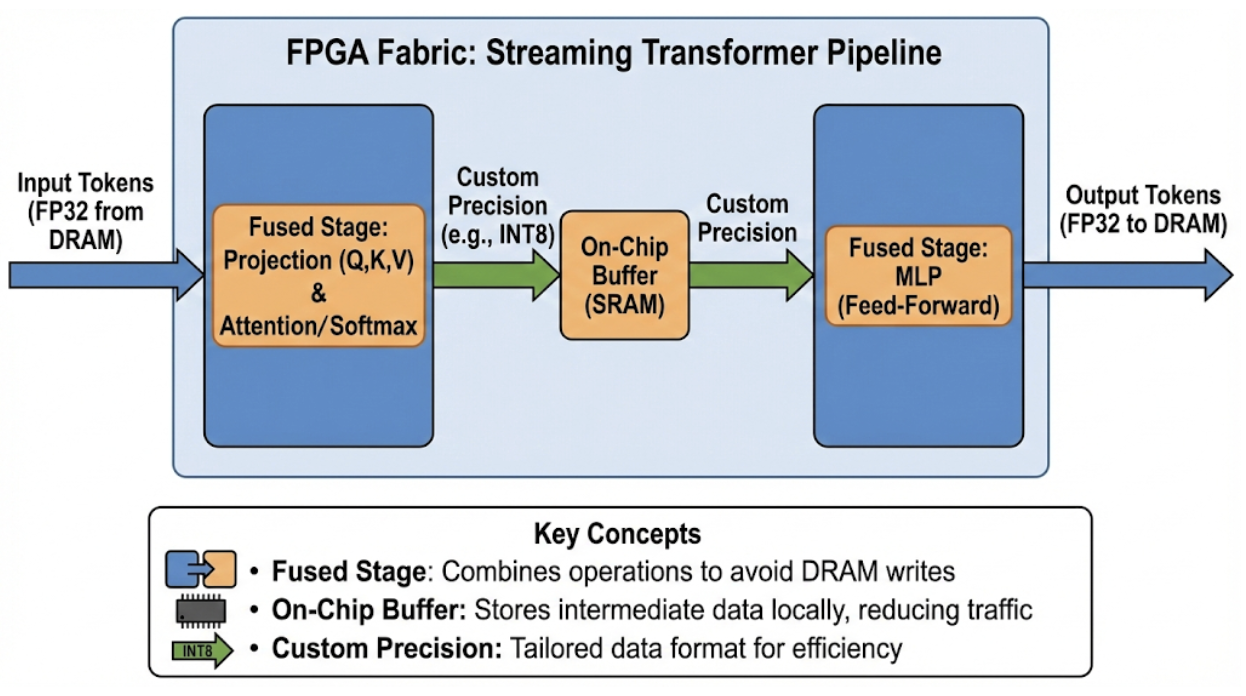}
\caption{FPGA acceleration pipeline for Transformers, demonstrating how streaming dataflows, custom precision, and fused operators are used to handle attention mechanisms and reduce latency.}
\label{fig:section20}
\end{figure}

FPGA acceleration(Fig.\ref{fig:section20}) for Transformers focuses on building streaming pipelines for projections/MLP and (when feasible) attention/softmax, using custom precision and on-chip buffering to reduce off-chip traffic \cite{FPGATrans,nurvitadhi2017can,boutros2024fpga}. The FPGA approach mirrors classical CNN FPGA acceleration: exploit spatial pipelining and determinism, tailor datapath precision, and fuse adjacent stages so that intermediate tensors do not spill to DRAM \cite{venieris2017fpgaconvnet,williams2009roofline}.

In practice, attention and KV-cache pressure external bandwidth, and limited on-chip BRAM can force frequent off-chip accesses, reducing throughput \cite{FPGATrans,kwon2023vllm,williams2009roofline}. This is especially challenging for decode, where the computation per token is small but the KV-cache traffic is unavoidable. As a result, FPGA designs often explore structured sparsity or block-sparse attention patterns, and they rely on careful buffering and tiling to keep memory access regular \cite{FPGATrans}.

System integration is a major determinant of end-to-end benefit. Host I/O (e.g., PCIe), data marshaling, and batching policies can dominate latency unless the serving pipeline is engineered so that the FPGA remains compute-fed \cite{nurvitadhi2017can}. Consequently, FPGA designs often target fixed model shapes and low-latency regimes where deterministic pipelines and tight I/O integration offset lower peak compute, while toolchain support and memory-system design remain central to achieving robust speedups \cite{venieris2017fpgaconvnet,nurvitadhi2017can}.

More broadly, recent Transformer/LLM platforms reflect a converging set of system requirements: (i) very high memory bandwidth and capacity (to sustain KV-cache and large activations), (ii) hardware support for low-precision tensor math (to increase throughput and reduce bandwidth pressure), and (iii) high-bandwidth interconnect to scale training and to avoid host bottlenecks in serving \cite{williams2009roofline,kwon2023vllm}. In datacenters, this has driven larger, LLM-focused accelerator complexes: GPU systems emphasizing large-scale inference and deployment \cite{nvidia2023llm_inference_platforms}, TPU clusters designed for large-scale training/inference in pods (e.g., recent TPU generations reported as Trillium and Ironwood) \cite{tpu_wikipedia_2025}, and alternative form factors such as wafer-scale systems (e.g., CS-3/WSE-3) that reduce distributed communication by concentrating more compute and memory on a single device \cite{cerebras_wikipedia_2024}. In parallel, specialized serving accelerators that prioritize deterministic low-latency token generation remain an active direction, reflecting that decode is often dominated by memory movement and dispatch overhead rather than peak GEMM throughput \cite{groq2024lpu,kwon2023vllm}. For practitioners, the implication is consistent with the mapping principles above: match the model’s execution regime (training vs.\ serving; prefill vs.\ decode; long-context vs.\ short-context) to a platform whose \emph{memory system} (HBM capacity/bandwidth and interconnect) and \emph{software stack} (attention kernels, KV-cache management, quantization support) can sustain end-to-end throughput and tail latency \cite{dao2022flashattention,kwon2023vllm}.

Beyond mainstream GPU/TPU deployments, the 2023--2025 landscape also includes a growing set of alternative datacenter accelerators (e.g., AMD Instinct MI300-class systems, Intel Gaudi-class accelerators, and AWS Trainium-class instances) \cite{amd_instinct_wikipedia_2025,intel_gaudi_wikipedia_2025,aws_trainium_wikipedia_2025,silvano2025survey}. On the research side, there is active work on transformer-specific accelerators and system co-design: dynamic sparsity/activation pruning in transformer blocks (e.g., AccelTran) \cite{wang2023acceltran}, 3D heterogeneous manycore organizations for end-to-end transformer execution (e.g., HeTraX) \cite{zhang2024hetrax}, FPGA toolflow-driven LLM inference (e.g., HLS-based transformer implementations) \cite{li2024hlstransform}, and heterogeneous processing-in-memory (PIM) approaches that couple SRAM/HBM compute to reduce bandwidth bottlenecks \cite{kim2025hpim}. These efforts reinforce the same overarching lesson: as models scale and contexts grow, end-to-end speedups come from jointly optimizing kernels, memory systems, and runtimes---not from compute throughput alone \cite{williams2009roofline,kwon2023vllm}.

\section{Evaluation}
\label{sec:evaluation}

Because accelerator results are highly sensitive to \emph{model regime} (training vs.\ inference; prefill vs.\ decode), \emph{software stack} (kernels, compiler, runtime scheduling), and \emph{hardware} (compute, memory hierarchy, and interconnect), fair comparison requires reporting both \textbf{what was measured} and \textbf{how it was measured}. We emphasize end-to-end, user-facing metrics---including TTFT/ITL latency, throughput vs.\ goodput under concurrency, energy efficiency, roofline limits, KV-cache footprint, and scaling efficiency---to make results comparable across platforms (CPU, GPU, TPU/NPU, FPGA, ASIC, and emerging PIM-style systems).

\subsection{Workload decomposition: prefill vs.\ decode}
\label{sec:eval_prefill_decode}

For decoder-only LLM serving, it is often necessary to evaluate \emph{prefill} and \emph{decode} separately because they stress different resources. Let \(L_{\text{in}}\) be prompt length (tokens) and \(L_{\text{out}}\) be generated length. Define measured wall-clock times \(T_{\text{prefill}}\) and \(T_{\text{decode}}\). A common decomposition is:
\begin{equation}
T_{\text{total}} = T_{\text{prefill}} + T_{\text{decode}}.
\end{equation}
Prefill tends to be GEMM-heavy (benefits from batching), while decode is commonly bandwidth/latency dominated due to KV-cache traffic \cite{kwon2023vllm,williams2009roofline}.

\subsection{Core performance metrics}
\label{sec:eval_metrics}

\paragraph{Latency (per request).}
Report mean and tail latency (e.g., \(p50/p95/p99\)) for interactive settings, and state whether measurements include warmup, compilation, and cache effects. For \(N\) requests with end-to-end times \(\{T_i\}_{i=1}^{N}\), define percentile latency:
\begin{equation}
T_{p} = \mathrm{percentile}_{p}\left(\{T_i\}_{i=1}^{N}\right), \;\; p \in \{50,95,99\}.
\end{equation}
For LLM serving, also report \emph{time-to-first-token} (TTFT) and \emph{inter-token latency} (ITL) because users perceive responsiveness primarily through TTFT, while sustained throughput is largely determined by ITL. One simple decomposition is:
\begin{equation}
T_{\text{TTFT}} \approx T_{\text{queue}} + T_{\text{prefill}} + T_{\text{first-decode}}.
\end{equation}
For a request producing \(L_{\text{out}}\) tokens, the average ITL is:
\begin{equation}
\overline{T}_{\text{ITL}} = \frac{T_{\text{decode}}}{L_{\text{out}}}, \qquad
\mathrm{TP}_{\text{decode}} = \frac{1}{\overline{T}_{\text{ITL}}}.
\end{equation}
When reporting tail behavior, it is often useful to provide \(p99\) for TTFT and ITL separately, since queueing and scheduling can dominate tails under high concurrency \cite{kwon2023vllm}.

\paragraph{Throughput.}
Throughput is tokens processed per second. For prefill and decode:
\begin{equation}
\mathrm{TP}_{\text{prefill}} = \frac{L_{\text{in}}}{T_{\text{prefill}}}, \qquad
\mathrm{TP}_{\text{decode}}  = \frac{L_{\text{out}}}{T_{\text{decode}}}.
\end{equation}
In serving, also report \emph{system throughput} under concurrency \(C\): \(\mathrm{TP}_{\text{sys}} = \frac{\sum_{i=1}^{C} L_{\text{out},i}}{T_{\text{window}}}\) over a fixed measurement window \(T_{\text{window}}\), since scheduler and batching policies strongly affect realized tokens/s \cite{kwon2023vllm}. To connect throughput to service quality, define \emph{goodput} as the throughput of tokens (or requests) that satisfy a latency SLO (e.g., \(T_{\text{TTFT}} \le \tau_{\text{TTFT}}\) and/or \(T_{p99} \le \tau\)):
\begin{equation}
\mathrm{Goodput} = \frac{\sum_{i=1}^{C} L_{\text{out},i}\cdot \mathbf{1}[\text{SLO}(i)]}{T_{\text{window}}}.
\end{equation}
Goodput can decrease even as raw tokens/s increases when batching and queueing inflate tail latency; reporting both clarifies this trade-off.

\paragraph{Batching efficiency and scheduling overhead.}
Because many accelerators reach higher utilization at larger effective batch sizes, it is useful to quantify how efficiently throughput scales with batch size \(B\). Let \(\mathrm{TP}(B)\) be throughput at batch size \(B\); a simple batching efficiency is:
\begin{equation}
\eta_{\text{batch}}(B) = \frac{\mathrm{TP}(B)}{B \cdot \mathrm{TP}(1)}.
\end{equation}
For online serving, also report the fraction of time spent outside accelerator kernels (e.g., runtime dispatch, padding, memory copies). If \(T_{\text{kernel}}\) is time executing device kernels and \(T_{\text{total}}\) is end-to-end time, an overhead ratio is:
\begin{equation}
\rho_{\text{overhead}} = 1 - \frac{T_{\text{kernel}}}{T_{\text{total}}}.
\end{equation}

\paragraph{Energy and power (efficiency).}
Measure average power \(P\) (watts) and energy \(E\) (joules) over the measurement interval \(T\):
\begin{equation}
E = \int_{0}^{T} P(t)\,dt \approx P_{\text{avg}} \cdot T.
\end{equation}
Energy per token (useful across platforms) is:
\begin{equation}
\mathrm{EPT} = \frac{E}{L_{\text{in}} + L_{\text{out}}}.
\end{equation}
A widely used combined metric is energy--delay product (EDP):
\begin{equation}
\mathrm{EDP} = E \cdot T, \qquad \mathrm{ED}^2\mathrm{P} = E \cdot T^2.
\end{equation}
For easier cross-platform comparison, also report normalized efficiency such as tokens per joule and tokens per watt:
\begin{equation}
\mathrm{TPJ} = \frac{L_{\text{in}} + L_{\text{out}}}{E}, \qquad
\mathrm{TPW} = \frac{L_{\text{in}} + L_{\text{out}}}{P_{\text{avg}} \cdot T}.
\end{equation}

\paragraph{Utilization and achieved performance.}
Report achieved throughput relative to theoretical limits. If a run performs \(F\) floating-point operations in time \(T\), achieved FLOP/s is:
\begin{equation}
\mathrm{FLOP/s}_{\text{ach}} = \frac{F}{T}, \qquad
\mathrm{Utilization} = \frac{\mathrm{FLOP/s}_{\text{ach}}}{\mathrm{FLOP/s}_{\text{peak}}}.
\end{equation}
For bandwidth-limited regimes, similarly report achieved memory bandwidth \(B_{\text{ach}}\) and \(\frac{B_{\text{ach}}}{B_{\text{peak}}}\).

\paragraph{Roofline analysis (compute vs.\ memory bound).}
Roofline connects operational intensity \(I\) (FLOPs per byte moved from main memory) to achievable performance \(\mathcal{P}\):
\begin{equation}
I = \frac{F}{Q}, \qquad
\mathcal{P} \le \min\left(\mathcal{P}_{\text{peak}},\; I \cdot B_{\text{peak}}\right),
\end{equation}
where \(Q\) is bytes transferred and \(B_{\text{peak}}\) is peak memory bandwidth \cite{williams2009roofline}. This is particularly informative for decode, where KV-cache reads can push \(I\) low, making bandwidth the dominant limiter.

\paragraph{Memory footprint and KV-cache capacity.}
For long-context or high-concurrency LLM serving, peak memory often limits batch size or concurrency before compute saturates. A practical quantity to report is the KV-cache footprint as a function of context length. For a decoder-only transformer with \(N_{\ell}\) layers, KV heads \(H_{\text{kv}}\), head dimension \(d\), context length \(L\), and element size \(s\) bytes (e.g., \(s=2\) for FP16/BF16), a rough KV-cache size is:
\begin{equation}
M_{\text{KV}} \approx N_{\ell} \cdot 2 \cdot L \cdot H_{\text{kv}} \cdot d \cdot s,
\end{equation}
and scales linearly with the number of concurrent sequences held in memory. Reporting \(M_{\text{KV}}\) (and whether paging/compression is used) helps explain why two platforms with similar peak FLOP/s can behave very differently at the same concurrency \cite{kwon2023vllm}.

\paragraph{Scaling efficiency (multi-device).}
For multi-accelerator setups, report speedup and parallel efficiency as a function of device count \(n\):
\begin{equation}
S(n) = \frac{T(1)}{T(n)}, \qquad
E(n) = \frac{S(n)}{n}.
\end{equation}
Because communication patterns differ across data-, tensor-, and pipeline-parallel schemes, it is important to state the parallelization strategy and whether communication is overlapped with compute; otherwise speedup numbers are not comparable across platforms or software stacks \cite{mattson2020mlperf}.

\section{Conclusion}
This survey has reviewed how neural-network acceleration has progressed from simply increasing peak compute to optimizing end-to-end execution under memory, communication, energy, and software constraints. Across model families and deployment settings, the dominant costs increasingly come from moving and managing data (activations, weights, and state) rather than from arithmetic alone. As a result, the most successful accelerator platforms are those that treat compute, memory hierarchy, interconnect, and the software stack as a single co-designed system.

We organized the hardware landscape into several major architectural families, each reflecting a different point in the design space. GPUs provide broad programmability and remain central for training and general-purpose inference, while TPU/NPU-style designs emphasize dense tensor throughput and efficiency via structured datapaths. FPGAs occupy a niche where determinism, custom precision, and streaming pipelines are valuable, and ASIC inference engines pursue maximum performance-per-watt at the cost of flexibility. Emerging architectures---including LPU-like designs for predictable token generation, in-/near-memory and analog approaches aimed at reducing data movement, and neuromorphic/event-driven processors for sparse workloads---highlight continued experimentation as models and serving requirements evolve.

Several optimization levers recur throughout the survey: precision reduction and quantization to improve effective bandwidth and energy efficiency; sparsity and pruning to reduce compute and memory footprint (when supported end-to-end by kernels and scheduling); and compilation, fusion, and layout transformations that reduce intermediate materialization and kernel overheads. For LLM serving in particular, system performance is often governed by memory capacity and bandwidth, KV-cache management, and request scheduling, making runtime policies as important as kernel-level efficiency.

Looking ahead, progress will likely be driven by three system-level needs. First, scalable LLM serving must sustain high concurrency and long contexts without becoming bandwidth- or capacity-limited. Second, dynamic and conditional computation (e.g., routed experts, tool-augmented pipelines, and heterogeneous operators) requires robust support for irregular control flow and load balance. Third, evaluation practice must mature: meaningful comparisons should report end-to-end metrics and configurations that reflect real deployment constraints, rather than relying on peak throughput alone.

In conclusion, the next generation of neural acceleration will be defined less by raw compute density and more by efficient data orchestration and hardware--software co-design. As workloads diversify and deployment expands from datacenters to edge devices, architectures that combine adaptable software stacks with well-balanced memory and communication subsystems will remain the most impactful.

\section*{Acknowledgements}
This work was partially funded by DARPA (AMP, N6600120C4020; FIRE, P000050426), the NSF (FDT-Biotech, 2436801), and the Helmsley Charitable Trust (2-SRA-2017-503-M-B).

\appendix

\section{Acronyms}
\begin{longtable}{l|l}
\caption{List of Acronyms.} \label{tab:acronyms} \\
\toprule
\textbf{Acronym} & \textbf{Definition} \\
\midrule
\endfirsthead
\toprule
\textbf{Acronym} & \textbf{Definition} \\
\midrule
\endhead
\bottomrule
\endfoot
ADC & Analog-to-Digital Converter \\
ANN & Artificial Neural Network \\
ASIC & Application-Specific Integrated Circuit \\
BERT & Bidirectional Encoder Representations from Transformers \\
BF16 & Brain Floating Point (16-bit) \\
BNN & Binarized Neural Network \\
BRAM & Block RAM \\
CMOS & Complementary Metal-Oxide-Semiconductor \\
CNN & Convolutional Neural Network \\
DAC & Digital-to-Analog Converter \\
DRAM & Dynamic Random Access Memory \\
DSL & Domain-Specific Language \\
DSP & Digital Signal Processor \\
ECC & Error Correction Code \\
EDP & Energy-Delay Product \\
FLOPs & Floating Point Operations \\
FP16 & 16-bit Floating Point \\
FP32 & 32-bit Floating Point \\
FPGA & Field-Programmable Gate Array \\
GEMM & General Matrix Multiplication \\
GNN & Graph Neural Network \\
GPT & Generative Pre-trained Transformer \\
GPU & Graphics Processing Unit \\
GRU & Gated Recurrent Unit \\
HBM & High-Bandwidth Memory \\
HLS & High-Level Synthesis \\
INT4 & 4-bit Integer \\
INT8 & 8-bit Integer \\
IO & Input/Output \\
ISA & Instruction Set Architecture \\
ITL & Inter-Token Latency \\
JIT & Just-In-Time \\
KV-cache & Key-Value cache \\
LLM & Large Language Model \\
LPU & Language Processing Unit \\
LSTM & Long Short-Term Memory \\
LUT & Look-Up Table \\
MAC & Multiply-Accumulate \\
MIG & Multi-Instance GPU \\
MLP & Multilayer Perceptron \\
MoE & Mixture-of-Experts \\
NIC & Network Interface Card \\
NoC & Network-on-Chip \\
NPU & Neural Processing Unit \\
OS & Operating System \\
PCM & Phase-Change Memory \\
PHY & Physical Layer \\
QKV & Query-Key-Value \\
ReRAM & Resistive RAM \\
RNN & Recurrent Neural Network \\
RTL & Register Transfer Level \\
SIMD & Single Instruction, Multiple Data \\
SIMT & Single Instruction, Multiple Threads \\
SLO & Service Level Objective \\
SM & Streaming Multiprocessor \\
SoC & System-on-Chip \\
SRAM & Static Random Access Memory \\
SSD & Solid State Drive \\
STDP & Spike-Timing-Dependent Plasticity \\
TOPS & Tera Operations Per Second \\
TPU & Tensor Processing Unit \\
TTFT & Time-To-First-Token \\
\end{longtable}

\bibliographystyle{ACM-Reference-Format}
\bibliography{ref}


\begin{thebibliography}{88}

%%% the Web address.

\ifx \showCODEN    \undefined \def \showCODEN     #1{\unskip}     \fi
\ifx \showISBNx    \undefined \def \showISBNx     #1{\unskip}     \fi
\ifx \showISBNxiii \undefined \def \showISBNxiii  #1{\unskip}     \fi
\ifx \showISSN     \undefined \def \showISSN      #1{\unskip}     \fi
\ifx \showLCCN     \undefined \def \showLCCN      #1{\unskip}     \fi
\ifx \shownote     \undefined \def \shownote      #1{#1}          \fi
\ifx \showarticletitle \undefined \def \showarticletitle #1{#1}   \fi
\ifx \showURL      \undefined \def \showURL       {\relax}        \fi
% invisible to TeX
\providecommand\bibfield[2]{#2}
\providecommand\bibinfo[2]{#2}
\providecommand\natexlab[1]{#1}
\providecommand\showeprint[2][]{arXiv:#2}

\bibitem[Abadi et~al\mbox{.}(2016)]%
        {abadi2016tensorflow}
\bibfield{author}{\bibinfo{person}{Mart{\'\i}n Abadi}, \bibinfo{person}{Paul Barham}, \bibinfo{person}{Jianmin Chen}, \bibinfo{person}{Zhifeng Chen}, \bibinfo{person}{Andy Davis}, \bibinfo{person}{Jeffrey Dean}, \bibinfo{person}{Matthieu Devin}, \bibinfo{person}{Sanjay Ghemawat}, \bibinfo{person}{Geoffrey Irving}, \bibinfo{person}{Michael Isard}, {et~al\mbox{.}}} \bibinfo{year}{2016}\natexlab{}.
\newblock \showarticletitle{Tensorflow: A system for large-scale machine learning}. In \bibinfo{booktitle}{\emph{12th USENIX Symposium on Operating Systems Design and Implementation (OSDI 16)}}. \bibinfo{pages}{265--283}.
\newblock


\bibitem[Ainslie et~al\mbox{.}(2023)]%
        {ainslie2023gqa}
\bibfield{author}{\bibinfo{person}{Joshua Ainslie}, \bibinfo{person}{James Lee-Thorp}, \bibinfo{person}{Michiel de Jong}, \bibinfo{person}{Yury Zemlyanskiy}, \bibinfo{person}{Federico Lebr{\'o}n}, {and} \bibinfo{person}{Sumit Sanghai}.} \bibinfo{year}{2023}\natexlab{}.
\newblock \showarticletitle{GQA: Training Generalized Multi-Query Transformer Models from Multi-Head Checkpoints}. In \bibinfo{booktitle}{\emph{Proceedings of the 2023 Conference on Empirical Methods in Natural Language Processing (EMNLP)}}.
\newblock


\bibitem[AMD(2024)]%
        {amd_mi300x_2024}
AMD \bibinfo{year}{2024}\natexlab{}.
\newblock \bibinfo{booktitle}{\emph{AMD Instinct MI300 series microarchitecture}}.
\newblock AMD.
\newblock
\urldef\tempurl%
\url{https://instinct.docs.amd.com/develop/gpu-arch/mi300.html}
\showURL{%
\tempurl}


\bibitem[Ankit et~al\mbox{.}(2019)]%
        {anzaroot2019puma}
\bibfield{author}{\bibinfo{person}{Anirudh Ankit}, \bibinfo{person}{Imad Hajj}, \bibinfo{person}{Sreenivas Chalamalasetti}, \bibinfo{person}{Gilbert Ndu}, \bibinfo{person}{Michael Foltin}, \bibinfo{person}{Robert~S Williams}, \bibinfo{person}{Paolo Faraboschi}, \bibinfo{person}{Wen-mei Hwu}, \bibinfo{person}{Mircea~R Stan}, \bibinfo{person}{John~Paul Strachan}, {et~al\mbox{.}}} \bibinfo{year}{2019}\natexlab{}.
\newblock \showarticletitle{PUMA: A programmable ultra-efficient memristor-based accelerator for machine learning inference}. In \bibinfo{booktitle}{\emph{Proceedings of the Twenty-Fourth International Conference on Architectural Support for Programming Languages and Operating Systems}}. \bibinfo{pages}{715--731}.
\newblock


\bibitem[Beltagy et~al\mbox{.}(2020)]%
        {beltagy2020longformer}
\bibfield{author}{\bibinfo{person}{Iz Beltagy}, \bibinfo{person}{Matthew~E. Peters}, {and} \bibinfo{person}{Arman Cohan}.} \bibinfo{year}{2020}\natexlab{}.
\newblock \showarticletitle{Longformer: The Long-Document Transformer}. In \bibinfo{booktitle}{\emph{Proceedings of the 58th Annual Meeting of the Association for Computational Linguistics (ACL)}}.
\newblock


\bibitem[Boutros et~al\mbox{.}(2024)]%
        {boutros2024fpga}
\bibfield{author}{\bibinfo{person}{Andrew Boutros}, \bibinfo{person}{Aman Arora}, {and} \bibinfo{person}{Vaughn Betz}.} \bibinfo{year}{2024}\natexlab{}.
\newblock \showarticletitle{Field-Programmable Gate Array Architecture for Deep Learning: Survey \& Future Directions}.
\newblock \bibinfo{journal}{\emph{arXiv preprint arXiv:2404.10076}} (\bibinfo{year}{2024}).
\newblock
\urldef\tempurl%
\url{https://arxiv.org/abs/2404.10076}
\showURL{%
\tempurl}


\bibitem[Brown et~al\mbox{.}(2020)]%
        {brown2020gpt3}
\bibfield{author}{\bibinfo{person}{Tom~B. Brown}, \bibinfo{person}{Benjamin Mann}, \bibinfo{person}{Nick Ryder}, \bibinfo{person}{Melanie Subbiah}, \bibinfo{person}{Jared Kaplan}, \bibinfo{person}{Prafulla Dhariwal}, \bibinfo{person}{Arvind Neelakantan}, \bibinfo{person}{Pranav Shyam}, \bibinfo{person}{Girish Sastry}, \bibinfo{person}{Amanda Askell}, {et~al\mbox{.}}} \bibinfo{year}{2020}\natexlab{}.
\newblock \showarticletitle{Language Models are Few-Shot Learners}. In \bibinfo{booktitle}{\emph{Advances in Neural Information Processing Systems (NeurIPS)}}.
\newblock


\bibitem[Canis et~al\mbox{.}(2013)]%
        {canis2013legup}
\bibfield{author}{\bibinfo{person}{Andrew Canis}, \bibinfo{person}{Jongsok Choi}, \bibinfo{person}{Mark Aldham}, \bibinfo{person}{Victor Zhang}, \bibinfo{person}{Ahmed Kammoona}, \bibinfo{person}{Tomasz Czajkowski}, \bibinfo{person}{Stephen~D Brown}, {and} \bibinfo{person}{Jason~H Anderson}.} \bibinfo{year}{2013}\natexlab{}.
\newblock \showarticletitle{LegUp: An open-source high-level synthesis tool for FPGA-based processor/accelerator systems}.
\newblock \bibinfo{journal}{\emph{ACM Transactions on Embedded Computing Systems (TECS)}} \bibinfo{volume}{13}, \bibinfo{number}{2} (\bibinfo{year}{2013}), \bibinfo{pages}{1--27}.
\newblock


\bibitem[Chavan et~al\mbox{.}(2024)]%
        {chavan2024faster}
\bibfield{author}{\bibinfo{person}{Arnav Chavan}, \bibinfo{person}{Raghav Magazine}, \bibinfo{person}{Shubham Kushwaha}, \bibinfo{person}{Merouane Debbah}, {and} \bibinfo{person}{Deepak Gupta}.} \bibinfo{year}{2024}\natexlab{}.
\newblock \showarticletitle{Faster and Lighter LLMs: A Survey on Current Challenges and Way Forward}. In \bibinfo{booktitle}{\emph{Proceedings of the Thirty-Third International Joint Conference on Artificial Intelligence}}. \bibinfo{pages}{7980--7988}.
\newblock
\href{https://doi.org/10.24963/ijcai.2024/883}{doi:\nolinkurl{10.24963/ijcai.2024/883}}


\bibitem[Chen et~al\mbox{.}(2018)]%
        {chen2018tvm}
\bibfield{author}{\bibinfo{person}{Tianqi Chen}, \bibinfo{person}{Thierry Moreau}, \bibinfo{person}{Ziheng Jiang}, \bibinfo{person}{Leyuan Zheng}, \bibinfo{person}{Eddie Yan}, \bibinfo{person}{Haichen Shen}, \bibinfo{person}{Meghan Cowan}, \bibinfo{person}{Leyuan Wang}, \bibinfo{person}{Yuwei Hu}, \bibinfo{person}{Luis Ceze}, {et~al\mbox{.}}} \bibinfo{year}{2018}\natexlab{}.
\newblock \showarticletitle{TVM: An automated end-to-end optimizing compiler for deep learning}. In \bibinfo{booktitle}{\emph{13th USENIX Symposium on Operating Systems Design and Implementation (OSDI 18)}}. \bibinfo{pages}{578--594}.
\newblock


\bibitem[Chen et~al\mbox{.}(2014)]%
        {chen2014dadiannao}
\bibfield{author}{\bibinfo{person}{Yunji Chen}, \bibinfo{person}{Tao Luo}, \bibinfo{person}{Shaoli Liu}, \bibinfo{person}{Shijin Zhang}, \bibinfo{person}{Liqiang He}, \bibinfo{person}{Jia Wang}, \bibinfo{person}{Ling Li}, \bibinfo{person}{Tianshi Chen}, \bibinfo{person}{Zhiwei Xu}, \bibinfo{person}{Ninghui Sun}, {et~al\mbox{.}}} \bibinfo{year}{2014}\natexlab{}.
\newblock \showarticletitle{Dadiannao: A machine-learning supercomputer}. In \bibinfo{booktitle}{\emph{2014 47th Annual IEEE/ACM International Symposium on Microarchitecture}}. IEEE, \bibinfo{pages}{609--622}.
\newblock


\bibitem[Chen et~al\mbox{.}(2016)]%
        {chen2016eyeriss}
\bibfield{author}{\bibinfo{person}{Yu-Hsin Chen}, \bibinfo{person}{Tushar Krishna}, \bibinfo{person}{Joel~S Emer}, {and} \bibinfo{person}{Vivienne Sze}.} \bibinfo{year}{2016}\natexlab{}.
\newblock \showarticletitle{Eyeriss: An energy-efficient reconfigurable accelerator for deep convolutional neural networks}. In \bibinfo{booktitle}{\emph{2016 IEEE International Solid-State Circuits Conference (ISSCC)}}. IEEE, \bibinfo{pages}{262--263}.
\newblock


\bibitem[Chetlur et~al\mbox{.}(2014)]%
        {chetluar2014cudnn}
\bibfield{author}{\bibinfo{person}{Sharan Chetlur}, \bibinfo{person}{Cliff Woolley}, \bibinfo{person}{Philippe Vandermersch}, \bibinfo{person}{Jonathan Cohen}, \bibinfo{person}{John Tran}, \bibinfo{person}{Bryan Catanzaro}, {and} \bibinfo{person}{Evan Shelhamer}.} \bibinfo{year}{2014}\natexlab{}.
\newblock \showarticletitle{cuDNN: Efficient primitives for deep learning}. In \bibinfo{booktitle}{\emph{NIPS Deep Learning Workshop}}.
\newblock


\bibitem[Chi et~al\mbox{.}(2016)]%
        {chi2016prime}
\bibfield{author}{\bibinfo{person}{Peng Chi}, \bibinfo{person}{Shuangchen Li}, \bibinfo{person}{Chong Xu}, \bibinfo{person}{Tianzhou Zhang}, \bibinfo{person}{Jishen Zhao}, \bibinfo{person}{Yu Liu}, \bibinfo{person}{Yiran Wang}, {and} \bibinfo{person}{Yuan Xie}.} \bibinfo{year}{2016}\natexlab{}.
\newblock \showarticletitle{PRIME: A novel processing-in-memory architecture for neural network computation in ReRAM-based main memory}. In \bibinfo{booktitle}{\emph{Proceedings of the 43rd Annual International Symposium on Computer Architecture}}. \bibinfo{pages}{27--39}.
\newblock


\bibitem[Choromanski et~al\mbox{.}(2021)]%
        {choromanski2021performer}
\bibfield{author}{\bibinfo{person}{Krzysztof Choromanski}, \bibinfo{person}{Valerii Likhosherstov}, \bibinfo{person}{David Dohan}, \bibinfo{person}{Xingyou Song}, \bibinfo{person}{Andreea Gane}, \bibinfo{person}{Tamas Sarlos}, \bibinfo{person}{Peter Hawkins}, \bibinfo{person}{Jared Davis}, \bibinfo{person}{Afroz Mohiuddin}, \bibinfo{person}{Lukasz Kaiser}, {and} \bibinfo{person}{David Belanger}.} \bibinfo{year}{2021}\natexlab{}.
\newblock \showarticletitle{Rethinking Attention with Performers}. In \bibinfo{booktitle}{\emph{International Conference on Learning Representations (ICLR)}}.
\newblock


\bibitem[Coleman et~al\mbox{.}(2017)]%
        {coleman2019dawnbench}
\bibfield{author}{\bibinfo{person}{Cody Coleman}, \bibinfo{person}{Daniel Kang}, \bibinfo{person}{Deepak Narayanan}, \bibinfo{person}{Luigi Nardi}, {and} \bibinfo{person}{Matei Zaharia}.} \bibinfo{year}{2017}\natexlab{}.
\newblock \showarticletitle{DAWNBench: An end-to-end deep learning benchmark and competition}. In \bibinfo{booktitle}{\emph{NIPS ML Systems Workshop}}.
\newblock


\bibitem[Dai et~al\mbox{.}(2019)]%
        {dai2019transformerxl}
\bibfield{author}{\bibinfo{person}{Zihang Dai}, \bibinfo{person}{Zhilin Yang}, \bibinfo{person}{Yiming Yang}, \bibinfo{person}{Jaime Carbonell}, \bibinfo{person}{Quoc Le}, {and} \bibinfo{person}{Ruslan Salakhutdinov}.} \bibinfo{year}{2019}\natexlab{}.
\newblock \showarticletitle{Transformer-XL: Attentive Language Models Beyond a Fixed-Length Context}. In \bibinfo{booktitle}{\emph{Proceedings of the 57th Annual Meeting of the Association for Computational Linguistics (ACL)}}.
\newblock


\bibitem[Dao(2023)]%
        {dao2023flashattention2}
\bibfield{author}{\bibinfo{person}{Tri Dao}.} \bibinfo{year}{2023}\natexlab{}.
\newblock \showarticletitle{FlashAttention-2: Faster Attention with Better Parallelism and Work Partitioning}. In \bibinfo{booktitle}{\emph{Advances in Neural Information Processing Systems (NeurIPS)}}.
\newblock


\bibitem[Dao et~al\mbox{.}(2022)]%
        {dao2022flashattention}
\bibfield{author}{\bibinfo{person}{Tri Dao}, \bibinfo{person}{Daniel~Y Fu}, \bibinfo{person}{Stefano Ermon}, \bibinfo{person}{Atri Rudra}, {and} \bibinfo{person}{Christopher R{\'e}}.} \bibinfo{year}{2022}\natexlab{}.
\newblock \showarticletitle{FlashAttention: Fast and memory-efficient exact attention with IO-awareness}. In \bibinfo{booktitle}{\emph{Advances in Neural Information Processing Systems (NeurIPS)}}.
\newblock


\bibitem[Davies et~al\mbox{.}(2018)]%
        {davies2018loihi}
\bibfield{author}{\bibinfo{person}{Mike Davies}, \bibinfo{person}{Narayan Srinivasa}, \bibinfo{person}{Tsung-Han Lin}, \bibinfo{person}{Girish Chinya}, \bibinfo{person}{Yongqiang Cao}, \bibinfo{person}{Shailendra Choday}, \bibinfo{person}{George Dimou}, \bibinfo{person}{Prasad Joshi}, \bibinfo{person}{Nabil Imam}, \bibinfo{person}{Shweta Jain}, {et~al\mbox{.}}} \bibinfo{year}{2018}\natexlab{}.
\newblock \showarticletitle{Loihi: A neuromorphic manycore processor with on-chip learning}.
\newblock \bibinfo{journal}{\emph{IEEE Micro}} \bibinfo{volume}{38}, \bibinfo{number}{1} (\bibinfo{year}{2018}), \bibinfo{pages}{82--99}.
\newblock


\bibitem[Devlin et~al\mbox{.}(2019)]%
        {devlin2019bert}
\bibfield{author}{\bibinfo{person}{Jacob Devlin}, \bibinfo{person}{Ming-Wei Chang}, \bibinfo{person}{Kenton Lee}, {and} \bibinfo{person}{Kristina Toutanova}.} \bibinfo{year}{2019}\natexlab{}.
\newblock \showarticletitle{BERT: Pre-training of Deep Bidirectional Transformers for Language Understanding}. In \bibinfo{booktitle}{\emph{Proceedings of the 2019 Conference of the North American Chapter of the Association for Computational Linguistics: Human Language Technologies (NAACL-HLT)}}.
\newblock


\bibitem[Fang et~al\mbox{.}(2024)]%
        {fang2024maskllm}
\bibfield{author}{\bibinfo{person}{Gongfan Fang}, \bibinfo{person}{Hongxu Yin}, \bibinfo{person}{Saurav Muralidharan}, \bibinfo{person}{Greg Heinrich}, \bibinfo{person}{Jeff Pool}, \bibinfo{person}{Jan Kautz}, \bibinfo{person}{Pavlo Molchanov}, {and} \bibinfo{person}{Xinchao Wang}.} \bibinfo{year}{2024}\natexlab{}.
\newblock \showarticletitle{MaskLLM: Learnable Semi-Structured Sparsity for Large Language Models}.
\newblock \bibinfo{journal}{\emph{arXiv preprint arXiv:2409.17481}} (\bibinfo{year}{2024}).
\newblock
\urldef\tempurl%
\url{https://arxiv.org/abs/2409.17481}
\showURL{%
\tempurl}


\bibitem[Fedus et~al\mbox{.}(2022)]%
        {fedus2022switchtransformer}
\bibfield{author}{\bibinfo{person}{William Fedus}, \bibinfo{person}{Barret Zoph}, {and} \bibinfo{person}{Noam Shazeer}.} \bibinfo{year}{2022}\natexlab{}.
\newblock \showarticletitle{Switch Transformers: Scaling to Trillion Parameter Models with Simple and Efficient Sparsity}.
\newblock \bibinfo{journal}{\emph{Journal of Machine Learning Research}} \bibinfo{volume}{23}, \bibinfo{number}{120} (\bibinfo{year}{2022}), \bibinfo{pages}{1--39}.
\newblock


\bibitem[Fowers et~al\mbox{.}(2018)]%
        {fowers2018brainwave}
\bibfield{author}{\bibinfo{person}{Jeremy Fowers}, \bibinfo{person}{Kalin Ovtcharov}, \bibinfo{person}{Michael Papamichael}, \bibinfo{person}{Todd Massengill}, \bibinfo{person}{Ming Liu}, \bibinfo{person}{Shlomi Lo}, \bibinfo{person}{Shachar Alkalay}, \bibinfo{person}{Michael Haselman}, \bibinfo{person}{Larry Adams}, \bibinfo{person}{Michael Gschwind}, {et~al\mbox{.}}} \bibinfo{year}{2018}\natexlab{}.
\newblock \showarticletitle{A configurable cloud-scale DNN processor for real-time AI}. In \bibinfo{booktitle}{\emph{Proceedings of the 45th Annual International Symposium on Computer Architecture}}. \bibinfo{pages}{1--14}.
\newblock


\bibitem[Frantar and Alistarh(2023)]%
        {frantar2022gptq}
\bibfield{author}{\bibinfo{person}{Elias Frantar} {and} \bibinfo{person}{Dan Alistarh}.} \bibinfo{year}{2023}\natexlab{}.
\newblock \showarticletitle{GPTQ: Accurate Post-Training Quantization for Generative Pre-trained Transformers}. In \bibinfo{booktitle}{\emph{International Conference on Learning Representations (ICLR)}}.
\newblock


\bibitem[Gale et~al\mbox{.}(2019)]%
        {gale2019state}
\bibfield{author}{\bibinfo{person}{Trevor Gale}, \bibinfo{person}{Erich Elsen}, {and} \bibinfo{person}{Sara Hooker}.} \bibinfo{year}{2019}\natexlab{}.
\newblock \showarticletitle{The state of sparsity in deep neural networks}.
\newblock \bibinfo{journal}{\emph{arXiv preprint arXiv:1902.09574}} (\bibinfo{year}{2019}).
\newblock


\bibitem[Gong et~al\mbox{.}(2024)]%
        {gong2024survey}
\bibfield{author}{\bibinfo{person}{Ruihao Gong}, \bibinfo{person}{Yifu Ding}, \bibinfo{person}{Zining Wang}, \bibinfo{person}{Chengtao Lv}, \bibinfo{person}{Xingyu Zheng}, \bibinfo{person}{Jinyang Du}, \bibinfo{person}{Haotong Qin}, \bibinfo{person}{Jinyang Guo}, \bibinfo{person}{Michele Magno}, {and} \bibinfo{person}{Xianglong Liu}.} \bibinfo{year}{2024}\natexlab{}.
\newblock \showarticletitle{A Survey of Low-bit Large Language Models: Basics, Systems, and Algorithms}.
\newblock \bibinfo{journal}{\emph{arXiv preprint arXiv:2409.16694}} (\bibinfo{year}{2024}).
\newblock
\urldef\tempurl%
\url{https://arxiv.org/abs/2409.16694}
\showURL{%
\tempurl}


\bibitem[{Groq}(2024)]%
        {groq2024lpu}
\bibfield{author}{\bibinfo{person}{{Groq}}.} \bibinfo{year}{2024}\natexlab{}.
\newblock \bibinfo{title}{Groq {LPU}: Language Processing Unit for low-latency {LLM} inference}.
\newblock \bibinfo{howpublished}{Technical report / product documentation}.
\newblock
\newblock
\shownote{Accessed: 2025-12-28}.


\bibitem[Gunter et~al\mbox{.}(2024)]%
        {gunter2024apple}
\bibfield{author}{\bibinfo{person}{Tom Gunter}, \bibinfo{person}{Zirui Wang}, \bibinfo{person}{Chong Wang}, \bibinfo{person}{Ruoming Pang}, \bibinfo{person}{Andy Narayanan}, \bibinfo{person}{Aonan Zhang}, \bibinfo{person}{Bowen Zhang}, \bibinfo{person}{Chen Chen}, \bibinfo{person}{Chung-Cheng Chiu}, \bibinfo{person}{David Qiu}, {et~al\mbox{.}}} \bibinfo{year}{2024}\natexlab{}.
\newblock \showarticletitle{Apple intelligence foundation language models}.
\newblock \bibinfo{journal}{\emph{arXiv preprint arXiv:2407.21075}} (\bibinfo{year}{2024}).
\newblock


\bibitem[Han et~al\mbox{.}(2016a)]%
        {han2016eie}
\bibfield{author}{\bibinfo{person}{Song Han}, \bibinfo{person}{Xingyu Liu}, \bibinfo{person}{Huizi Mao}, \bibinfo{person}{Jing Pu}, \bibinfo{person}{Ardavan Pedram}, \bibinfo{person}{Mark~A Horowitz}, {and} \bibinfo{person}{William~J Dally}.} \bibinfo{year}{2016}\natexlab{a}.
\newblock \showarticletitle{EIE: Efficient Inference Engine on compressed deep neural network}. In \bibinfo{booktitle}{\emph{Proceedings of the 43rd Annual International Symposium on Computer Architecture}}. \bibinfo{pages}{243--254}.
\newblock


\bibitem[Han et~al\mbox{.}(2016b)]%
        {han2016deepcompression}
\bibfield{author}{\bibinfo{person}{Song Han}, \bibinfo{person}{Huizi Mao}, {and} \bibinfo{person}{William~J Dally}.} \bibinfo{year}{2016}\natexlab{b}.
\newblock \showarticletitle{Deep compression: Compressing deep neural networks with pruning, trained quantization and Huffman coding}.
\newblock \bibinfo{journal}{\emph{International Conference on Learning Representations}} (\bibinfo{year}{2016}).
\newblock


\bibitem[Hennessy and Patterson(2019)]%
        {hennessy2019hwds}
\bibfield{author}{\bibinfo{person}{John~L Hennessy} {and} \bibinfo{person}{David~A Patterson}.} \bibinfo{year}{2019}\natexlab{}.
\newblock \showarticletitle{A new golden age for computer architecture: domain-specific hardware/software co-design, enhanced security, open instruction sets, and agile chip development}.
\newblock \bibinfo{journal}{\emph{Commun. ACM}} \bibinfo{volume}{62}, \bibinfo{number}{2} (\bibinfo{year}{2019}), \bibinfo{pages}{48--60}.
\newblock


\bibitem[Horowitz(2014)]%
        {horowitz2014energy}
\bibfield{author}{\bibinfo{person}{Mark Horowitz}.} \bibinfo{year}{2014}\natexlab{}.
\newblock \showarticletitle{Computing's energy problem (and what we can do about it)}. In \bibinfo{booktitle}{\emph{2014 IEEE International Solid-State Circuits Conference Digest of Technical Papers (ISSCC)}}. IEEE, \bibinfo{pages}{10--14}.
\newblock


\bibitem[Huang(2024)]%
        {nvidia_blackwell_2024}
\bibfield{author}{\bibinfo{person}{Jensen Huang}.} \bibinfo{year}{2024}\natexlab{}.
\newblock \showarticletitle{‘We Created a Processor for the Generative AI Era,’ NVIDIA CEO Says}.
\newblock \bibinfo{journal}{\emph{NVIDIA Blog}} (\bibinfo{year}{2024}).
\newblock
\urldef\tempurl%
\url{https://blogs.nvidia.com/blog/2024-gtc-keynote/}
\showURL{%
\tempurl}


\bibitem[Jacob et~al\mbox{.}(2018)]%
        {jacob2018quantization}
\bibfield{author}{\bibinfo{person}{Benoit Jacob}, \bibinfo{person}{Skirmantas Kligys}, \bibinfo{person}{Bo Chen}, \bibinfo{person}{Menglong Zhu}, \bibinfo{person}{Matthew Tang}, \bibinfo{person}{Andrew Howard}, \bibinfo{person}{Hartwig Adam}, {and} \bibinfo{person}{Dmitry Kalenichenko}.} \bibinfo{year}{2018}\natexlab{}.
\newblock \showarticletitle{Quantization and training of neural networks for efficient integer-arithmetic-only inference}. In \bibinfo{booktitle}{\emph{Proceedings of the IEEE conference on computer vision and pattern recognition}}. \bibinfo{pages}{2704--2713}.
\newblock


\bibitem[Jouppi et~al\mbox{.}(2021)]%
        {jouppi2021tpuv4}
\bibfield{author}{\bibinfo{person}{Norman~P Jouppi}, \bibinfo{person}{Doe~Hyun Yoon}, \bibinfo{person}{George Kurian}, \bibinfo{person}{Sheng Li}, \bibinfo{person}{Nishant Patil}, \bibinfo{person}{James Laudon}, \bibinfo{person}{Cliff Young}, {and} \bibinfo{person}{David Patterson}.} \bibinfo{year}{2021}\natexlab{}.
\newblock \showarticletitle{A domain-specific supercomputer for training deep neural networks}.
\newblock \bibinfo{journal}{\emph{Commun. ACM}} \bibinfo{volume}{64}, \bibinfo{number}{7} (\bibinfo{year}{2021}), \bibinfo{pages}{56--68}.
\newblock


\bibitem[Jouppi et~al\mbox{.}(2017)]%
        {jouppi2017tpu}
\bibfield{author}{\bibinfo{person}{Norman~P Jouppi}, \bibinfo{person}{Cliff Young}, \bibinfo{person}{Nishant Patil}, \bibinfo{person}{David Patterson}, \bibinfo{person}{Gaurav Agrawal}, \bibinfo{person}{Raminder Bajwa}, \bibinfo{person}{Sarah Bates}, \bibinfo{person}{Suresh Bhatia}, \bibinfo{person}{Nan Boden}, \bibinfo{person}{Al Borchers}, {et~al\mbox{.}}} \bibinfo{year}{2017}\natexlab{}.
\newblock \showarticletitle{In-datacenter performance analysis of a tensor processing unit}. In \bibinfo{booktitle}{\emph{Proceedings of the 44th Annual International Symposium on Computer Architecture}}. \bibinfo{pages}{1--12}.
\newblock


\bibitem[Kim et~al\mbox{.}(2025)]%
        {kim2025hpim}
\bibfield{author}{\bibinfo{person}{J. Kim} {et~al\mbox{.}}} \bibinfo{year}{2025}\natexlab{}.
\newblock \showarticletitle{HPIM: A Heterogeneous Processing-in-Memory Accelerator for Large Language Model Inference}.
\newblock \bibinfo{journal}{\emph{arXiv preprint arXiv:2509.12993}} (\bibinfo{year}{2025}).
\newblock


\bibitem[Krizhevsky et~al\mbox{.}(2012)]%
        {krizhevsky2012imagenet}
\bibfield{author}{\bibinfo{person}{Alex Krizhevsky}, \bibinfo{person}{Ilya Sutskever}, {and} \bibinfo{person}{Geoffrey~E Hinton}.} \bibinfo{year}{2012}\natexlab{}.
\newblock \showarticletitle{Imagenet classification with deep convolutional neural networks}.
\newblock \bibinfo{journal}{\emph{Advances in neural information processing systems}}  \bibinfo{volume}{25} (\bibinfo{year}{2012}).
\newblock


\bibitem[Kung(1982)]%
        {kung1982systolic}
\bibfield{author}{\bibinfo{person}{H~T Kung}.} \bibinfo{year}{1982}\natexlab{}.
\newblock \showarticletitle{Why systolic architectures?}
\newblock \bibinfo{journal}{\emph{Computer}} \bibinfo{volume}{15}, \bibinfo{number}{1} (\bibinfo{year}{1982}), \bibinfo{pages}{37--46}.
\newblock


\bibitem[Kwon et~al\mbox{.}(2023)]%
        {kwon2023vllm}
\bibfield{author}{\bibinfo{person}{Woosuk Kwon}, \bibinfo{person}{Zhuohan Li}, \bibinfo{person}{Sheng Zhuang}, \bibinfo{person}{Ying Sheng}, \bibinfo{person}{Xin Zhao}, \bibinfo{person}{Joseph Gonzalez}, \bibinfo{person}{Ion Stoica}, {and} \bibinfo{person}{Hao Zhang}.} \bibinfo{year}{2023}\natexlab{}.
\newblock \showarticletitle{Efficient Memory Management for Large Language Model Serving with PagedAttention}. In \bibinfo{booktitle}{\emph{Proceedings of the 29th Symposium on Operating Systems Principles (SOSP)}}.
\newblock


\bibitem[Leviathan et~al\mbox{.}(2023)]%
        {leviathan2023speculative}
\bibfield{author}{\bibinfo{person}{Yaniv Leviathan}, \bibinfo{person}{Matan Kalman}, {and} \bibinfo{person}{Yossi Matias}.} \bibinfo{year}{2023}\natexlab{}.
\newblock \showarticletitle{Fast Inference from Transformers via Speculative Decoding}. In \bibinfo{booktitle}{\emph{International Conference on Machine Learning (ICML)}}.
\newblock


\bibitem[Li et~al\mbox{.}(2024b)]%
        {li2024comprehensive}
\bibfield{author}{\bibinfo{person}{Changhao Li}, \bibinfo{person}{Haoling Li}, \bibinfo{person}{Mengqi Xue}, \bibinfo{person}{Gongfan Fang}, \bibinfo{person}{Sheng Zhou}, \bibinfo{person}{Zunlei Feng}, \bibinfo{person}{Huiqiong Wang}, \bibinfo{person}{Mingli Song}, {and} \bibinfo{person}{Jie Song}.} \bibinfo{year}{2024}\natexlab{b}.
\newblock \showarticletitle{A Comprehensive Study of Structural Pruning for Vision Models}.
\newblock \bibinfo{journal}{\emph{arXiv preprint arXiv:2406.12315}} (\bibinfo{year}{2024}).
\newblock
\urldef\tempurl%
\url{https://arxiv.org/abs/2406.12315}
\showURL{%
\tempurl}


\bibitem[Li et~al\mbox{.}(2024a)]%
        {li2024hlstransform}
\bibfield{author}{\bibinfo{person}{Yiming Li} {et~al\mbox{.}}} \bibinfo{year}{2024}\natexlab{a}.
\newblock \showarticletitle{HLSTransform: Energy-Efficient LLM Inference on FPGAs via High-Level Synthesis}.
\newblock \bibinfo{journal}{\emph{arXiv preprint arXiv:2405.00738}} (\bibinfo{year}{2024}).
\newblock


\bibitem[Lin et~al\mbox{.}(2024)]%
        {lin2023awq}
\bibfield{author}{\bibinfo{person}{Ji Lin}, \bibinfo{person}{Jiaming Tang}, \bibinfo{person}{Haotian Tang}, \bibinfo{person}{Shang Yang}, \bibinfo{person}{Wei Wang}, {and} \bibinfo{person}{Song Han}.} \bibinfo{year}{2024}\natexlab{}.
\newblock \showarticletitle{AWQ: Activation-aware Weight Quantization for LLM Compression and Acceleration}. In \bibinfo{booktitle}{\emph{Proceedings of the 7th Conference on Machine Learning and Systems (MLSys)}}.
\newblock


\bibitem[Liu et~al\mbox{.}(2015)]%
        {liu2015shidiannao}
\bibfield{author}{\bibinfo{person}{Tianshi Liu}, \bibinfo{person}{Zhenman Du}, \bibinfo{person}{Yaoyu Tao}, \bibinfo{person}{Jing Wang}, \bibinfo{person}{Sheng Wen}, \bibinfo{person}{Ling Li}, \bibinfo{person}{Tao Luo}, {and} \bibinfo{person}{Yunji Chen}.} \bibinfo{year}{2015}\natexlab{}.
\newblock \showarticletitle{ShiDianNao: Shifting vision processing closer to the sensor}. In \bibinfo{booktitle}{\emph{Proceedings of the 42nd Annual International Symposium on Computer Architecture}}. \bibinfo{pages}{92--104}.
\newblock


\bibitem[Ma et~al\mbox{.}(2024)]%
        {ma2024era}
\bibfield{author}{\bibinfo{person}{Shuming Ma}, \bibinfo{person}{Hongyu Wang}, \bibinfo{person}{Lingxiao Ma}, \bibinfo{person}{Lei Wang}, \bibinfo{person}{Wenhui Wang}, \bibinfo{person}{Shaohan Huang}, \bibinfo{person}{Li Dong}, \bibinfo{person}{Ruiping Wang}, \bibinfo{person}{Jilong Xue}, {and} \bibinfo{person}{Furu Wei}.} \bibinfo{year}{2024}\natexlab{}.
\newblock \showarticletitle{The Era of 1-bit LLMs: All Large Language Models Are in 1.58 Bits}.
\newblock \bibinfo{journal}{\emph{CoRR}}  \bibinfo{volume}{abs/2402.17764} (\bibinfo{year}{2024}).
\newblock
\showeprint[arXiv]{2402.17764}
\urldef\tempurl%
\url{https://arxiv.org/abs/2402.17764}
\showURL{%
\tempurl}


\bibitem[Ma et~al\mbox{.}(2018)]%
        {ma2018optimizing}
\bibfield{author}{\bibinfo{person}{Yufei Ma}, \bibinfo{person}{Yu Cao}, \bibinfo{person}{Sarma Vrudhula}, {and} \bibinfo{person}{Jae-sun Seo}.} \bibinfo{year}{2018}\natexlab{}.
\newblock \showarticletitle{Optimizing the convolution operation to accelerate deep neural networks on FPGA}.
\newblock \bibinfo{journal}{\emph{IEEE Transactions on Very Large Scale Integration (VLSI) Systems}} \bibinfo{volume}{26}, \bibinfo{number}{7} (\bibinfo{year}{2018}), \bibinfo{pages}{1354--1367}.
\newblock


\bibitem[Mattson et~al\mbox{.}(2020)]%
        {mattson2020mlperf}
\bibfield{author}{\bibinfo{person}{Peter Mattson}, \bibinfo{person}{Christine Cheng}, \bibinfo{person}{Cody Coleman}, \bibinfo{person}{Gregory Diamos}, \bibinfo{person}{Paulius Micikevicius}, \bibinfo{person}{David Patterson}, \bibinfo{person}{Hanlin Tang}, \bibinfo{person}{Gu-Yeon Wei}, {et~al\mbox{.}}} \bibinfo{year}{2020}\natexlab{}.
\newblock \showarticletitle{MLPerf: An industry standard benchmark suite for machine learning performance}.
\newblock \bibinfo{journal}{\emph{IEEE Micro}} \bibinfo{volume}{40}, \bibinfo{number}{2} (\bibinfo{year}{2020}), \bibinfo{pages}{8--16}.
\newblock


\bibitem[Mazumder et~al\mbox{.}(2021)]%
        {mazumder2021survey}
\bibfield{author}{\bibinfo{person}{Arnab~Neelim Mazumder}, \bibinfo{person}{Jian Meng}, \bibinfo{person}{Hasib-Al Rashid}, \bibinfo{person}{Utteja Kallakuri}, \bibinfo{person}{Xin Zhang}, \bibinfo{person}{Jae-Sun Seo}, {and} \bibinfo{person}{Tinoosh Mohsenin}.} \bibinfo{year}{2021}\natexlab{}.
\newblock \showarticletitle{A survey on the optimization of neural network accelerators for micro-ai on-device inference}.
\newblock \bibinfo{journal}{\emph{IEEE Journal on Emerging and Selected Topics in Circuits and Systems}} \bibinfo{volume}{11}, \bibinfo{number}{4} (\bibinfo{year}{2021}), \bibinfo{pages}{532--547}.
\newblock


\bibitem[Merolla et~al\mbox{.}(2014)]%
        {merolla2014truenorth}
\bibfield{author}{\bibinfo{person}{Paul~A Merolla}, \bibinfo{person}{John~V Arthur}, \bibinfo{person}{Rodrigo Alvarez-Icaza}, \bibinfo{person}{Andrew~S Cassidy}, \bibinfo{person}{Jun Sawada}, \bibinfo{person}{Filipp Akopyan}, \bibinfo{person}{Bryan~L Jackson}, \bibinfo{person}{Nabil Imam}, \bibinfo{person}{Chen Guo}, \bibinfo{person}{Yutaka Nakamura}, {et~al\mbox{.}}} \bibinfo{year}{2014}\natexlab{}.
\newblock \showarticletitle{A million spiking-neuron integrated circuit with a scalable communication network and interface}.
\newblock \bibinfo{journal}{\emph{Science}} \bibinfo{volume}{345}, \bibinfo{number}{6197} (\bibinfo{year}{2014}), \bibinfo{pages}{668--673}.
\newblock


\bibitem[Micikevicius et~al\mbox{.}(2018)]%
        {micikevicius2018mixedprecision}
\bibfield{author}{\bibinfo{person}{Paulius Micikevicius}, \bibinfo{person}{Sharan Narang}, \bibinfo{person}{Jonah Alben}, \bibinfo{person}{Gregory Diamos}, \bibinfo{person}{Erich Elsen}, \bibinfo{person}{David Garcia}, \bibinfo{person}{Boris Ginsburg}, \bibinfo{person}{Michael Houston}, \bibinfo{person}{Oleksii Kuchaiev}, {and} \bibinfo{person}{Ganesh Venkatesh}.} \bibinfo{year}{2018}\natexlab{}.
\newblock \showarticletitle{Mixed precision training}. In \bibinfo{booktitle}{\emph{International Conference on Learning Representations}}.
\newblock


\bibitem[Muralidharan et~al\mbox{.}(2024)]%
        {muralidharan2024compact}
\bibfield{author}{\bibinfo{person}{Saurav Muralidharan}, \bibinfo{person}{Sharath~Turuvekere Sreenivas}, \bibinfo{person}{Raviraj Joshi}, \bibinfo{person}{Marcin Chochowski}, \bibinfo{person}{Mostofa Patwary}, \bibinfo{person}{Mohammad Shoeybi}, \bibinfo{person}{Bryan Catanzaro}, \bibinfo{person}{Jan Kautz}, {and} \bibinfo{person}{Pavlo Molchanov}.} \bibinfo{year}{2024}\natexlab{}.
\newblock \showarticletitle{Compact Language Models via Pruning and Knowledge Distillation}.
\newblock \bibinfo{journal}{\emph{arXiv preprint arXiv:2407.14679}} (\bibinfo{year}{2024}).
\newblock
\urldef\tempurl%
\url{https://arxiv.org/abs/2407.14679}
\showURL{%
\tempurl}


\bibitem[Nurvitadhi et~al\mbox{.}(2017)]%
        {nurvitadhi2017can}
\bibfield{author}{\bibinfo{person}{Eriko Nurvitadhi}, \bibinfo{person}{Ganesh Venkatesh}, \bibinfo{person}{Jaewoong Sim}, \bibinfo{person}{Debbie Marr}, \bibinfo{person}{Randy Huang}, \bibinfo{person}{Jason Ong Gee~Hock}, \bibinfo{person}{Yeong~Tat Liew}, \bibinfo{person}{Krishnan Srivatsan}, \bibinfo{person}{Duncan Moss}, \bibinfo{person}{Suchit Subhaschandra}, {et~al\mbox{.}}} \bibinfo{year}{2017}\natexlab{}.
\newblock \showarticletitle{Can FPGAs beat GPUs in accelerating next-generation deep neural networks?}. In \bibinfo{booktitle}{\emph{Proceedings of the 2017 ACM/SIGDA international symposium on field-programmable gate arrays}}. \bibinfo{pages}{5--14}.
\newblock


\bibitem[{NVIDIA}(2023)]%
        {nvidia2023llm_inference_platforms}
\bibfield{author}{\bibinfo{person}{{NVIDIA}}.} \bibinfo{year}{2023}\natexlab{}.
\newblock \bibinfo{title}{NVIDIA launches inference platforms for large language models and generative AI workloads}.
\newblock \bibinfo{howpublished}{NVIDIA Newsroom}.
\newblock
\urldef\tempurl%
\url{https://nvidianews.nvidia.com/news/nvidia-launches-inference-platforms-for-large-language-models-and-generative-ai-workloads}
\showURL{%
\tempurl}
\newblock
\shownote{Accessed 2025-12-28}.


\bibitem[Parashar et~al\mbox{.}(2017)]%
        {parashar2017scnn}
\bibfield{author}{\bibinfo{person}{Angshuman Parashar}, \bibinfo{person}{Minsoo Rhu}, \bibinfo{person}{Anurag Mukkara}, \bibinfo{person}{Anthony Puglielli}, \bibinfo{person}{Rangharajan Venkatesan}, \bibinfo{person}{Brucek Khaitan}, \bibinfo{person}{Jianxin Shao}, \bibinfo{person}{Yu-Hsin Chen}, \bibinfo{person}{Joel Emer}, \bibinfo{person}{Stephen~W Keckler}, {et~al\mbox{.}}} \bibinfo{year}{2017}\natexlab{}.
\newblock \showarticletitle{SCNN: An accelerator for compressed-sparse convolutional neural networks}. In \bibinfo{booktitle}{\emph{Proceedings of the 44th Annual International Symposium on Computer Architecture}}. \bibinfo{pages}{27--40}.
\newblock


\bibitem[Raffel et~al\mbox{.}(2020)]%
        {raffel2020t5}
\bibfield{author}{\bibinfo{person}{Colin Raffel}, \bibinfo{person}{Noam Shazeer}, \bibinfo{person}{Adam Roberts}, \bibinfo{person}{Katherine Lee}, \bibinfo{person}{Sharan Narang}, \bibinfo{person}{Michael Matena}, \bibinfo{person}{Yanqi Zhou}, \bibinfo{person}{Wei Li}, {and} \bibinfo{person}{Peter~J. Liu}.} \bibinfo{year}{2020}\natexlab{}.
\newblock \showarticletitle{Exploring the Limits of Transfer Learning with a Unified Text-to-Text Transformer}.
\newblock \bibinfo{journal}{\emph{Journal of Machine Learning Research}} \bibinfo{volume}{21}, \bibinfo{number}{140} (\bibinfo{year}{2020}), \bibinfo{pages}{1--67}.
\newblock


\bibitem[Rajbhandari et~al\mbox{.}(2020)]%
        {rajbhandari2020zero}
\bibfield{author}{\bibinfo{person}{Samyam Rajbhandari}, \bibinfo{person}{Olatunji Ruwase}, \bibinfo{person}{Jeff Rasley}, \bibinfo{person}{Shaden Smith}, {and} \bibinfo{person}{Yuxiong He}.} \bibinfo{year}{2020}\natexlab{}.
\newblock \showarticletitle{ZeRO: Memory optimizations toward training trillion parameter models}. In \bibinfo{booktitle}{\emph{SC20: International Conference for High Performance Computing, Networking, Storage and Analysis}}. IEEE, \bibinfo{pages}{1--16}.
\newblock


\bibitem[Shafiee et~al\mbox{.}(2016)]%
        {shafiee2016isaac}
\bibfield{author}{\bibinfo{person}{Armin Shafiee}, \bibinfo{person}{Aniruddha Nag}, \bibinfo{person}{Naveen Muralimanohar}, \bibinfo{person}{Rajeev Balasubramonian}, \bibinfo{person}{John~Paul Strachan}, \bibinfo{person}{Miao Hu}, \bibinfo{person}{R~Stanley Williams}, {and} \bibinfo{person}{Vivek Srikumar}.} \bibinfo{year}{2016}\natexlab{}.
\newblock \showarticletitle{ISAAC: A convolutional neural network accelerator with in-situ analog arithmetic in crossbars}. In \bibinfo{booktitle}{\emph{Proceedings of the 43rd Annual International Symposium on Computer Architecture}}. \bibinfo{pages}{14--26}.
\newblock


\bibitem[Shazeer et~al\mbox{.}(2017)]%
        {shazeer2017outrageously}
\bibfield{author}{\bibinfo{person}{Noam Shazeer}, \bibinfo{person}{Azalia Mirhoseini}, \bibinfo{person}{Krzysztof Maziarz}, \bibinfo{person}{Andy Davis}, \bibinfo{person}{Quoc Le}, \bibinfo{person}{Geoffrey Hinton}, {and} \bibinfo{person}{Jeff Dean}.} \bibinfo{year}{2017}\natexlab{}.
\newblock \showarticletitle{Outrageously large neural networks: The sparsely-gated mixture-of-experts layer}. In \bibinfo{booktitle}{\emph{International Conference on Learning Representations}}.
\newblock


\bibitem[Shoeybi et~al\mbox{.}(2019)]%
        {shoeybi2019megatronlm}
\bibfield{author}{\bibinfo{person}{Mohammad Shoeybi}, \bibinfo{person}{Mostofa Patwary}, \bibinfo{person}{Raul Puri}, \bibinfo{person}{Patrick LeGresley}, \bibinfo{person}{Jared Casper}, {and} \bibinfo{person}{Bryan Catanzaro}.} \bibinfo{year}{2019}\natexlab{}.
\newblock \showarticletitle{Megatron-LM: Training multi-billion parameter language models using model parallelism}.
\newblock \bibinfo{journal}{\emph{arXiv preprint arXiv:1909.08053}} (\bibinfo{year}{2019}).
\newblock


\bibitem[Silvano et~al\mbox{.}(2025)]%
        {silvano2025survey}
\bibfield{author}{\bibinfo{person}{Cristina Silvano}, \bibinfo{person}{Daniele Ielmini}, \bibinfo{person}{Fabrizio Ferrandi}, \bibinfo{person}{Leandro Fiorin}, \bibinfo{person}{Serena Curzel}, \bibinfo{person}{Luca Benini}, \bibinfo{person}{Francesco Conti}, \bibinfo{person}{Angelo Garofalo}, \bibinfo{person}{Cristian Zambelli}, \bibinfo{person}{Enrico Calore}, {et~al\mbox{.}}} \bibinfo{year}{2025}\natexlab{}.
\newblock \showarticletitle{A Survey on Deep Learning Hardware Accelerators for Heterogeneous HPC Platforms}.
\newblock \bibinfo{journal}{\emph{arXiv preprint arXiv:2306.15552}} (\bibinfo{year}{2025}).
\newblock
\urldef\tempurl%
\url{https://arxiv.org/abs/2306.15552}
\showURL{%
\tempurl}


\bibitem[Spiridonov and Ji(2024)]%
        {google_tpu_v5p_2024}
\bibfield{author}{\bibinfo{person}{Alex Spiridonov} {and} \bibinfo{person}{Gang Ji}.} \bibinfo{year}{2024}\natexlab{}.
\newblock \showarticletitle{What’s new with Google Cloud’s AI Hypercomputer architecture}.
\newblock \bibinfo{journal}{\emph{Google Cloud Blog}} (\bibinfo{year}{2024}).
\newblock
\urldef\tempurl%
\url{https://cloud.google.com/blog/products/compute/whats-new-with-google-clouds-ai-hypercomputer-architecture}
\showURL{%
\tempurl}


\bibitem[Tillet et~al\mbox{.}(2019)]%
        {tillet2019triton}
\bibfield{author}{\bibinfo{person}{Philippe Tillet}, \bibinfo{person}{H~T Kung}, {and} \bibinfo{person}{David Cox}.} \bibinfo{year}{2019}\natexlab{}.
\newblock \showarticletitle{Triton: an intermediate language and compiler for tiled neural network computations}. In \bibinfo{booktitle}{\emph{Proceedings of the 3rd ACM SIGPLAN International Workshop on Machine Learning and Programming Languages (MAPL)}}.
\newblock


\bibitem[Touvron et~al\mbox{.}(2023)]%
        {touvron2023llama}
\bibfield{author}{\bibinfo{person}{Hugo Touvron}, \bibinfo{person}{Thibaut Lavril}, \bibinfo{person}{Gautier Izacard}, \bibinfo{person}{Xavier Martinet}, \bibinfo{person}{Marie-Anne Lachaux}, \bibinfo{person}{Timoth{\'e}e Lacroix}, \bibinfo{person}{Baptiste Rozi{\`e}re}, \bibinfo{person}{Naman Goyal}, \bibinfo{person}{Eric Hambro}, \bibinfo{person}{Faisal Azhar}, {et~al\mbox{.}}} \bibinfo{year}{2023}\natexlab{}.
\newblock \showarticletitle{LLaMA: Open and Efficient Foundation Language Models}.
\newblock \bibinfo{journal}{\emph{arXiv preprint arXiv:2302.13971}} (\bibinfo{year}{2023}).
\newblock


\bibitem[Umuroglu et~al\mbox{.}(2017)]%
        {umuroglu2017finn}
\bibfield{author}{\bibinfo{person}{Yaman Umuroglu}, \bibinfo{person}{Nicholas~J Fraser}, \bibinfo{person}{Giulio Gambardella}, \bibinfo{person}{Michaela Blott}, \bibinfo{person}{Philip Leong}, \bibinfo{person}{Magnus Jahre}, {and} \bibinfo{person}{Kees Vissers}.} \bibinfo{year}{2017}\natexlab{}.
\newblock \showarticletitle{Finn: A framework for fast, scalable binarized neural network inference}. In \bibinfo{booktitle}{\emph{Proceedings of the 2017 ACM/SIGDA international symposium on field-programmable gate arrays}}. \bibinfo{pages}{65--74}.
\newblock


\bibitem[Vaswani et~al\mbox{.}(2017)]%
        {vaswani2017attention}
\bibfield{author}{\bibinfo{person}{Ashish Vaswani}, \bibinfo{person}{Noam Shazeer}, \bibinfo{person}{Niki Parmar}, \bibinfo{person}{Jakob Uszkoreit}, \bibinfo{person}{Llion Jones}, \bibinfo{person}{Aidan~N Gomez}, \bibinfo{person}{{\L}ukasz Kaiser}, {and} \bibinfo{person}{Illia Polosukhin}.} \bibinfo{year}{2017}\natexlab{}.
\newblock \showarticletitle{Attention is all you need}. In \bibinfo{booktitle}{\emph{Advances in Neural Information Processing Systems}}, Vol.~\bibinfo{volume}{30}.
\newblock


\bibitem[Venieris and Bouganis(2016)]%
        {venieris2017fpgaconvnet}
\bibfield{author}{\bibinfo{person}{Stylianos~I Venieris} {and} \bibinfo{person}{Christos-Savvas Bouganis}.} \bibinfo{year}{2016}\natexlab{}.
\newblock \showarticletitle{fpgaConvNet: A toolflow for mapping diverse convolutional neural networks on embedded FPGAs}. In \bibinfo{booktitle}{\emph{2016 International Conference on Field-Programmable Technology (FPT)}}.
\newblock


\bibitem[Wang et~al\mbox{.}(2023)]%
        {wang2023acceltran}
\bibfield{author}{\bibinfo{person}{Yang Wang} {et~al\mbox{.}}} \bibinfo{year}{2023}\natexlab{}.
\newblock \showarticletitle{AccelTran: A Sparsity-Aware Accelerator for Dynamic Inference of Transformer Models}. In \bibinfo{booktitle}{\emph{60th ACM/IEEE Design Automation Conference (DAC)}}.
\newblock


\bibitem[Wang et~al\mbox{.}(2024)]%
        {wang2024art}
\bibfield{author}{\bibinfo{person}{Yanshu Wang}, \bibinfo{person}{Tong Yang}, \bibinfo{person}{Xiyan Liang}, \bibinfo{person}{Guoan Wang}, \bibinfo{person}{Hanning Lu}, \bibinfo{person}{Xu Zhe}, \bibinfo{person}{Yaoming Li}, {and} \bibinfo{person}{Li Weitao}.} \bibinfo{year}{2024}\natexlab{}.
\newblock \showarticletitle{Art and Science of Quantizing Large-Scale Models: A Comprehensive Overview}.
\newblock \bibinfo{journal}{\emph{arXiv preprint arXiv:2409.11650}} (\bibinfo{year}{2024}).
\newblock
\urldef\tempurl%
\url{https://arxiv.org/abs/2409.11650}
\showURL{%
\tempurl}


\bibitem[{Wikipedia contributors}(2024)]%
        {cerebras_wikipedia_2024}
\bibfield{author}{\bibinfo{person}{{Wikipedia contributors}}.} \bibinfo{year}{2024}\natexlab{}.
\newblock \bibinfo{title}{Cerebras}.
\newblock \bibinfo{howpublished}{Wikipedia}.
\newblock
\urldef\tempurl%
\url{https://en.wikipedia.org/wiki/Cerebras}
\showURL{%
\tempurl}
\newblock
\shownote{Accessed 2025-12-28}.


\bibitem[{Wikipedia contributors}(2025a)]%
        {amd_instinct_wikipedia_2025}
\bibfield{author}{\bibinfo{person}{{Wikipedia contributors}}.} \bibinfo{year}{2025}\natexlab{a}.
\newblock \bibinfo{title}{AMD Instinct}.
\newblock \bibinfo{howpublished}{Wikipedia}.
\newblock
\urldef\tempurl%
\url{https://en.wikipedia.org/wiki/AMD_Instinct}
\showURL{%
\tempurl}
\newblock
\shownote{Accessed 2025-12-28}.


\bibitem[{Wikipedia contributors}(2025b)]%
        {aws_trainium_wikipedia_2025}
\bibfield{author}{\bibinfo{person}{{Wikipedia contributors}}.} \bibinfo{year}{2025}\natexlab{b}.
\newblock \bibinfo{title}{AWS Trainium}.
\newblock \bibinfo{howpublished}{Wikipedia}.
\newblock
\urldef\tempurl%
\url{https://en.wikipedia.org/wiki/AWS_Trainium}
\showURL{%
\tempurl}
\newblock
\shownote{Accessed 2025-12-28}.


\bibitem[{Wikipedia contributors}(2025c)]%
        {intel_gaudi_wikipedia_2025}
\bibfield{author}{\bibinfo{person}{{Wikipedia contributors}}.} \bibinfo{year}{2025}\natexlab{c}.
\newblock \bibinfo{title}{Intel Gaudi}.
\newblock \bibinfo{howpublished}{Wikipedia}.
\newblock
\urldef\tempurl%
\url{https://en.wikipedia.org/wiki/Intel_Gaudi}
\showURL{%
\tempurl}
\newblock
\shownote{Accessed 2025-12-28}.


\bibitem[{Wikipedia contributors}(2025d)]%
        {tpu_wikipedia_2025}
\bibfield{author}{\bibinfo{person}{{Wikipedia contributors}}.} \bibinfo{year}{2025}\natexlab{d}.
\newblock \bibinfo{title}{Tensor Processing Unit}.
\newblock \bibinfo{howpublished}{Wikipedia}.
\newblock
\urldef\tempurl%
\url{https://en.wikipedia.org/wiki/Tensor_Processing_Unit}
\showURL{%
\tempurl}
\newblock
\shownote{Accessed 2025-12-28}.


\bibitem[Williams et~al\mbox{.}(2009)]%
        {williams2009roofline}
\bibfield{author}{\bibinfo{person}{Samuel Williams}, \bibinfo{person}{Andrew Waterman}, {and} \bibinfo{person}{David Patterson}.} \bibinfo{year}{2009}\natexlab{}.
\newblock \showarticletitle{Roofline: An insightful visual performance model for multicore architectures}.
\newblock \bibinfo{journal}{\emph{Commun. ACM}} \bibinfo{volume}{52}, \bibinfo{number}{4} (\bibinfo{year}{2009}), \bibinfo{pages}{65--76}.
\newblock


\bibitem[Wulf and McKee(1995)]%
        {wulf1995memorywall}
\bibfield{author}{\bibinfo{person}{Wm.~A. Wulf} {and} \bibinfo{person}{Sally~A. McKee}.} \bibinfo{year}{1995}\natexlab{}.
\newblock \showarticletitle{Hitting the memory wall: Implications of the obvious}.
\newblock \bibinfo{journal}{\emph{ACM SIGARCH Computer Architecture News}} \bibinfo{volume}{23}, \bibinfo{number}{1} (\bibinfo{year}{1995}), \bibinfo{pages}{20--24}.
\newblock


\bibitem[Xiao et~al\mbox{.}(2023)]%
        {xiao2022smoothquant}
\bibfield{author}{\bibinfo{person}{Guangxuan Xiao}, \bibinfo{person}{Ji Lin}, \bibinfo{person}{Matthieu Seznec}, \bibinfo{person}{Julien Demouth}, \bibinfo{person}{Luka Alvarez}, {and} \bibinfo{person}{Tom Goldstein}.} \bibinfo{year}{2023}\natexlab{}.
\newblock \showarticletitle{SmoothQuant: Accurate and Efficient Post-Training Quantization for Large Language Models}. In \bibinfo{booktitle}{\emph{International Conference on Machine Learning (ICML)}}.
\newblock


\bibitem[Xu et~al\mbox{.}(2025a)]%
        {xu2025enabling}
\bibfield{author}{\bibinfo{person}{Bin Xu}, \bibinfo{person}{Ayan Banerjee}, {and} \bibinfo{person}{Sandeep Gupta}.} \bibinfo{year}{2025}\natexlab{a}.
\newblock \showarticletitle{Enabling Physical AI at the Edge: Hardware-Accelerated Recovery of System Dynamics}.
\newblock \bibinfo{journal}{\emph{arXiv preprint arXiv:2512.23767}} (\bibinfo{year}{2025}).
\newblock


\bibitem[Xu et~al\mbox{.}(2025b)]%
        {xu2025hardware}
\bibfield{author}{\bibinfo{person}{Bin Xu}, \bibinfo{person}{Ayan Banerjee}, {and} \bibinfo{person}{Sandeep Gupta}.} \bibinfo{year}{2025}\natexlab{b}.
\newblock \showarticletitle{Hardware Software Optimizations for Fast Model Recovery on Reconfigurable Architectures}.
\newblock \bibinfo{journal}{\emph{arXiv preprint arXiv:2512.06113}} (\bibinfo{year}{2025}).
\newblock


\bibitem[Xu et~al\mbox{.}(2025c)]%
        {xu2025fast}
\bibfield{author}{\bibinfo{person}{B Xu}, \bibinfo{person}{A Banerjee}, {and} \bibinfo{person}{Sandeep~KS Gupta}.} \bibinfo{year}{2025}\natexlab{c}.
\newblock \showarticletitle{Fast Online Digital Twinning on FPGA for Mission Critical Applications}. IEEE Military Communication Conference.
\newblock


\bibitem[Xu et~al\mbox{.}(2025d)]%
        {xu2025model}
\bibfield{author}{\bibinfo{person}{Bin Xu}, \bibinfo{person}{Ayan Banerjee}, {and} \bibinfo{person}{Sandeep~KS Gupta}.} \bibinfo{year}{2025}\natexlab{d}.
\newblock \showarticletitle{Model Recovery at the Edge Under Resource Constraints for Physical AI}.
\newblock  (\bibinfo{year}{2025}).
\newblock


\bibitem[Xu et~al\mbox{.}(2025e)]%
        {xu2025accelerated}
\bibfield{author}{\bibinfo{person}{Bin Xu}, \bibinfo{person}{Ayan Banerjee}, \bibinfo{person}{Midhat Urooj}, {and} \bibinfo{person}{Sandeep~KS Gupta}.} \bibinfo{year}{2025}\natexlab{e}.
\newblock \showarticletitle{Accelerated Digital Twin Learning for Edge AI: A Comparison of FPGA and Mobile GPU}. In \bibinfo{booktitle}{\emph{2025 IEEE 68th International Midwest Symposium on Circuits and Systems (MWSCAS)}}. IEEE, \bibinfo{pages}{149--153}.
\newblock


\bibitem[Yan et~al\mbox{.}(2024)]%
        {yan2024fpga}
\bibfield{author}{\bibinfo{person}{Feng Yan}, \bibinfo{person}{Andreas Koch}, {and} \bibinfo{person}{Oliver Sinnen}.} \bibinfo{year}{2024}\natexlab{}.
\newblock \showarticletitle{A survey on FPGA-based accelerator for ML models}.
\newblock \bibinfo{journal}{\emph{arXiv preprint arXiv:2412.15666}} (\bibinfo{year}{2024}).
\newblock
\urldef\tempurl%
\url{https://arxiv.org/abs/2412.15666}
\showURL{%
\tempurl}


\bibitem[Zaheer et~al\mbox{.}(2020)]%
        {zaheer2020bigbird}
\bibfield{author}{\bibinfo{person}{Manzil Zaheer}, \bibinfo{person}{Guru Guruganesh}, \bibinfo{person}{Kumar~Avinava Dubey}, \bibinfo{person}{Joshua Ainslie}, \bibinfo{person}{Chris Alberti}, \bibinfo{person}{Santiago Onta{\~n}{\'o}n}, \bibinfo{person}{Philip Pham}, \bibinfo{person}{Anirudh Ravula}, \bibinfo{person}{Qifan Wang}, \bibinfo{person}{Li Yang}, {and} \bibinfo{person}{Amr Ahmed}.} \bibinfo{year}{2020}\natexlab{}.
\newblock \showarticletitle{Big Bird: Transformers for Longer Sequences}. In \bibinfo{booktitle}{\emph{Advances in Neural Information Processing Systems (NeurIPS)}}.
\newblock


\bibitem[Zhang et~al\mbox{.}(2015)]%
        {zhang2015optimizing}
\bibfield{author}{\bibinfo{person}{Chen Zhang}, \bibinfo{person}{Peng Li}, \bibinfo{person}{Guangyu Sun}, \bibinfo{person}{Yijin Guan}, \bibinfo{person}{Bingjun Xiao}, {and} \bibinfo{person}{Jason Cong}.} \bibinfo{year}{2015}\natexlab{}.
\newblock \showarticletitle{Optimizing FPGA-based accelerator design for deep convolutional neural networks}. In \bibinfo{booktitle}{\emph{Proceedings of the 2015 ACM/SIGDA international symposium on field-programmable gate arrays}}. \bibinfo{pages}{161--170}.
\newblock


\bibitem[Zhang et~al\mbox{.}(2024a)]%
        {FPGATrans}
\bibfield{author}{\bibinfo{person}{Manting Zhang}, \bibinfo{person}{Jialin Cao}, \bibinfo{person}{Kejia Shi}, \bibinfo{person}{Keqing Zhao}, \bibinfo{person}{Genhao Zhang}, \bibinfo{person}{Jun Yu}, {and} \bibinfo{person}{Kun Wang}.} \bibinfo{year}{2024}\natexlab{a}.
\newblock \showarticletitle{FNM-Trans: Efficient FPGA-based Transformer Architecture with Full N:M Sparsity}. In \bibinfo{booktitle}{\emph{Proceedings of the 61st ACM/IEEE Design Automation Conference}} (San Francisco, CA, USA) \emph{(\bibinfo{series}{DAC '24})}. \bibinfo{publisher}{Association for Computing Machinery}, \bibinfo{address}{New York, NY, USA}, Article \bibinfo{articleno}{191}, \bibinfo{numpages}{6}~pages.
\newblock
\showISBNx{9798400706011}
\href{https://doi.org/10.1145/3649329.3656497}{doi:\nolinkurl{10.1145/3649329.3656497}}


\bibitem[Zhang et~al\mbox{.}(2024b)]%
        {zhang2024hetrax}
\bibfield{author}{\bibinfo{person}{Yiming Zhang} {et~al\mbox{.}}} \bibinfo{year}{2024}\natexlab{b}.
\newblock \showarticletitle{HeTraX: A 3D Heterogeneous Manycore Architecture for End-to-End Transformer Acceleration}.
\newblock \bibinfo{journal}{\emph{arXiv preprint arXiv:2408.03397}} (\bibinfo{year}{2024}).
\newblock


\end{thebibliography}

\end{document}